\documentclass[11p,a4paper]{article}
\pdfoutput=1
\usepackage{jheppub}
\usepackage[utf8]{inputenc}
\usepackage{latexsym,amsfonts,amsmath,amscd,amssymb,epsf}
\usepackage{hyperref}
\usepackage{soul,xcolor}
\usepackage{natbib}
\usepackage{graphicx}
\usepackage[normalem]{ulem}

\usepackage{bm}
\let\vec=\bm

\newcommand{\beq}{\begin{equation}}
\newcommand{\eeq}{\end{equation}}

\def\td{\mathrm{d}}

\title{Full energy fraction and angular dependence of medium-induced splittings in the large-$N_c$ limit}

\author[a]{Carlota Andres,}
\author[a,b]{Fabio Dominguez}
\emailAdd{carlota.andres@polytechnique.edu}
\emailAdd{fabio-alejandro.dominguez-gonzalez@polytechnique.edu}

\affiliation[a]{CPHT, CNRS, École polytechnique, Institut Polytechnique de Paris,
91120 Palaiseau, France}
\affiliation[b]{Instituto Galego de F\'isica de Altas Enerx\'ias IGFAE, Universidade de Santiago de Compostela, Santiago de Compostela, 15782 (Spain)}
\begin{document}

\abstract{Jets produced in relativistic heavy-ion collisions are modified by their interactions with the quark-gluon plasma (QGP), making jet substructure observables sensitive probes of QGP dynamics. A quantitative description of these modifications requires understanding how the medium affects elementary parton splittings with full dependence on both their energy fraction $z$ and splitting angle $\theta$, beyond the widely used soft emitted-gluon approximation.
Here, we study  medium-induced $1 \to 2$ splittings double-differential in $z$ and $\theta$, with full resummation of multiple scatterings, and show that in the large-$N_c$
limit and under the harmonic oscillator (HO) approximation, all path integrals can be evaluated analytically for any splitting channel, providing a computationally efficient semi-analytical result. We also revisit the semi-hard approximation (SHA), extending it to include leading corrections in inverse powers of the partons energies, which we denote the improved semi-hard approximation (ISHA), and assess its validity through a comparison with the large-$N_c$-HO results.  Our analysis shows that while the SHA is found to be unreliable across most of phase space, even for high-energy emitters, the ISHA provides a robust approximation for splittings where all partons are sufficiently energetic.
}

\maketitle

\section{Introduction}

Relativistic heavy-ion collisions at the Relativistic Heavy Ion Collider (RHIC) and at the Large Hadron Collider (LHC) produce droplets of quark–gluon plasma (QGP), a phase of QCD matter that, at long wavelengths, behaves as a strongly coupled liquid \cite{Gyulassy:2004zy, Busza:2018rrf,ALICE:2022wpn,Harris:2023tti,CMS:2024krd,Arslandok:2023utm}. In addition to creating this hot and dense medium, these violent collisions occasionally generate hard scatterings that give rise to highly energetic jets. As these jets propagate through the  QCD matter, they undergo modifications driven by their interactions with the medium. This phenomenon, known as jet quenching, has become a well-established tool to probe the QGP dynamics \cite{Wang:2025lct,Cao:2020wlm,Apolinario:2022vzg,Cunqueiro:2021wls,Connors:2017ptx}.

In recent years, increasing attention has been devoted to understanding how the medium modifies the internal structure of jets---see, for instance,~\cite{Larkoski:2017jix,Marzani:2019hun,Apolinario:2022vzg,Cunqueiro:2021wls,Connors:2017ptx}. Jet substructure observables are particularly attractive because they are sensitive to the dynamics of the QGP across a wide range of length scales,  and thus may provide insight into the microscopic structure of the medium.  To disentangle the various sources of QGP-driven substructure modifications in heavy-ion collisions, such as medium response \cite{Cao:2020wlm} and color decoherence dynamics \cite{Mehtar-Tani:2010ebp,Mehtar-Tani:2011hma,Casalderrey-Solana:2011ule,Mehtar-Tani:2012mfa,Casalderrey-Solana:2012evi}, it is essential to calculate how the medium affects elementary parton splittings, which form the fundamental building blocks of  parton showers in any perturbative framework. On the theory side, however, a detailed description of such in-medium splittings, including their full energy and angular  dependence, has only recently begun to emerge. 

Earlier jet quenching studies, both at strong \cite{Casalderrey-Solana:2011dxg} and weak coupling \cite{Baier:1996kr,Baier:1996sk,Zakharov:1996fv,Zakharov:1997uu,Gyulassy:2000fs,Gyulassy:2000er,Arnold:2002ja,Guo:2000nz},  focused primarily on  energy loss, as it was both the first predicted signature  of jet quenching \cite{Bjorken:1982tu} and the first to be experimentally observed through the suppression of high-$p_T$ hadrons and di-hadron correlations at RHIC in the early 2000s \cite{PHENIX:2001hpc,STAR:2002svs,STAR:2003pjh}. In this context, most theoretical calculations targeted the energy lost by high-energy partons, usually relegating the angular distribution of the radiation to a secondary role. In perturbative approaches, this early focus on hadron suppression led to a series of standard approximations in the calculations of medium-induced $1 \to 2$ splittings. First of all, partons are assumed to be highly energetic with longitudinal momentum much larger than any transverse scale. Interactions with the QGP are assumed to transfer only transverse momentum, leaving the emitter energy unchanged, and emissions occur in the small-angle/collinear regime. Second, the dominant energy-loss process is soft gluon emission, where the emitted gluons are much softer than the leading parton, which, actually, is the only way in which a leading parton can be unambiguously defined. For a more detailed discussion of these general approximations, we refer the reader to \cite{Casalderrey-Solana:2007knd,Arnold:2009mr,Mehtar-Tani:2013pia,Mehtar-Tani:2025rty}. 

Beyond these general approximations, early calculations relied on additional method-specific simplifications to facilitate the computation of the energy spectrum of the emitted gluon. For instance, some approaches truncated parton–medium interactions to a single scattering at first order in the opacity expansion \cite{Gyulassy:2000fs,Gyulassy:2000er}, others resummed multiple scatterings assuming the transverse momentum transfers follow a gaussian profile (the harmonic oscillator approximation), and yet others assumed an infinitely long medium \cite{Arnold:2002ja}. While significant progress has been made in relaxing many of these approach-specific approximations \cite{Zakharov:2004vm,Ovanesyan:2011kn,Blaizot:2012fh,Apolinario:2014csa,Feal:2018sml,Sievert:2018imd,Mehtar-Tani:2019tvy,Mehtar-Tani:2019ygg,Andres:2020vxs,Andres:2020kfg,Andres:2023jao,Isaksen:2022pkj},  relaxing the standard  high-energy or/and soft- approximations while retaining the full angular dependence remains considerably more challenging. In this manuscript, we focus specifically on relaxing the soft-gluon approximation for $1 \to 2$ in-medium splittings.

While it could be also interesting to relax the high-energy approximation \cite{Sadofyev:2021ohn,Barata:2022krd,Andres:2022ndd,Kuzmin:2023hko}, it remains appropriate for jet studies, as we are primarily interested in the collinear regime and its corresponding medium modifications. In contrast, the soft approximation becomes insufficient when moving beyond energy-loss observables toward jet substructure measurements. Indeed, many modern jet substructure observables, such as groomed observables \cite{Larkoski:2014wba},  are explicitly designed to reduce sensitivity to soft particles. Moving beyond the soft limit has long been known  to be relatively straightforward when one integrates out all transverse dynamics. However, keeping both the energy fraction $z$ and the emission angle $\theta$ of the splitting, is considerably more challenging. To date,  results differential in $z$ and $\theta$ are available within the first-order opacity approximation, where the fast-moving parton interacts with the medium a single time \cite{Ovanesyan:2011kn}. Extending this framework to include several scatterings has proven extremely difficult \cite{Sievert:2019cwq} and does not appear to offer clear advantages compared to approaches that attempt an all-order resummation in the number of scatterings. Within these all-order resummed frameworks, the double-differential splitting cross section was formally derived in \cite{Blaizot:2012fh,Apolinario:2014csa}, but explicit evaluations have so far only been achieved under restrictive approximations, such as the so-called semi-hard approximation \cite{Dominguez:2019ges, Isaksen:2020npj}, which has been used, for instance, in semi-analytical calculations of energy correlator observables \cite{Andres:2022ovj,Andres:2023xwr,Barata:2023bhh,Barata:2024wsu}. A notable exception is the recent calculation in~\cite{Isaksen:2023nlr}, where the medium-induced double-differential splitting function and multiple-scattering resummation was obtained. However, this approach is  currently limited to the $\gamma \to q\bar q$ channel, and due to its numerical cost, results in this channel are limited to static and relatively thin media (with lengths up to 2 fm).

In this manuscript, we work under the BDMPS-Z framework  \cite{Baier:1996kr,Baier:1996sk,Zakharov:1996fv,Zakharov:1997uu}, which is the standard approach to medium-induced radiation accounting for an arbitrary number of interactions \cite{Baier:1996kr,Baier:1996sk,Zakharov:1996fv,Zakharov:1997uu}. Following \cite{Blaizot:2012fh,Apolinario:2014csa,Isaksen:2023nlr}, we derive the splitting spectrum at large-$N_c$ within the harmonic approximation for all partonic channels. In this setup, the path integrals arsing in the resummation of multiple scatterings can be computed analytically, yielding a final expression that involves  only two time integrals, which are straightforward to evaluate numerically.

Beyond the harmonic oscillator approximation, analytical solutions for the double-differential splitting spectrum are currently unknown. In \cite{Dominguez:2019ges,Isaksen:2020npj}, the semi-hard approximation, applicable beyond the harmonic oscillator, was derived. However, its accuracy is not fully established, with results for the $\gamma \to q \bar q $ channel suggesting that it may considerably overestimate the spectrum \cite{Isaksen:2023nlr}. Here, we introduce the improved semi-hard approximation (ISHA), which incorporates the  finite-energy corrections missing in the the semi-hard approximation. To assess its performance, we compute the splitting spectrum, differential in $z$ and in $\theta$, in the harmonic oscillator approximation using the ISHA and compare it to the full large-$N_c$ result, finding remarkable agreement when all partons involved in the splitting process are sufficiently energetic.

This manuscript is organized as follows:  in section~\ref{sec:splittings}, we first introduce the framework for medium-induced double-differential $1 \to 2$ splittings resumming all multiple scatterings. In the same section, we derive the general expression for the double-differential splitting spectrum in the large-$N_c$ limit. In section~\ref{sec:HO}, we present explicit expressions for this spectrum  for all partonic channels within the harmonic approximation. We then introduce the ISHA in section~\ref{sec:ISHA}, where we identify all finite energy corrections and provide formulas valid for arbitrary dipole cross sections. Numerical results, comparing the large-$N_c$, SHA, and ISHA approaches using the the harmonic oscillator dipole cross section, are presented in section~\ref{sec:evaluation}, allowing us to evaluate the validity of the SHA and ISHA across different regions of phase space. Finally, in section~\ref{sec:conclusions}, we discuss the implications of these findings and present our conclusions.

%%%%%%%%%%%%%%%%%%%%%%%%%%%%%%%%%%%%%%%%%%%%%
%%%%%%%%%%%%%%%%%%%%%%%%%%%%%%%%%%%%%%%%%%%%%

\section{In-medium splittings}
\label{sec:splittings}

We start by summarizing the derivation of the in-medium splitting spectrum, both differential in the splitting energy fraction and the splitting angle (or, equivalently, in the relative transverse momentum of the daughter particles) within the BDMPS-Z framework. In this approach  multiple scatterings are resummed by considering the splitting process occurring in the presence of a strong background field.

As discussed above, we adopt the standard BDMPS-Z setup \cite{Baier:1996kr,Baier:1996sk,Zakharov:1996fv,Zakharov:1997uu}, in which the emitter is assumed to be highly energetic, allowing for several simplifying approximations.  First, we assume that the hard process producing the initial parton occurs on a timescale much shorter than any other timescale relevant to the splitting. This justifies treating the hard process as occurring in vacuum and factorizing it from the subsequent in-medium evolution. Second, the longitudinal momenta (defined along the direction of the parent parton) are taken to be much larger than the transverse momenta involved in the process. This separation effectively decouples longitudinal and transverse dynamics and restricts the calculation to the small-angle (DGLAP) regime.

Interactions with the background field are treated in the eikonal limit, where the large longitudinal component of the parton's momenta $p^+$ couples only to the $A^-(x^+,\vec x)$ component of the background field, which is treated as ``static'' being independent of $x^-$.\footnote{Light-cone coordinates are defined as $x^{\pm}\equiv(x^0\pm x^3)/\sqrt{2}$ and $\vec{x}\equiv (x^1,x^2)$.}  Under these conditions, the background field does not transfer any longitudinal momentum to the traversing partons, thereby preserving helicity. The transverse dynamics is then governed by classical Brownian motion, with the light-cone time   $t\equiv x^+$ as the evolution variable.

Observables in this formalism are computed through cross sections calculated in the presence of a fixed background field, which is then averaged over all possible field configurations. A common approach to perform  this averaging assumes that the color charges are uniformly distributed in transverse space and exhibit only non-trivial two-point correlations \cite{McLerran:1993ni}. In terms of the field $A^-$, the  non-trivial averages are given by
\beq
\label{eq:Average}
\left\langle A^{-a}(t_1,\vec{x}_1)A^{-b}(t_2,\vec{x}_2)\right\rangle = n(t_1)\,\delta^{ab}\,\delta(t_2-t_1)\,\gamma(\vec{x}_2-\vec{x}_1,t_1)\,,
\eeq
where  $a,b$ are adjoint color indices, $n$ denotes the linear density of scatterings,  and $\gamma$ is the Fourier transform of the collision cross section. 

This prescription for medium averaging has several important consequences. First, correlations are local in time, which is equivalent to assuming that the duration of a scattering is much shorter than the time between successive scatterings. Second, there is no color flow out of the system, allowing one to restrict the analysis to color-singlet configurations when considering all partons in amplitude and conjugate amplitude. Finally, since the correlations in \eqref{eq:Average} depend only on the difference of the transverse positions, the total transverse momentum  is conserved.

\subsection{In-medium propagators}

Within this framework, the effect of the background field on any high-energy parton traversing the medium can be encoded in an in-medium propagator, which describes the time evolution of the parton's  transverse dynamics at fixed light-cone energy $E\equiv p^+$.  For a parton in representation $R$, the propagator in transverse coordinate space reads \cite{Casalderrey-Solana:2007knd}:
\beq
{\cal G}_R(\vec{x}_2,t_2;\vec{x}_1,t_1;E) = \int_{\vec{r}(t_1)=\vec{x}_1}^{\vec{r}(t_2)=\vec{x}_2} {\cal D}\vec{r}\,\exp\left\{\frac{iE}{2}\int_{t_1}^{t_2}\td s\,\dot{\vec{r}}^2(s)\right\}\,U_R(t_2,t_1;[\vec{r}])\,,
\label{eq:prop}
\eeq
where the multiple interactions with the background field are resummed into the Wilson line
\beq
U_R(t_2,t_1;[\vec{r}]) = {\cal P}\exp\left\{ig_s\int_{t_1}^{t_2} \td s\, A^{-a}(s,\vec{r}(s))\,T^a_R\right\}\,,\label{eq:Wilson}
\eeq
with $g_s$ the strong coupling constant and $T^a_R$  the generators of $SU(N_c)$ in the representation $R$. 

These propagators satisfy convolution properties that allows them to be sliced and combined when needed (see, for instance, \cite{Casalderrey-Solana:2007knd}). In particular, for $t_1 < t_2 < t_3$, we have 
\beq
\label{eq:convol}
{\cal G}_R(\vec{x}_3,t_3;\vec{x}_1,t_1;E) = \int \td^2\vec{x}_2\,{\cal G}_R(\vec{x}_3,t_3;\vec{x}_2,t_2;E)\,{\cal G}_R(\vec{x}_2,t_2;\vec{x}_1,t_1;E)\,,
\eeq
and an analogous relation also holds when the propagators are Fourier transformed into transverse momentum space.

In the absence of a background field, $A^-=0$, the in-medium propagator in eq.~\eqref{eq:prop} reduces to the vacuum propagator in  two-dimensional transverse space, which in coordinate space reads
\beq
{\cal G}_0(\vec{x}_2,t_2;\vec{x}_1,t_1;E) = \frac{E}{2\pi i(t_2-t_1)}\, e^{i\frac{E(\vec{x}_2-\vec{x}_1)^2}{2(t_2-t_1)}}\,,
\label{eq:prop0}
\eeq
while in momentum space takes the form
\beq
{\cal G}_0(\vec{p}_2,t_2;\vec{p}_1,t_1;E) = (2\pi)^2\,\delta^{(2)}(\vec{p}_2-\vec{p}_1)\,e^{-i\frac{\vec{p}_2^2}{2E}(t_2-t_1)}\,.
\eeq
For convenience, we  adopt the following notation for the propagator of a final-state parton:
\beq
{\cal G}_R(\vec{p}_2,\infty;\vec{p}_1,t_1;E) =
\begin{cases}
{\cal G}_R(\vec{p}_2,L;\vec{p}_1,t_1;E)\,e^{i\frac{\vec{p}_2^2}{2E}L} & t_1 < L\,,\\
(2\pi)^2\delta^{(2)}(\vec{p}_2-\vec{p}_1)\,e^{i\frac{\vec{p}_2^2}{2E}t_1} & t_1 > L\,,
\end{cases}\label{eq:propinf}
\eeq
where $L$ is the length of the medium, and we have used $A^-(t, \vec{x}) = 0$ for $t>L$.

\subsection{In-medium amplitude}

\begin{figure}
\centering
\includegraphics[scale=0.48]{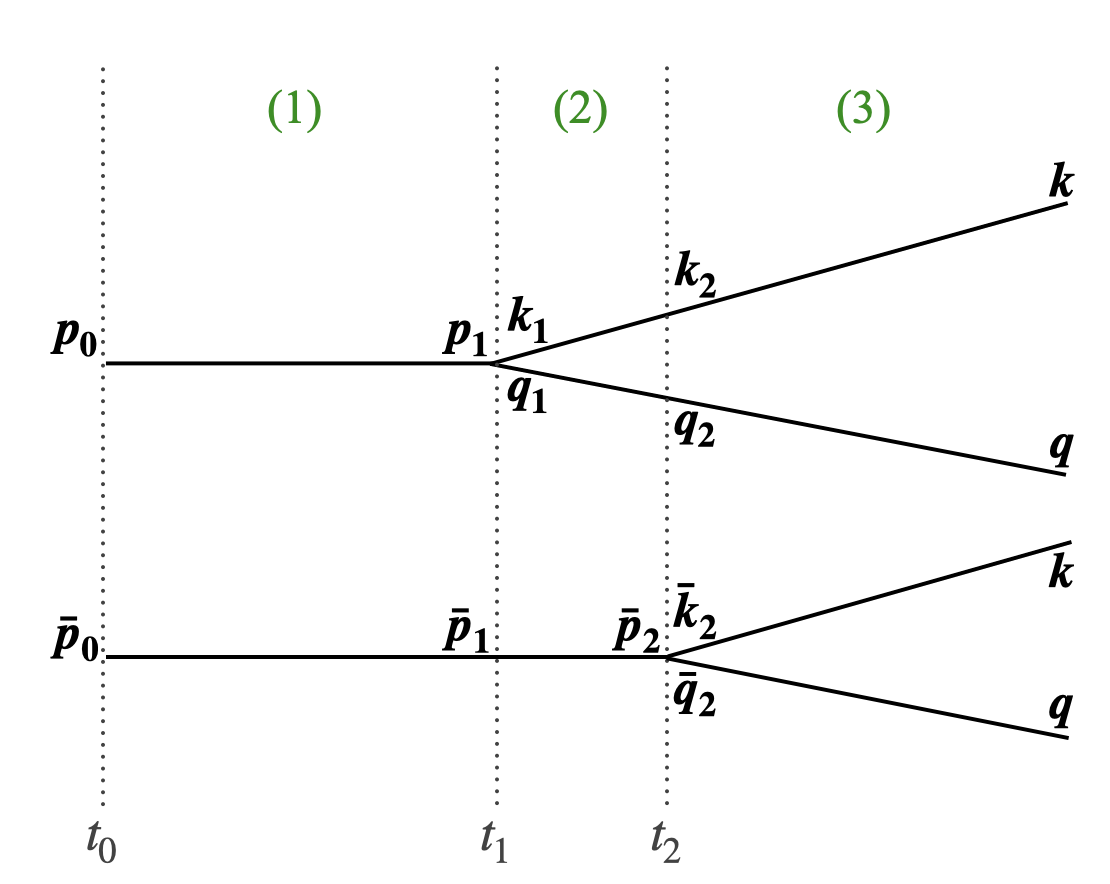}
\caption{Graphical illustration of the transverse momenta flow in $ 1\to 2$ in-medium splittings. Time runs from left to right and the amplitude (upper part) is depicted together with the complex conjugate amplitude (lower part).}
\label{fig:diagram}
\end{figure}

We consider one-to-two in-medium splittings, $a \rightarrow b + c$, where the initial parton is in representation $R_a$ and splits into daughter partons in representations $R_b$ and $R_c$. In terms of the in-medium propagators discussed earlier, the amplitude for this process in transverse momentum space is given by \cite{Blaizot:2012fh,Apolinario:2014csa,Mehtar-Tani:2017ypq}:
\begin{align}
{\cal M}^{\alpha\beta}_{\lambda,\sigma} &= \frac{1}{2E}\int_{\vec{p}_0\vec{p}_1\vec{k}_1\vec{q}_1}\int_{t_0}^\infty \td t_1 \,(2\pi)^2\delta^{(2)}(\vec{p}_1-\vec{k}_1-\vec{q}_1)\,{\cal G}^{\alpha\alpha_1}_{R_b}(\vec{k},\infty;\vec{k}_1,t_1;zE)\nonumber\\
&\quad\times{\cal G}^{\beta\beta_1}_{R_c}(\vec{q},\infty;\vec{q}_1,t_1;(1-z)E)\,V_{\lambda,\sigma,\rho}(\vec{k}_1-z\vec{p}_1,z)\,T^{\alpha_1\beta_1\gamma_1}_{abc}\,{\cal G}^{\gamma_1\gamma}_{R_a}(\vec{p}_1,t_1;\vec{p}_0,t_0;E)\,{\cal M}^{\gamma}_{0\rho}(E,\vec{p}_0)\,,\label{eq:ampl}
\end{align}
where we have adopted the shorthand $\int_{\vec p}=\int d^2 \vec{p}/(2\pi)^2$ for transverse  momentum integrals. Here, $zE,\vec{k}$ and $(1-z)E,\vec{q}$ are respectively the final energy and transverse momenta of the daughter partons; $\alpha,\beta,\dots$ denote color indices; and $\lambda,\sigma,\rho$  refer to polarization indices.  The term ${\cal M}^{\gamma}_{0\rho}(E,\vec{p}_0)$  represents the amplitude for the production of the initial hard parton, which has been factored out. The vertex contributions are included through the explicit color matrix $T_{abc}$ and the vertex factor $V_{\lambda,\sigma,\rho}(\vec{k}_1-z\vec{p}_1,z)$ which can be computed using the  Dirac algebra corresponding to the specific $a \rightarrow b + c$ splitting considered. For any one-to-two splittings, this vertex factor depends on the transverse momenta through the combination $\vec{l} = \vec{k}_1-z\vec{p}_1=(1-z)\vec{k}_1-z\vec{q}_1$. A visual representation of the transverse momentum flow in the amplitude of one-to-two splittings can be seen in the top half of figure~\ref{fig:diagram}.

\subsection{Splitting spectrum}
\label{sec:spec_der}

To compute the spectrum for in-medium splittings, we start by squaring the amplitude given in eq.~\eqref{eq:ampl} and summing over final colors and polarizations. This results in
\begin{align}
\left|{\cal M}^{\alpha\beta}_{\lambda,\sigma}\right|^2 &= \frac{1}{(2E)^2}\text{Re}\int_{\vec{p}_0\vec{p}_1\vec{k}_1\vec{q}_1\bar{\vec{p}}_0\bar{\vec{p}}_2\bar{\vec{q}}_2\bar{\vec{k}}_2}\int_{t_0}^\infty \td t_1\int_{t_1}^\infty \td t_2\,(2\pi)^4\delta^{(2)}(\vec{p}_1-\vec{k}_1-\vec{q}_1)\delta^{(2)}(\bar{\vec{p}}_2-\bar{\vec{k}}_2-\bar{\vec{q}}_2)\nonumber\\
&\quad\times{\cal G}^{\alpha\alpha_1}_{R_b}(\vec{k},\infty;\vec{k}_1,t_1;zE)\,
{\cal G}^{\beta\beta_1}_{R_c}(\vec{q},\infty;\vec{q}_1,t_1;(1-z)E)\,{\cal G}^{\dagger\bar\alpha_2\alpha}_{R_b}(\bar{\vec{k}}_2,t_2;\vec{k},\infty;zE)\nonumber\\
&\quad\times{\cal G}^{\dagger\bar\beta_2\beta}_{R_c}(\bar{\vec{q}}_2,t_2;\vec{q},\infty;(1-z)E)\,V_{\lambda,\sigma,\rho}(\vec{k}_1-z\vec{p}_1,z)\,T^{\alpha_1\beta_1\gamma_1}\,V_{\lambda,\sigma,\bar\rho}(\bar{\vec{k}}_2-z\bar{\vec{p}}_2,z)\,T^{\dagger\bar\alpha_2\bar\beta_2\bar\gamma_2}\nonumber\\
&\quad\times{\cal G}^{\gamma_1\gamma}_{R_a}(\vec{p}_1,t_1;\vec{p}_0,t_0;E)\,{\cal G}^{\dagger\bar\gamma\bar\gamma_2}_{R_a}(\bar{\vec{p}}_0,t_0;\bar{\vec{p}}_2,t_2;E)\,{\cal M}^{\gamma}_{0\rho}(E,\vec{p}_0)\,{\cal M}^{\dagger\bar\gamma}_{0\bar\rho}(E,\bar{\vec{p}}_0)\,.
\label{eq:suared_ampl}
\end{align}
See figure~\ref{fig:diagram} for a visualization of all transverse momenta.

We now proceed by taking the medium average of this expression. Using the properties of the medium averages introduced in \eqref{eq:Average}, we can simplify the color and momentum structure, as previously shown in \cite{Blaizot:2012fh}. First, we exploit the locality in time of the medium averages, in combination  with the convolution property of the in-medium propagators from \eqref{eq:convol}, to divide the calculation into three regions corresponding to two, three, and four propagators --- labelled as regions (1), (2), and (3) in figure~\ref{fig:diagram}, respectively. Next, we simplify the color algebra by noting that, at all times, the partons must form an overall color singlet state. This constraint allows us to project into the relevant singlet space, a process that is straightforward for regions (1) and (2) in figure~\ref{fig:diagram}.

Starting from region (3) in figure~\ref{fig:diagram}, the average of propagators of interest is: 
\begin{align}
&\left\langle{\cal G}^{\alpha\alpha_2}_{R_b}(\vec{k},\infty;\vec{k}_2,t_2;zE)\,
{\cal G}^{\beta\beta_2}_{R_c}(\vec{q},\infty;\vec{q}_2,t_2;(1-z)E)\,{\cal G}^{\dagger\bar\alpha_2\alpha}_{R_b}(\bar{\vec{k}}_2,t_2;\vec{k},\infty;zE)\right.\nonumber\\
&\qquad\left.\times\,{\cal G}^{\dagger\bar\beta_2\beta}_{R_c}(\bar{\vec{q}}_2,t_2;\vec{q},\infty;(1-z)E)\right\rangle T^{\alpha_2\beta_2\bar\gamma_2}T^{\bar\alpha_2\bar\beta_2\bar\gamma_2}\nonumber\\
&\equiv T^{\alpha_2\beta_2\bar\gamma_2}T^{\alpha_2\beta_2\bar\gamma_2}\,(2\pi)^2\delta^{(2)}(\vec{k}_2+\vec{q}_2-\bar{\vec{k}}_2-\bar{\vec{q}}_2)\nonumber\\
&\qquad\times{\cal S}^{(4)}\left((1-z)\vec{k}-z\vec{q},\infty;(1-z)\vec{k}_2-z\vec{q}_2,(1-z)\bar{\vec{k}}_2-z\bar{\vec{q}}_2,t_2;\vec{k}+\vec{q}-\vec{k}_2-\vec{q}_2,z \right)\,,
\label{eq:S4}
\end{align}
where we have used the overall momentum conservation property of the averages to explicitly extract a $\delta$-function from the definition of ${\cal S}^{(4)}$. Additionally, we choose to express the dependence of ${\cal S}^{(4)}$ on the transverse momenta through the relative momenta of the daughter partons, as these  momenta are directly involved in the vertex factors. The dependence of ${\cal S}^{(4)}$ on the total transverse momenta arises only from the total transfer of transverse momentum ($\vec{k}+\vec{q}-\vec{k}_2-\vec{q}_2$ ) due to translational invariance. 

The $\delta$-function appearing in eq.~\eqref{eq:S4} can be combined with one of the $\delta$-functions coming from the vertices in \eqref{eq:suared_ampl}  to set $\bar{\vec{p}}_2= \bar{\vec{k}}_2+ \bar{ \vec{q}}_2= \vec{k}_2+\vec{q}_2$. This allows us to express the medium average in region (2) of figure~\ref{fig:diagram} as follows
\begin{align}
&\left\langle{\cal G}^{\alpha_2\alpha_1}_{R_b}(\vec{k}_2,t_2;\vec{k}_1,t_1;zE)\,{\cal G}^{\beta_2\beta_1}_{R_c}\left(\vec{q}_2,t_2;\vec{q}_1,t_1;(1-z)E\right)\,{\cal G}^{\dagger\gamma_1\bar\gamma_2}_{R_a}(\bar{\vec{p}}_1,t_1;\bar{\vec{p}}_2,t_2;E)\right\rangle T^{\alpha_1\beta_1\gamma_1}T^{\alpha_2\beta_2\bar\gamma_2}\nonumber\\
&\equiv (2\pi)^2\delta^{(2)}(\vec{k}_1+\vec{q}_1-\bar{\vec{p}}_1)\,{\cal K}^{(3)}\left((1-z)\vec{k}_2-z\vec{q}_2,t_2;(1-z)\vec{k}_1-z\vec{q}_1,t_1;\bar{\vec{p}}_2-\bar{\vec{p}}_1,z\right)
\,T^{\alpha_2\beta_2\bar\gamma_2}T^{\alpha_2\beta_2\bar\gamma_2}\,,
\label{eq:K3}
\end{align}
where the same considerations as those for region (3) apply: a momentum-conserving $\delta$-function is explicitly extracted from the definition of ${\cal K}^{(3)}$, and the momentum dependence of ${\cal K}^{(3)}$ is expressed in terms of the relative transverse momenta and the total transverse momentum transfer ($\bar{\vec{p}}_2-\bar{\vec{p}}_1$).

As in the case of region~(3), we now combine the extracted $\delta$-function from \eqref{eq:K3} with the other $\delta$-function arising from the vertices in \eqref{eq:suared_ampl}  to set $\vec{p}_1=\vec{k}_1+\vec{q}_1=\bar{\vec{p}}_1$. We can then write the medium average for region (1) as
\begin{align}
\left\langle\text{Tr}\left[{\cal G}_{R_a}(\vec{p}_1,t_1;\vec{p}_0,t_0;E)\,{\cal G}^{\dagger}_{R_a}(\bar{\vec{p}}_0,t_0;\vec{p}_1,t_1;E)\right]\right\rangle /d_{R_a}= (2\pi)^2\delta^{(2)}(\vec{p}_0-\bar{\vec{p}}_0)\,{\cal P}_{R_a}(\vec{p}_1-\vec{p}_0;t_1,t_0)\,,
\end{align}
where $d_{R_a}$ is the dimension of representation $R_a$.

We can now collect the remaining color and vertex factors in \eqref{eq:suared_ampl}, to obtain \cite{Mehtar-Tani:2017ypq}
\begin{align}
&V_{\lambda,\sigma,\rho}(\vec{k}_1-z\vec{p}_1,z)V_{\lambda,\sigma,\bar\rho}(\bar{\vec{k}}_2-z\bar{\vec{p}}_2,z)\,T^{\alpha_2\beta_2\bar\gamma_2}T^{\alpha_2\beta_2\bar\gamma_2}/d_{R_a}\nonumber\\
&= \,\frac{4g_s^2\delta^{\rho\bar\rho}}{z(1-z)}P_{a\to bc}(z)\,(\vec{k}_1-z\vec{p}_1)\cdot(\bar{\vec{k}}_2-z\bar{\vec{p}}_2)\,.
\end{align}
where $P_{a\to bc}(z)$ is the Altarelli-Parisi splitting function of the $a \to b+c$ splitting \cite{Altarelli:1977zs}.

Putting everything together, we obtain the following expression for the double-differential cross section:
\begin{align}
\frac{\td \sigma}{\td \Omega_k \td \Omega_q} &= \frac{g_s^2}{z(1-z)E^2}P_{a\to bc}(z)\,\text{Re}\int_{\vec{p}_0\vec{p}_1\bar{\vec{p}}_2\vec{k}_1\vec{k}_2\bar{\vec{k}}_2}\int_{t_0}^\infty \td t_1\int_{t_1}^\infty \td t_2\, (\vec{k}_1-z\vec{p}_1)\cdot(\bar{\vec{k}}_2-z\bar{\vec{p}}_2)\nonumber\\
&\quad\times{\cal S}^{(4)}((1-z)\vec{k}-z\vec{q},\infty;\vec{k}_2-z\bar{\vec{p}}_2,\bar{\vec{k}}_2-z\bar{\vec{p}}_2,t_2;\vec{k}+\vec{q}-\bar{\vec{p}}_2,z)\nonumber\\
&\quad\times{\cal K}^{(3)}(\vec{k}_2-z\bar{\vec{p}}_2,t_2;\vec{k}_1-z\vec{p}_1,t_1;\bar{\vec{p}}_2-\vec{p}_1,z)\,{\cal P}_{R_a}(\vec{p}_1-\vec{p}_0;t_1,t_0)\frac{\td \sigma_{\rm hard}}{\td \Omega_{p_0}}\,,
\end{align}
where $\td \Omega_k \equiv (2\pi)^{-3}\td^2\vec{k}\td k^+/(2k^+)$, $p_0^+\equiv E=k^++q^+$, $z\equiv k^+/E$, and $\td\sigma_{\rm hard}/\td\Omega_{p_0}$ is the cross section for production of the initial hard parton.

We are interested in obtaining the splitting spectrum differential in the energy fraction $z$ and the angle $\theta$ between the outgoing partons. It is then convenient to make a change of momentum variables
\begin{equation}
\vec{l} = (1-z)\vec{k} - z\vec{q}\,, \qquad\qquad \text{and} \qquad\qquad \vec{P} = \vec{k} + \vec{q}\,,
\end{equation}
where the relative transverse momentum $\vec{l}$ is directly related to the splitting angle by
\begin{equation}
\theta = \frac{|\vec{l}|}{z(1-z)E}\,.\label{eq:theta}
\end{equation}
We then proceed to integrate over the total momentum $\vec{P}$. This yields
\begin{align}
E\frac{\td \sigma}{\td z \td E \td^2\vec{l}} &= \frac{g_s^2}{2(2\pi)^3(z(1-z)E)^2}\,P_{a\to bc}(z)\,\text{Re}\int_{\vec{l}_1\vec{l}_2\bar{\vec{l}}_2}\int_{t_0}^\infty \td t_1\int_{t_1}^\infty \td t_2 \,(\vec{l}_1\cdot\bar{\vec{l}}_2)\nonumber\\
&\quad\times\tilde{\cal S}^{(4)}(\vec{l},\infty;\vec{l}_2,\bar{\vec{l}}_2,t_2;z)\,\tilde{\cal K}^{(3)}(\vec{l}_2,t_2;\vec{l}_1,t_1;z)\,E\frac{\td \sigma_{\rm hard}}{\td E}\,,
\label{eq:redsplit}
\end{align}
where
\begin{align}
\tilde{\cal S}^{(4)}(\vec{l},\infty;\vec{l}_2,\bar{\vec{l}}_2,t_2;z) &= \int_{\vec{P}} {\cal S}^{(4)}(\vec{l},\infty;\vec{l}_2,\bar{\vec{l}}_2,t_2;\vec{P},z)\,,\\
\tilde{\cal K}^{(3)}(\vec{l}_2,t_2;\vec{l}_1,t_1;z) &= \int_{\vec{P}} {\cal K}^{(3)}(\vec{l}_2,t_2;\vec{l}_1,t_1;\vec{P},z)\,.
\end{align}
Integrating over the total transverse momentum has the effect of removing the overall broadening of the multi-parton system, as  seen from the disappearance of the initial broadening from $t_0$ to $t_1$ in this final expression.

By removing the dependence on the hard cross section, we obtain the  emission spectrum
\begin{align}
\frac{\td I}{\td z \td^2\vec{l}} &= \frac{\alpha_s}{(2\pi)^2\omega^2}\,P_{a\to bc}(z)\,\text{Re}\int_{\vec{l}_1\vec{l}_2\bar{\vec{l}}_2}\int_{t_0}^\infty \td t_1\int_{t_1}^\infty \td t_2 \,(\vec{l}_1\cdot\bar{\vec{l}}_2)\nonumber\\
&\quad\times\tilde{\cal S}^{(4)}(\vec{l},\infty;\vec{l}_2,\bar{\vec{l}}_2,t_2;z)\,\tilde{\cal K}^{(3)}(\vec{l}_2,t_2;\vec{l}_1,t_1;z)\,.
\label{eq:specsplit}
\end{align}
where $\alpha_s=g^2_s/(4\pi)$ and $\omega=z(1-z)E$.

Since having the time integrations go up to infinity in \eqref{eq:specsplit} is impractical for numerical evaluations, we separate them into regions corresponding to emissions inside and outside of a medium fo finite length $L$. In particular, we distinguish the three following cases: $t_1<t_2<L$, $t_1<L<t_2$, and $L<t_1<t_2$, which we referred to as \textit{in-in}, \textit{in-out}, and \textit{out-out}, respectively.

Let us  now examine the behavior of $\tilde{\cal S}^{(4)}$ and $\tilde{\cal K}^{(3)}$ in each of these regions, using~\eqref{eq:propinf} to simplify the corresponding expressions. If $t_2>L$ all four propagators  in region (3) of figure~\ref{fig:diagram} reduce to vacuum propagators whose phases cancel pairwise against those of their complex conjugates, leading to
\beq
\tilde{\cal S}^{(4)}(\vec{l},\infty;\vec{l}_2,\bar{\vec{l}}_2,t_2;z) = (2\pi)^4\delta^{(2)}(\vec{l}-\vec{l}_2)\delta^{(2)}(\vec{l}-\bar{\vec{l}}_2)\,.
\eeq
This expression applies to both the in-out and out-out contributions. For the in-in region ($t_2<L$), the propagators are non-trivial, remaining sensitive to the medium only up to its finite extent, such that
\beq
\tilde{\cal S}^{(4)}(\vec{l},\infty;\vec{l}_2,\bar{\vec{l}}_2,t_2;z) = \tilde{\cal S}^{(4)}(\vec{l},L;\vec{l}_2,\bar{\vec{l}}_2,t_2;z)\,.
\eeq

A similar analysis applies to the three-point function $\tilde{\cal K}^{(3)}$. When $t_1>L$ (the out-out region), all propagators become vacuum propagators. However, unlike in the case of $\tilde{\cal S}^{(4)}$ with $t_2>L$, the phases do not cancel; instead, they combine into the following way:
\beq
\tilde{\cal K}^{(3}(\vec{l}_2,t_2;\vec{l}_1,t_1;z) = (2\pi)^2\delta^{(2)}(\vec{l}_2-\vec{l}_1)\,e^{-i\frac{\vec{l}_2^2}{2\omega}(t_2-t_1)}\,.
\eeq
For the in-out term ($t_1<L<t_2$), one obtains the same vacuum phase accumulated from $L$ to $t_2$, multiplied by the in-medium three-point function from $t_1$ to $L$:
\beq
\tilde{\cal K}^{(3}(\vec{l}_2,t_2;\vec{l}_1,t_1;z) = \tilde{\cal K}^{(3}(\vec{l}_2,L;\vec{l}_1,t_1;z)\,e^{-i\frac{\vec{l}_2^2}{2\omega}(t_2-L)}\,.
\eeq
Finally, the in-in term ($t_1<t_2<L$) cannot be further simplified at this stage.

Putting everything together, the expressions for the out-out and in-out contributions simplify considerably. For the out-out term, one finds
\begin{align}
\frac{\td I^{\text{out-out}}}{\td z \td^2\vec{l}} &= \frac{\alpha_s}{(2\pi)^2\omega^2}P_{a\to bc}(z)\,\text{Re}\int_{L}^\infty \td t_1\int_{t_1}^\infty \td t_2 \,\vec{l}^2\,e^{-i\frac{\vec{l}^2}{2\omega}(t_2-t_1)}\,e^{-\epsilon(t_2+t_1)}\nonumber\\
&= \frac{\alpha_s}{2\pi^2\,\vec{l}^2}P_{a\to bc}(z)\,,
\label{eq:outout}
\end{align}
thereby recovering the well-known vacuum spectrum, as expected. Here, we have introduced an adiabatic switching-off factor at infinity and taken the limit $\epsilon\to 0$. 

For the in-out contribution, only the four-point function reduces to a trivial delta structure, and one obtains
\begin{align}
\frac{\td I^{\text{in-out}}}{\td z \td^2\vec{l}} &= \frac{\alpha_s}{(2\pi)^2\omega^2}P_{a\to bc}(z)\,\text{Re}\int_{\vec{l}_1}\int_{t_0}^L \td t_1\int_{L}^\infty \td t_2 \,(\vec{l}_1\cdot\vec{l}) \,\tilde{\cal K}^{(3)}(\vec{l},L;\vec{l}_1,t_1;z)\,e^{-i\frac{\vec{l}^2}{2\omega}(t_2-L)}\nonumber\\
&= -\frac{\alpha_s}{2\pi^2 \omega\,\vec{l}^2}P_{a\to bc}(z)\,\text{Re}\,i\int_{\vec{l}_1}\int_{t_0}^L \td t_1\,(\vec{l}_1\cdot\vec{l})\,\tilde{\cal K}^{(3)}(\vec{l},L;\vec{l}_1,t_1;z)\,. 
\label{eq:inout}
\end{align}

Finally, the in-in term has the same form as \eqref{eq:specsplit}, but with the time integrals restricted to the medium length $L$
\begin{align}
\frac{\td I^{\text{in-in}}}{\td z \td^2\vec{l}} &= \frac{\alpha_s}{(2\pi)^2\omega^2}P_{a\to bc}(z)\,\text{Re}\int_{\vec{l}_1\vec{l}_2\bar{\vec{l}}_2}\int_{t_0}^L \td t_1\int_{t_1}^L \td t_2 \,(\vec{l}_1\cdot\bar{\vec{l}}_2)\nonumber\\
&\quad\times\tilde{\cal S}^{(4)}(\vec{l},L;\vec{l}_2,\bar{\vec{l}}_2,t_2;z)\,\tilde{\cal K}^{(3)}(\vec{l}_2,t_2;\vec{l}_1,t_1;z)\,.
\label{eq:inin}
\end{align}

To proceed with the evaluation of the splitting spectrum, one must therefore compute the three- and four-point functions, defined as medium averages of products of three and four in-medium propagators, respectively.

\subsection{Averages of products of propagators}
\label{sec:averages}

The medium averages of products of propagators are computed by repeatedly applying the average in~\eqref{eq:Average} to all possible pairs of background field insertions. Since the background field enters the propagators through the Wilson lines defined in~\eqref{eq:Wilson}, we first  consider the averages of two, three, and four Wilson lines over arbitrary paths. In particular, we will need the results for products of two and three Wilson lines, which can be obtained as a straightforward generalization to arbitrary transverse paths of  the expressions presented in \cite{Fukushima:2007dy}. For the average of two Wilson lines, we have
\begin{align}
\left\langle \text{Tr}\left[U_R(t_2,t_1;[\vec{r}])\,U_R^\dagger(t_1,t_2;[\bar{\vec{r}}])\right]\right\rangle/d_R &= \exp\left\{-C_R\int_{t_1}^{t_2}\td s\,n(s)\,\sigma\left(\vec{r}(s)-\bar{\vec{r}}(s)\right)\right\}\nonumber\\
&\equiv \mathcal{C}^{(2)}_R(t_2,t_1;[\vec{r}-\bar{\vec{r}}])
\,,
\label{eq:2Us}
\end{align}
where $\sigma(\vec{x})\equiv g^2(\gamma(0)-\gamma(\vec{x}))$ is the dipole cross section, with $\gamma$ defined in \eqref{eq:Average}. For the average of three Wilson lines, we obtain 
\begin{align}
&\left\langle U^{\alpha_2\alpha_1}_{R_b}(t_2,t_1;[\vec{r}_b])\,U^{\beta_2\beta_1}_{R_c}(t_2,t_1;[\vec{r}_c])\,U^{\dagger\gamma_1\bar\gamma_2}_{R_a}(t_1,t_2;[\bar{\vec{r}}_a])\right\rangle T^{\alpha_1\beta_1\gamma_1}T^{\alpha_2\beta_2\bar\gamma_2}/\left(T^{\alpha\beta\gamma}T^{\alpha\beta\gamma}\right) \nonumber\\
&= \exp\left\{-\frac{1}{2}\int_{t_1}^{t_2}\td s\,n(s)\big[C_{bca}\,\sigma(\vec{r}_b(s)-\vec{r}_c(s))+C_{abc}\,\sigma(\vec{r}_b(s)-\bar{\vec{r}}_a(s))+C_{cab}\,\sigma\left(\vec{r}_c(s)-\bar{\vec{r}}_a(s) \right)\big]\right\}\nonumber\\
&\equiv \mathcal{C}^{(3)}(t_2,t_1;[\vec{r}_c-\vec{r}_b,(1-z)\vec{r}_b+z\vec{r}_c-\bar{\vec{r}}_a])\,,
\label{eq:3Us}
\end{align}
where $C_{abc}=C_{R_a}+C_{R_b}-C_{R_c}$. We have explicitly used that the dependence on the trajectories enters only through their differences, allowing us to write $C^{(3)}$ as a function of only two independent combinations,
\begin{align}
\vec{u} &= \vec{r}_c - \vec{r}_b\,,\\
\vec{v} &= (1-z)\vec{r}_b + z\vec{r}_c - \bar{\vec{r}}_a\,.
\end{align}
From \eqref{eq:3Us}, the corresponding 3-point function can be calculated. Following the steps outlined in Appendix B.2 of \cite{Blaizot:2012fh}, we obtain
\begin{align}
{\cal K}^{(3)}(\vec{l}_2,t_2;\vec{l}_1,t_1;\vec{P},z) &= \int \td^2\vec{u}_1\, \td^2\vec{u}_2\,\td^2\vec{v}\,e^{-i\vec{l}_2\cdot\vec{u}_2+i\vec{l}_1\cdot\vec{u}_1-i\vec{P}\cdot\vec{v}}\int_{\vec{u}(t_1)=\vec{u}_1}^{\vec{u}(t_2)=\vec{u}_2}{\cal D}\vec{u}\nonumber\\
&\qquad\times\,\exp\left\{\frac{i\omega}{2}\int_{t_1}^{t_2}\td s\,\dot{\vec{u}}^2(s)\right\}\,\mathcal{C}^{(3)}(t_2,t_1;[\vec{u},\vec{v}])\,,
\end{align}
where $\vec{v}$ is now constant. Integrating over $\vec P$ enforces $\vec{v}=0$, yielding
\begin{align}
\tilde{\cal K}^{(3)}(\vec{l}_2,t_2;\vec{l}_1,t_1;z) &= \int \td^2\vec{u}_1\,\td^2\vec{u}_2\,e^{-i\vec{l}_2\cdot\vec{u}_2+i\vec{l}_1\cdot\vec{u}_1}\int_{\vec{u}(t_1)=\vec{u}_1}^{\vec{u}(t_2)=\vec{u}_2}{\cal D}\vec{u}\,\exp\left\{\frac{i\omega}{2}\int_{t_1}^{t_2}\td s\,\dot{\vec{u}}^2(s)\right\}\nonumber\\
&\qquad\times\,\mathcal{C}^{(3)}(t_2,t_1;[\vec{u},\vec{0}])\,.
\label{eq:K3tilde}
\end{align}
To proceed beyond this compact expression and evaluate the remaining path integral, it is necessary to specify the dipole cross section $\sigma$ or to introduce further approximations. These scenarios will be explored in sections~\ref{sec:HO} and \ref{sec:ISHA}.

Next, we consider the following average of four Wilson lines:
\begin{align}
&\left\langle
U^{\alpha\alpha_2}_{R_b}(L,t_2;[\vec{r}_b])\,
U^{\beta\beta_2}_{R_c}(L,t_2;[\vec{r}_c])\,
U^{\dagger\bar\alpha_2\alpha}_{R_b}(t_2,L;[\bar{\vec{r}}_b])\,
U^{\dagger\bar\beta_2\beta}_{R_c}(t_2,L;[\bar{\vec{r}}_c])
\right\rangle
\nonumber\\
&\quad\times T^{\alpha_2\beta_2\bar\gamma_2}
T^{\bar\alpha_2\bar\beta_2\bar\gamma_2}
/\left(T^{\alpha_1\beta_1\gamma_1}
T^{\alpha_1\beta_1\gamma_1}\right)
\nonumber\\
&\equiv
\mathcal{C}^{(4)}(L,t_2;[\vec{r}_c-\vec{r}_b,
\bar{\vec{r}}_c-\bar{\vec{r}}_b,
(1-z)(\vec{r}_b-\bar{\vec{r}}_b)
+z(\vec{r}_c-\bar{\vec{r}}_c)])\,.
\end{align}
where, as before, we have used the fact that the average depends only on differences of trajectories rather than on the trajectories themselves. This motivates the introduction of the following set of variables:
\begin{align}
\vec{u} &= \vec{r}_c - \vec{r}_b\,, \\
\bar{\vec{u}} &= \bar{\vec{r}}_c - \bar{\vec{r}}_b\,, \\
\vec{v} &= (1-z)(\vec{r}_b-\bar{\vec{r}}_b) + z(\vec{r}_c-\bar{\vec{r}}_c)\,.
\end{align}
Following appendix B.3 of \cite{Blaizot:2012fh}, the relevant 4-point function can be written as
\begin{align}
&{\cal S}^{(4)}(\vec{l},L;\vec{l}_2,\bar{\vec{l}}_2,t_2;\vec{P},z) = \int \td^2\vec{v}\,\td^2\vec{u}_L \,\td^2\vec{u}_2\, \td^2\bar{\vec{u}}_L\, \td^2\bar{\vec{u}}_2\,e^{-i\vec{v}\cdot\vec{P}-i(\vec{u}_L-\bar{\vec{u}}_L)\cdot\vec{l}+i\vec{u}_2\cdot\vec{l}_2-i\bar{\vec{u}}_2\cdot\bar{\vec{l}}_2} \int_{\vec{u}(t_2)=\vec{u}_2}^{\vec{u}(L)=\vec{u}_L}{\cal D}\vec{u}\nonumber\\
&\qquad \times\int_{\bar{\vec{u}}(t_2)=\bar{\vec{u}}_2}^{\bar{\vec{u}}(L)=\bar{\vec{u}}_L}{\cal D}\bar{\vec{u}}\,\exp\left\{\frac{i\omega}{2}\int_{t_2}^{L}\td s\,\left(\dot{\vec{u}}^2(s)-\dot{\bar{\vec{u}}}^2(s)\right)\right\}\, \mathcal{C}^{(4)}(L,t_2;[\vec{u},\bar{\vec{u}},\vec{v}])\,,
\label{eq:S4_aux}
\end{align}
where $\vec{v}$ no longer represents a trajectory, but rather a fixed transverse coordinate conjugate to $\vec{P}$. Upon integrating eq.~\eqref{eq:S4_aux} over $\vec{P}$, $\vec{v}$ is set to zero, yielding
\begin{align}
&\tilde{\cal S}^{(4)}(\vec{l},L;\vec{l}_2,\bar{\vec{l}}_2,t_2;z) = \int \td^2\vec{u}_L\, \td^2\vec{u}_2 \,\td^2\bar{\vec{u}}_L \,\td^2\bar{\vec{u}}_2\,e^{-i(\vec{u}_L-\bar{\vec{u}}_L)\cdot\vec{l}+i\vec{u}_2\cdot\vec{l}_2-i\bar{\vec{u}}_2\cdot\bar{\vec{l}}_2}\int_{\vec{u}(t_2)=\vec{u}_2}^{\vec{u}(L)=\vec{u}_L}{\cal D}\vec{u}\nonumber\\
&\qquad \times\,\int_{\bar{\vec{u}}(t_2)=\bar{\vec{u}}_2}^{\bar{\vec{u}}(L)=\bar{\vec{u}}_L}{\cal D}\bar{\vec{u}}\, \exp\left\{\frac{i\omega}{2}\int_{t_2}^{L}\td s\,\left(\dot{\vec{u}}^2(s)-\dot{\bar{\vec{u}}}^2(s)\right)\right\}\,\mathcal{C}^{(4)}(L,t_2;[\vec{u},\bar{\vec{u}},\vec{0}])\,.
\label{eq:S4v0}
\end{align}

\subsubsection{Large-$N_c$ limit}
\label{subsubsec:largeNc}

The average of 4 Wilson lines $\mathcal{C}^{(4)}$ cannot be directly calculated in terms of the dipole cross section without additional approximations, in contrast to the cases involving two and three Wilson lines. The origin of this difference lies in the underlying color structure. For systems of two and three particles, the color configuration is trivial, as they always form  (and remain) in a unique singlet state. By contrast, a four-particle system admits multiple singlet configurations, as well as transitions between them. The analytical resummation of these transitions becomes impossible when the transverse separations between particles are allowed to vary continuously, as required for the calculation of the splitting spectrum.  Numerical evaluations of such four-point functions are computationally demanding. Some recent progress has been made in the simplest case of $\gamma \to q\bar{q}$, where only two singlet states contribute to the four-point function. However, even in this case, the results are currently limited to a rather restricted region of parameter space. \cite{Isaksen:2023nlr}.

In the following, we will recur to the large-$N_c$ limit, where transitions between different singlet states are suppressed by inverse powers of $N_c$, leading to a significant simplification of the color structure. In this framework, it is convenient to express all adjoint Wilson lines in terms of fundamental ones by repeatedly using the identity
\beq
U_A^{ab}([\vec{r}])=2\,\text{Tr}[U_F^\dagger([\vec{r}]) t^a U_F([\vec{r}]) t^b]\,,
\eeq
and subsequently eliminating explicit color matrices through the Fierz identity.
An additional simplification arises in the large-$N_c$ limit: averages of products of traces factorize into products of averages of individual traces. As a result, the relevant averages of four Wilson lines in any representation can be expressed in terms of dipoles $\mathcal{C}^{(2)}_F$ and quadrupoles \cite{Dominguez:2012ad,Isaksen:2020npj}, the latter corresponding to the trace of four Wilson lines in the fundamental representation. The quadrupole $\mathcal{Q}$ is defined as
\begin{align}
\mathcal{Q}(L,t_2;[\vec{r}_c-\vec{r}_b,\bar{\vec{r}}_c-&\bar{\vec{r}}_b,(1-z)(\vec{r}_b-\bar{\vec{r}}_b)+z(\vec{r}_c-\bar{\vec{r}}_c)])  \equiv  \nonumber \\
&\frac{1}{N_c}  \left\langle\text{Tr}\left[U_F^\dagger(t_2,L;[\bar{\vec{r}}_b])\,U_F(L,t_2;[\vec{r}_b])\,U_F^\dagger(t_2,L;[\vec{r}_c])\,U_F\left(L,t_2;[\bar{\vec{r}}_c]\right)\right]\right\rangle\,,
\label{eq:quadrupole}
\end{align}
where, once again, $\mathcal{Q}$ does depend on the variables $\vec{u}$, $\bar{\vec{u}}$, and $\vec{v}$, rather than on the trajectories themselves.

The evaluation of the average in \eqref{eq:quadrupole} involves only two singlet states. In the large-$N_c$ limit, transitions between these two states are suppressed by inverse powers of $N_c$ with the leading contribution allowing for only one of such transitions. In this limit, the two singlet states can be regarded as  the two possible ways of forming two separate dipoles. This leads to the following expression in terms of the dipole cross section \cite{Blaizot:2012fh}:
\begin{align}
\mathcal{Q}(L,t_2;[\vec{u},\bar{\vec{u}},\vec{v}]) &= \mathcal{C}^{(2)}_F(L,t_2;[\vec{v}-z(\vec{u}-\bar{\vec{u}})])\,\mathcal{C}^{(2)}_F(L,t_2;[\vec{v}+(1-z)(\vec{u}-\bar{\vec{u}})])\nonumber\\
&\quad + \int_{t_2}^L \td t_3\,\mathcal{C}^{(2)}_F(L,t_3;[\vec{v}-z(\vec{u}-\bar{\vec{u}})])\,\mathcal{C}^{(2)}_F(L,t_3;[\vec{v}+(1-z)(\vec{u}-\bar{\vec{u}})])\nonumber\\
&\qquad\times T\left(t_3;\vec{u}(t_3),\bar{\vec{u}}(t_3),\vec{v}(t_3)\right)\,\mathcal{C}^{(2)}_F(t_3,t_2;[\vec{u}])\,\mathcal{C}^{(2)}_F(t_3,t_2;[\bar{\vec{u}}])\,,
\label{eq:quad}
\end{align}
where $T$ denotes the transition factor given by
\begin{align}
T(t_3;\vec{u}_3,\bar{\vec{u}}_3,\vec{v}_3) &=- n(t_3)\frac{N_c}{2}\left[\sigma(\vec{u}_3)+\sigma(\bar{\vec{u}}_3)-\sigma(\vec{v}_3-z\vec{u}_3-(1-z)\bar{\vec{u}}_3)-\sigma(\vec{v}_3+(1-z)\vec{u}_3+z\bar{\vec{u}}_3)\right]\,.
\end{align}
The two contributions in eq.~\eqref{eq:quad} are commonly referred to as the factorizable (first line) and non-factorizable (second and third lines) terms \cite{Blaizot:2012fh}. Although the non-factorizable contribution has been argued to be parametrically suppressed in the final spectrum \cite{Blaizot:2012fh}, we retain both terms and show how they can be evaluated numerically.

The specific averages of four Wilson lines  relevant for the splittings considered here take the following form in the large-$N_c$ limit \cite{Isaksen:2020npj}:
\begin{align}
\mathcal{C}^{(4)}_{\gamma\to q\bar q}(L,t_2;[\vec{u},\bar{\vec{u}},\vec{v}]) &\simeq \mathcal{Q}(L,t_2;[\vec{u},\bar{\vec{u}},\vec{v}])\,,\label{eq:C4ph} \\
\mathcal{C}^{(4)}_{q\to qg}(L,t_2;[\vec{u},\bar{\vec{u}},\vec{v}]) &\simeq \mathcal{Q}(L,t_2;[\vec{u},\bar{\vec{u}},\vec{v}])\, \mathcal{C}^{(2)}_F(L,t_2;[\vec{v}+(1-z)(\vec{u}-\bar{\vec{u}})])\,, \\
\mathcal{C}^{(4)}_{g\to gg}(L,t_2;[\vec{u},\bar{\vec{u}},\vec{v}]) &\simeq \mathcal{Q}(L,t_2;[\vec{u},\bar{\vec{u}},\vec{v}])\,  \mathcal{C}^{(2)}_F(L,t_2;[\vec{v}-z(\vec{u}-\bar{\vec{u}})])\,\mathcal{C}^{(2)}_F(L,t_2;[\vec{v}+(1-z)(\vec{u}-\bar{\vec{u}})])\,, \\
\mathcal{C}^{(4)}_{g\to q\bar{q}}(L,t_2;[\vec{u},\bar{\vec{u}},\vec{v}]) &\simeq \mathcal{C}^{(2)}_F(L,t_2;[\vec{v}-z(\vec{u}-\bar{\vec{u}})])\,\mathcal{C}^{(2)}_F(L,t_2;[\vec{v}+(1-z)(\vec{u}-\bar{\vec{u}})])\,.
\label{eq:C4qqbar}
\end{align}
A noteworthy feature of the $g \to q \bar q$ splitting is that, in the large-$N_c$ limit, it is the only case in which the four-Wilson-line average does not involve a quadrupole. Instead, it can be expressed entirely in terms of dipoles, which can be evaluated directly for any choice of dipole cross section without further approximations. This significantly simplifies its computation compared to the other channels. Its evaluation within the harmonic oscillator approximation was first presented in \cite{Attems:2022ubu}.

In all other cases, eq.~\eqref{eq:quad} can be used  to decompose $C^{(4)}$ into factorizable and non-factorizable contributions (the latter  vanishes for the $g\to q\bar{q}$). These pieces share a similar structure, given by
\beq
\mathcal{C}^{(4)}_{\rm fac} (L,t_2;[\vec{u},\bar{\vec{u}},\vec{v}]) \simeq \left[\mathcal{C}^{(2)}_F(L,t_2;[\vec{v}-z(\vec{u}-\bar{\vec{u}})])\right]^n\,\left[\mathcal{C}^{(2)}_F(L,t_2;[\vec{v}+(1-z)(\vec{u}-\bar{\vec{u}})])\right]^m\,,
\label{eq:C4fac}
\eeq
and
\begin{align}
\mathcal{C}^{(4)}_{\rm non-fac} (L,t_2;[\vec{u},\bar{\vec{u}},\vec{v}]) &\simeq \int_{t_2}^{L}\td t_3\,\left[\mathcal{C}^{(2)}_F(L,t_3;[\vec{v}-z(\vec{u}-\bar{\vec{u}})])\right]^n\,\left[\mathcal{C}^{(2)}_F(L,t_3;[\vec{v}+(1-z)(\vec{u}-\bar{\vec{u}})])\right]^m\nonumber\\
&\quad\times\,T\left(t_3;\vec{u}(t_3),\bar{\vec{u}}(t_3),\vec{v}(t_3)\right)\,\mathcal{C}^{(2)}_F\left(t_3,t_2;[\vec{u}]\right)\,\mathcal{C}^{(2)}_F(t_3,t_2;[\bar{\vec{u}}])\nonumber\\
&\quad\times\left[\mathcal{C}^{(2)}_F\left(t_3,t_2;[\vec{v}-z(\vec{u}-\bar{\vec{u}})]\right)\right]^{n-1}\,\left[\mathcal{C}^{(2)}_F\left(t_3,t_2;[\vec{v}+(1-z)(\vec{u}-\bar{\vec{u}})]\right)\right]^{m-1}\,,
\label{eq:C4nonfac}
\end{align}
where the values of $(n,m)$ for each splitting process are can be directly read off eqs.~\eqref{eq:C4ph}-\eqref{eq:C4qqbar}, and are summarized  in table~\ref{tab:nm_values}.
\begin{table}[ht]
\centering
\begin{tabular}{lcc}
\hline
Splitting & $n$ & $m$ \\
\hline
$\gamma \to q\bar{q}$ & 1 & 1 \\
$q \to qg$ & 1 & 2 \\
$g \to gg$ & 2 & 2 \\
$g \to q\bar{q}$ & 1 & 1 \\
\hline
\end{tabular}
\caption{Values of $n$ and $m$ in eqs.~\eqref{eq:C4fac} and\eqref{eq:C4nonfac} for the different splitting processes. The last row applies only to eq.~\eqref{eq:C4fac} since in that case the non-factorizable part is zero.}
\label{tab:nm_values}
\end{table}

Even though ${\cal C}^{(2)}$ and ${\cal C}^{(3)}$ can be evaluated explicitly for finite $N_c$, we present here the result in the large-$N_c$ limit for consistency
\begin{equation}
{\cal C}^{(2)}(t_2,t_1;[\vec{r}]) \simeq \exp\left\{-\frac{N_c}{2}\int_{t_1}^{t_2}\td s\,n(s)\,\sigma(\vec{r}(s))\right\}\,,\label{eq:C2largeNc}
\end{equation}
\begin{equation}
{\cal C}^{(3)}(t_2,t_1;[\vec{u},\vec{v}]) \simeq \left[{\cal C}_F^{(2)}(t_2,t_1;[\vec{u}])\right]^p\,\left[{\cal C}_F^{(2)}(t_2,t_1;[\vec{v}-z\vec{u}])\right]^n\,\left[{\cal C}_F^{(2)}(t_2,t_1;[\vec{v}+(1-z)\vec{u}])\right]^m\,,\label{eq:C3largeNc}
\end{equation}
with $p=0$ for the $g\to q\bar{q}$ case and $p=1$ for all other splittings

We can now return to the computation of ${\cal S}^{(4)}$, separating it into factorizable and non-factorizable contributions. The factorizable part is obtained by substituting~\eqref{eq:C4fac} into~\eqref{eq:S4v0}:
\begin{align}
\tilde{\cal S}^{(4)}_{\rm fac}(\vec{l},L;\vec{l}_2,\bar{\vec{l}}_2,t_2;z) &= \int \td^2\vec{u}_L\, \td^2\vec{u}_2 \,\td^2\bar{\vec{u}}_L \,\td^2\bar{\vec{u}}_2\,e^{-i(\vec{u}_L-\bar{\vec{u}}_L)\cdot\vec{l}+i\vec{u}_2\cdot\vec{l}_2-i\bar{\vec{u}}_2\cdot\bar{\vec{l}}_2}\nonumber\\
&\quad \times\,\int_{\vec{u}(t_2)=\vec{u}_2}^{\vec{u}(L)=\vec{u}_L}{\cal D}\vec{u}\int_{\bar{\vec{u}}(t_2)=\bar{\vec{u}}_2}^{\bar{\vec{u}}(L)=\bar{\vec{u}}_L}{\cal D}\bar{\vec{u}} \,\exp\left\{\frac{i\omega}{2}\int_{t_2}^{L}\td s\,\left(\dot{\vec{u}}^2(s)-\dot{\bar{\vec{u}}}^2(s)\right)\right\}\nonumber\\
&\qquad\times\,\left[\mathcal{C}^{(2)}_F(L,t_2;[-z(\vec{u}-\bar{\vec{u}})])\right]^n\,\left[\mathcal{C}^{(2)}_F(L,t_2;[(1-z)(\vec{u}-\bar{\vec{u}})])\right]^m\,.
\label{eq:S4intfac}
\end{align}
Since the integrand depends only on the difference $\vec{u}-\bar{\vec{u}}$, the path integrals can be analytically evaluated by performing an appropriate change of variables, yielding
\begin{align}
\tilde{\cal S}^{(4)}_{\rm fac}(\vec{l},L;\vec{l}_2,\bar{\vec{l}}_2,t_2;z) &= (2\pi)^2\delta^{(2)}(\vec{l}_2-\bar{\vec{l}}_2) \int d^2\vec{w}_2 \, e^{-i\vec{w}_2\cdot(\vec{l}-\vec{l}_2)}\,\left[\mathcal{C}_F^{(2)}(L,t_2;[z\vec{w}_2])\right]^n\left[\mathcal{C}_F^{(2)}(L,t_2;[(1-z)\vec{w}_2])\right]^m\nonumber\\
&\equiv (2\pi)^2\delta^{(2)}(\vec{l}_2-\bar{\vec{l}}_2) \,{\cal P}_{\rm eff}(L,t_2;\vec{l}-\vec{l}_2)\,,
\label{eq:S4fac}
\end{align}
where we have introduced the effective broadening factor ${\cal P}_{\rm eff}$.

Similarly, the non-factorizable contribution is obtained by replacing~\eqref{eq:C4nonfac} into eq.~\eqref{eq:S4v0}
\begin{align}
&\tilde{\cal S}^{(4)}_{\rm non\text{-}fac}(\vec{l},L;\vec{l}_2,\bar{\vec{l}}_2,t_2;z) = \int_{\vec{l}_3}\int_{t_2}^L \td t_3\,{\cal P}_{\rm eff}(L,t_3;\vec{l}-\vec{l}_3)\int \td^2\vec{u}_3 \,\td^2\bar{\vec{u}}_3\, \td^2\vec{u}_2\, \td^2\bar{\vec{u}}_2\,e^{-i(\vec{u}_3-\bar{\vec{u}}_3)\cdot\vec{l}_3+i\vec{u}_2\cdot\vec{l}_2-i\bar{\vec{u}}_2\cdot\bar{\vec{l}}_2}\nonumber\\
&\qquad\qquad\times T(t_3;\vec{u}_3,\bar{\vec{u}}_3,z)\int_{\vec{u}(t_2)=\vec{u}_2}^{\vec{u}(t_3)=\vec{u}_3}{\cal D}\vec{u}\int_{\bar{\vec{u}}(t_2)=\bar{\vec{u}}_2}^{\bar{\vec{u}}(t_3)=\bar{\vec{u}}_3}{\cal D}\bar{\vec{u}}\,\exp\left\{\frac{i\omega}{2}\int_{t_2}^{t_3}\td s\,(\dot{\vec{u}}^2(s)-\dot{\bar{\vec{u}}}^2(s))\right\}\nonumber\\
&\qquad\qquad\times \mathcal{C}_F^{(2)}(t_3,t_2;[\vec{u}])\,\mathcal{C}_F^{(2)}(t_3,t_2;[\bar{\vec{u}}])\,\left[\mathcal{C}_F^{(2)}(t_3,t_2;[-z(\vec{u}-\bar{\vec{u}})])\right]^{n-1}\left[\mathcal{C}_F^{(2)}(t_3,t_2;[(1-z)(\vec{u}-\bar{\vec{u}})])\right]^{m-1}\,.
\label{eq:S4nonfac}
\end{align}

Further simplification of these expressions will require additional approximations or the adoption of specific models for the interaction with the medium, which will be discussed in the following section.

%%%%%%%%%%%%%%%%%%%%%%%%%%%%%%%%%%%%%%%%%%%%%
%%%%%%%%%%%%%%%%%%%%%%%%%%%%%%%%%%%%%%%%%%%%%

\section{Analytical evaluation in the harmonic approximation}
\label{sec:HO}

The main challenge in evaluating the 3-point and 4-point lies in the explicit computation of the path integrals appearing in $\mathcal{K}^{(3)}$ (see~\eqref{eq:K3tilde}) and in the non-factorizable component of $\mathcal{S}^{(4)}$ \eqref{eq:S4nonfac}, for which analytical solutions are generally hard to obtain. The notable exception arises when the integrals are Gaussian, corresponding to the so-called harmonic oscillator (HO) approximation. In this case, the dipole cross section becomes quadratic in coordinate space
\beq\label{eq:sigmaho}
N_c \,n(t)\sigma(\vec{r}) = \frac{1}{2}\hat q(t)\,\vec{r}^2\,,
\eeq
which describes a regime of multiple soft interactions between the probe and the medium.

In this section, we show that for all one-to-two splittings, the double-differential splitting spectrum at large $N_c$ derived in the previous section, becomes (semi)-analytical under the HO approximation. In particular, all remaining path integrals in eqs.~\eqref{eq:K3tilde}~and~\eqref{eq:S4nonfac} can be evaluated analytically.

The general solution for the in-medium energy spectrum (integrated in the transverse plane) in the HO approximation for an arbitrarily varying $\hat q$ was first obtained in \cite{Arnold:2008iy}. Here we follow the methods developed in \cite{Blaizot:2012fh,Apolinario:2014csa} to generalize these results to the splitting spectrum with full angle dependence. As a starting point, we adopt the presentation in section 3.2 of \cite{Mehtar-Tani:2019tvy} where it was shown that the following path integral
\beq
{\cal K}_\Omega(t_2,\vec{x}_2;t_1,\vec{x}_1) = \int_{\vec{r}(t_1)=\vec{x}_1}^{\vec{r}(t_2)=\vec{x}_2}{\cal D}\vec{r}\,\exp\left\{\frac{i\omega}{2}\int_{t_1}^{t_2}\td s\,\left(\dot{\vec{r}}^2(s)-\Omega^2(s)\vec{r}^2(s)\right)\right\}\,
\eeq
can be written as
\beq
{\cal K}_\Omega(t_2,\vec{x}_2;t_1,\vec{x}_1) = \frac{i\omega}{2\pi S_\Omega(t_2,t_1)}\exp\left\{\frac{i\omega}{2S_\Omega(t_2,t_1)}\big(C_\Omega(t_1,t_2)\vec{x}_2^2+C_\Omega(t_2,t_1)\vec{x}_1^2-2\vec{x}_1\cdot\vec{x}_2\big)\right\}\,,
\label{eq:pathintHO}
\eeq
where $S_{\Omega}$ satisfies
\beq
\partial_t^2 S_\Omega(t,t') + \Omega^2(t)S_\Omega(t,t') = 0\,,\label{eq:eqdiffho}
\eeq
with initial conditions $S_\Omega(t,t)=0$ and $\left.\partial_t S_\Omega(t,t')\right|_{t=t'}=1$. The function
\beq
C_\Omega(t,t')  =-\partial_{t'}S_\Omega(t,t')\,,
\eeq
is also a solution to \eqref{eq:eqdiffho}, satisfying $C_\Omega(t,t)=1$, and $\left.\partial_t C_\Omega(t,t')\right|_{t=t'}=0$. 

In the static case, where $\Omega(t)=\Omega_0$ is a constant, we get
\beq
S_{\Omega_0}(t,t')=\frac{1}{\Omega_0}\sin\left(\Omega_0(t-t')\right)\,,\qquad C_{\Omega_0}(t,t') = \cos\left(\Omega_0(t-t')\right)\,.
\eeq
In the general case $S_{\Omega}$ remains antisymmetric under $t \leftrightarrow t'$, while $C_{\Omega}$ is not necessarily symmetric (see~\cite{Arnold:2008iy}).

We can readily take the Fourier transform of eq.~\eqref{eq:pathintHO} to express it in transverse momentum space as
\beq
{\cal K}_\Omega(t_2,\vec{p}_2;t_1,\vec{p}_1) = \frac{2\pi i}{\omega \partial_{t_2}C_\Omega(t_2,t_1)}\exp\left\{-\frac{i}{2\omega \partial_{t_2}C_\Omega(t_2,t_1)}\left(C_\Omega(t_1,t_2)\vec{p}_1^2+C_\Omega(t_2,t_1)\vec{p}_2^2-2\vec{p}_1\cdot\vec{p}_2\right)\right\}\,,\label{eq:KOmmom}
\eeq
where we have used the fact that the Wronskian of the two independent solutions is constant,
\beq
\partial_{t_2}S_\Omega(t_2,t_1)C_\Omega(t_2,t_1)-S_\Omega(t_2,t_1)\partial_{t_2}C_\Omega(t_2,t_1) = C_\Omega(t_1,t_2)C_\Omega(t_2,t_1)-S_\Omega(t_2,t_1)\partial_{t_2}C_\Omega(t_2,t_1) = 1\,.
\eeq

We now return to the evaluation of the splitting spectrum and first consider the 3-point function in the HO approximation.
Within this approximation,~\eqref{eq:K3tilde} becomes
\begin{align}
\tilde{\cal K}^{(3)}(\vec{l}_2,t_2;\vec{l}_1,t_1;z) &= \int \td^2\vec{u}_1\,\td^2\vec{u}_2\,e^{-i\vec{l}_2\cdot\vec{u}_2+i\vec{l}_1\cdot\vec{u}_1}\int_{\vec{u}(t_1)=\vec{u}_1}^{\vec{u}(t_2)=\vec{u}_2}{\cal D}\vec{u}\,\exp\left\{\frac{i\omega}{2}\int_{t_1}^{t_2}\td s\,\dot{\vec{u}}^2(s)\right.\nonumber\\
&\quad\left.-\frac{1}{4}\int_{t_1}^{t_2}\td s\,\hat q(s)\frac{1}{N_c}\left(C_{bca}+C_{abc}z^2+C_{cab}(1-z)^2\right)\vec{u}^2(s)\right\}\nonumber\\
&= {\cal K}_{\Omega_1}(t_2,\vec{l}_2;t_1,\vec{l}_1)\,,
\end{align}
where $\Omega_1^2(s)=-\frac{i\hat q(s)}{2\omega}\frac{1}{N_c}(C_{bca}+C_{abc}z^2+C_{cab}(1-z)^2)$. We emphasize that in this expression all path integrals have been analytically evaluated. This result allows us to perform the remaining integrals in the in-out contribution to the spectrum given in \eqref{eq:inout}, which give
\begin{align}
\int_{t_0}^L \td t_1\int_{\vec{l}_1}(\vec{l}\cdot\vec{l}_1)\,{\cal K}_{\Omega_1}(\vec{l},L;\vec{l}_1,t_1) &= \int_{t_0}^L \td t_1 \frac{\vec{l}^2}{C_{\Omega_1}^2(t_1,L)}\exp\left\{-\frac{iS_{\Omega_1}(L,t_1)}{2\omega C_{\Omega_1}(t_1,L)}\vec{l}^2\right\}\nonumber\\
&= -2i\omega\left(1-\exp\left\{-\frac{iS_{\Omega_1}(L,t_0)}{2\omega C_{\Omega_1}(t_0,L)}\vec{l}^2\right\}\right)\,,\label{eq:intt1}
\end{align}
where we have used 
\beq
\partial_{t_1}\left(\frac{S_{\Omega_1}(L,t_1)}{C_{\Omega_1}(t_1,L)}\right)=-\frac{1}{C_{\Omega_1}^{ 2}(t_1,L)}\,,
\eeq
which follows from the Wronskian relation and the antisymmetry of $S_{\Omega_1}$. The in-out contribution to the splitting spectrum in the HO approximation is thus
\begin{align}
\frac{\td I^{\text{in-out}}_{\rm HO}}{\td z \td^2\vec{l}} &= -\frac{\alpha_s}{\pi^2 \vec{l}^2}P_{a\to bc}(z)\,\text{Re}\left(1-\exp\left\{-\frac{iS_{\Omega_1}(L,t_0)}{2\omega C_{\Omega_1}(t_0,L)}\vec{l}^2\right\}\right)\,. 
\label{eq:inoutho}
\end{align}

Writing explicit expressions for the in-in contribution requires more work, particularly for the non-factorizable part of the 4-point function. We begin with the factorizable contribution, for which the 4-point function can be expressed in terms of an effective broadening factor, as shown in~\eqref{eq:S4fac}. In the HO approximation, the broadening factor takes the form
\beq
{\cal P}_{\rm eff}(L,t_2;\vec{p}) = \frac{4\pi}{Q_{\rm eff}^2(L,t_2)}\exp\left\{-\frac{\vec{p}^2}{Q_{\rm eff}^2(L,t_2)}\right\}\,,
\label{eq:Peff}
\eeq
where $Q_{\rm eff}^2(t',t)=\int_t^{t'}\td s\,(nz^2+m(1-z)^2)\,\hat q(s)$. Using this, the contribution from the factorizable piece to the in-in term of the splitting spectrum in the HO approximation is
\begin{align}
\frac{\td I^{\text{in-in}}_{\rm fac, HO}}{\td z \td^2\vec{l}} &= \frac{\alpha_s}{(2\pi)^2\omega^2}P_{a\to bc}(z)\,\text{Re}\int_{\vec{l}_1\vec{l}_2}\int_{t_0}^L \td t_1\int_{t_1}^L \td t_2 \,(\vec{l}_1\cdot\vec{l}_2)\,{\cal P}_{\rm eff}(L,t_2;\vec{l}-\vec{l}_2)\,{\cal K}_{\Omega_1}(\vec{l}_2,t_2;\vec{l}_1,t_1;z)\nonumber\\
&= \frac{\alpha_s}{2\pi^2\omega}P_{a\to bc}(z)\,\text{Re}\,i\int_{\vec{l}_2}\int_{t_0}^L \td t_2\,{\cal P}_{\rm eff}(L,t_2;\vec{l}-\vec{l}_2)\exp\left\{-\frac{iS_{\Omega_1}(t_2,t_0)}{2\omega C_{\Omega_1}(t_0,t_2)}\vec{l}_2^2\right\}\nonumber\\
&= \frac{\alpha_s}{\pi^2}P_{a\to bc}(z)\,\text{Re}\int_{t_0}^L \td t_2\,\frac{iC_{\Omega_1}(t_0,t_2)}{2\omega C_{\Omega_1}(t_0,t_2)+iQ_{\rm eff}^2(L,t_2)S_{\Omega_1}(t_2,t_0)}\nonumber\\
&\qquad\qquad\qquad\qquad\qquad\times\exp\left\{-\frac{iS_{\Omega_1}(t_2,t_0)}{2\omega C_{\Omega_1}(t_0,t_2)+iQ_{\rm eff}^2(L,t_2)S_{\Omega_1}(t_2,t_0)}\vec{l}^2\right\}\,.
\label{eq:ininhofac}
\end{align}

The non-factorizable piece of the splitting spectrum is the most challenging part of the calculation due to the two coupled path integrals in \eqref{eq:S4nonfac}. In the HO approximation, this expression can be written as
\begin{align}
\tilde{\cal S}^{(4)}_{\rm non\text{-}fac,HO}(\vec{l},L;\vec{l}_2,\bar{\vec{l}}_2,t_2;z) &= -\int_{\vec{l}_3}\int_{t_2}^L \td t_3\,{\cal P}_{\rm eff}(L,t_3;\vec{l}-\vec{l}_3)\int d^2\vec{u}_3 \,\td^2\bar{\vec{u}}_3\, \td^2\vec{u}_2\, \td^2\bar{\vec{u}}_2\,e^{-i(\vec{u}_3-\bar{\vec{u}}_3)\cdot\vec{l}_3+i\vec{u}_2\cdot\vec{l}_2-i\bar{\vec{u}}_2\cdot\bar{\vec{l}}_2}\nonumber\\
&\quad\times \frac{z(1-z)\hat q(t_3)}{2}(\vec{u}_3-\bar{\vec{u}}_3)^2\,{\cal I}(\vec{u}_3,\bar{\vec{u}}_3,t_3;\vec{u}_2,\bar{\vec{u}}_2,t_2;\omega,z)\nonumber\\
&= \frac{z(1-z)}{2}\int_{\vec{l}_3}\int_{t_2}^L \td t_3\,\partial_{\vec{l}}^2{\cal P}_{\rm eff}(L,t_3;\vec{l}-\vec{l}_3)\,\hat q(t_3)\int \td^2\vec{u}_3 \,d^2\bar{\vec{u}}_3\, \td^2\vec{u}_2\, \td^2\bar{\vec{u}}_2\nonumber\\
&\quad\times e^{-i(\vec{u}_3-\bar{\vec{u}}_3)\cdot\vec{l}_3+i\vec{u}_2\cdot\vec{l}_2-i\bar{\vec{u}}_2\cdot\bar{\vec{l}}_2}\,{\cal I}(\vec{u}_3,\bar{\vec{u}}_3,t_3;\vec{u}_2,\bar{\vec{u}}_2,t_2;\omega,z)\,,
\label{eq:S4hononfac}
\end{align}
where
\begin{align}
{\cal I}(\vec{u}_3,\bar{\vec{u}}_3,t_3;\vec{u}_2,\bar{\vec{u}}_2,t_2;\omega,z) &= \int_{\vec{u}(t_2)=\vec{u}_2}^{\vec{u}(t_3)=\vec{u}_3}{\cal D}\vec{u}\int_{\bar{\vec{u}}(t_2)=\bar{\vec{u}}_2}^{\bar{\vec{u}}(t_3)=\bar{\vec{u}}_3}{\cal D}\bar{\vec{u}}\,\exp\left\{\frac{i\omega}{2}\int_{t_2}^{t_3}\td s\,(\dot{\vec{u}}^2(s)-\dot{\bar{\vec{u}}}^2(s))\right.\nonumber\\
&\qquad\left.-\frac{1}{4}\int_{t_2}^{t_3}\td s\,\hat q(s)\left(\vec{u}^2(s)+\bar{\vec{u}}^2(s)+A\,\left(\vec{u}(s)-\bar{\vec{u}}(s) \right)^2\right)\right\}\,,
\end{align}
with $A = (n-1)z^2+(m-1)(1-z)^2$.
This double path integral can be evaluated through a change of variables, as explained in appendix B.3 of \cite{Blaizot:2012fh} and appendix D.4 of \cite{Apolinario:2014csa}. Defining\footnote{For the case of the $\gamma\to q\bar q$ splitting $A=0$ and the two path integrals can be solved separately without changing variables, equivalent to taking $\beta=0$.}
\beq
\begin{pmatrix}
\vec{u} \\ \bar{\vec{u}}
\end{pmatrix}
= \gamma\begin{pmatrix}
1 & \beta \\ \beta & 1
\end{pmatrix}
\begin{pmatrix}
\vec{u}' \\ \bar{\vec{u}}'
\end{pmatrix}\,,
\eeq
with
\beq
\beta = \frac{1+A-\sqrt{1+2A}}{A}\,,\qquad \gamma = \frac{1}{\sqrt{1-\beta^2}}\,,
\eeq
the double path integral factorizes as
\begin{align}
{\cal I}(\vec{u}_3,\bar{\vec{u}}_3,t_3;\vec{u}_2,\bar{\vec{u}}_2,t_2;\omega,z) &= \int_{\vec{u}'(t_2)=\vec{u}'_2}^{\vec{u}'(t_3)=\vec{u}'_3}{\cal D}\vec{u}'\int_{\bar{\vec{u}}'(t_2)=\bar{\vec{u}}'_2}^{\bar{\vec{u}}'(t_3)=\bar{\vec{u}}'_3}{\cal D}\bar{\vec{u}}'\,\exp\left\{\frac{i\omega}{2}\int_{t_2}^{t_3}\td s\,\left(\dot{\vec{u}}^{\prime 2}(s)-\dot{\bar{\vec{u}}}^{\prime 2}(s)\right)\right.\nonumber\\
&\qquad\left.-\frac{1}{4}\int_{t_2}^{t_3}\td s\,\hat q(s)\sqrt{1+2A}\left(\vec{u}^{\prime 2}(s)+\bar{\vec{u}}^{\prime 2}(s)\right)\right\}\nonumber\\
&= {\cal K}_{\Omega_2}(\vec{u}'_3,t_3;\vec{u}'_2,t_2)\,{\cal K}^*_{\Omega_2}\left(\bar{\vec{u}}'_3,t_3;\bar{\vec{u}}'_2,t_2 \right)\,,
\end{align}
with $\Omega_2^2(s) = -\frac{i\hat q(s)}{2\omega}\sqrt{1+2A}$.
Using this result and introducing the conjugate momenta for the primed coordinates, the non-factorizable 4-point function in \eqref{eq:S4hononfac} becomes
\begin{align}
\tilde{\cal S}^{(4)}_{\rm non\text{-}fac}(\vec{l},L;\vec{l}_2,\bar{\vec{l}}_2,t_2;z) &= \frac{z(1-z)}{2}\int_{\vec{l}_3}\int_{t_2}^L \td t_3\,\partial_{\vec{l}}^2{\cal P}_{\rm eff}(L,t_3;\vec{l}-\vec{l}_3)\,\hat q(t_3)\nonumber\\
&\quad\times {\cal K}_{\Omega_2}\left(\gamma(1-\beta)\vec{l}_3,t_3;\gamma(\vec{l}_2-\beta\bar{\vec{l}}_2),t_2 \right)\,{\cal K}^*_{\Omega_2} \left(\gamma(1-\beta)\vec{l}_3,t_3;\gamma(\bar{\vec{l}}_2-\beta\vec{l}_2),t_2 \right)\,,
\label{eq:S4hononfac2}
\end{align}
Finally, the corresponding non-factorizable contribution to the in-in splitting spectrum is
\begin{align}
\frac{\td I^{\text{in-in}}_{\rm non\text{-}fac, HO}}{\td z \td^2\vec{l}} &= \frac{\alpha_s\,z(1-z)}{2(2\pi)^2\omega^2}P_{a\to bc}(z)\,\text{Re}\int_{\vec{l}_1\vec{l}_2\bar{\vec{l}}_2\vec{l}_3}\int_{t_0}^L \td t_1\int_{t_1}^L \td t_2\int_{t_2}^L\td t_3 \,(\vec{l}_1\cdot\bar{\vec{l}}_2)\,\partial_{\vec{l}}^2{\cal P}_{\rm eff}(L,t_3;\vec{l}-\vec{l}_3)\,\hat q(t_3)\nonumber\\
&\quad\times {\cal K}_{\Omega_2}\left(\gamma(1-\beta)\vec{l}_3,t_3;\gamma(\vec{l}_2-\beta\bar{\vec{l}}_2),t_2 \right)\,{\cal K}^*_{\Omega_2} \left(\gamma(1-\beta)\vec{l}_3,t_3;\gamma(\bar{\vec{l}}_2-\beta\vec{l}_2),t_2 \right)\,{\cal K}_{\Omega_1}(\vec{l}_2,t_2;\vec{l}_1,t_1;z)\,.
\label{eq:ininhOnonfact}
\end{align}

All the factors in \eqref{eq:ininhOnonfact} are either the broadening factor, given in~\eqref{eq:Peff}, or  kernels of the form given in~\eqref{eq:KOmmom}, which are gaussian in the transverse momenta. The integrations over $t_1$ and $\vec{l}_1$ can be performed analytically in an analogous manner to what was done in eq.~\eqref{eq:intt1}, and the remaining transverse-momentum integrals can then also be carried out analytically. This procedure leads to
\begin{align}
\frac{\td I^{\text{in-in}}_{\rm non\text{-}fac, HO}}{\td z \td^2\vec{l}} &= -\frac{4\alpha_s\,z(1-z)}{(2\pi)^2\omega} P_{a\to bc}(z)\,\text{Re}\int_{t_0}^L\td t_2\int_{t_2}^L\td t_3\frac{i(1-\beta^2)\,\hat{q}(t_3)}{\beta^2C_{23}\partial_{t_3}C_{32}^*-C_{23}^*\partial_{t_3}C_{32}}\nonumber\\
&\quad\times\left[\frac{B}{(BQ^2_{\rm eff}(L,t_3)+1)^2}\left(-\frac{\beta\partial_{t_3}C_{32}^*+\partial_{t_3}C_{32}}{C_{23}^*+\beta C_{23}}+\frac{\beta(C_{23}\partial_{t_3}C_{32}^*-C_{23}^*\partial_{t_3}C_{32})}{(1-\beta^2)|C_{23}|^2}\right)\right.\nonumber\\
&\qquad\quad\times\left(\frac{B\,\vec{l}^2}{BQ^2_{\rm eff}(L,t_3)+1}-1\right)\exp\left\{-\frac{B\,\vec{l}^2}{BQ^2_{\rm eff}(L,t_3)+1}\right\}\nonumber\\
&\qquad-\frac{D}{(DQ^2_{\rm eff}(L,t_3)+1)^2}\left(-\frac{\beta\partial_{t_3}C_{32}^*+\partial_{t_3}C_{32}}{C_{23}^*+\beta C_{23}}\right.\nonumber\\
&\qquad\qquad\qquad\qquad\left.+\frac{\beta(C_{23}\partial_{t_3}C_{32}^*-C_{23}^*\partial_{t_3}C_{32})}{(1-\beta^2)|C_{23}|^2-\frac{S_{20}}{C_{02}}\left(\beta^2C_{23}\partial_{t_3}C_{32}^*-C_{23}^*\partial_{t_3}C_{32} \right)}\right)\nonumber\\
&\qquad\left.\times\left(\frac{D\,\vec{l}^2}{DQ^2_{\rm eff}(L,t_3)+1}-1\right)\exp\left\{-\frac{D\,\vec{l}^2}{DQ^2_{\rm eff}(L,t_3)+1}\right\}\right]\,,
\label{eq:ininhononfacexpl}
\end{align}
where
\begin{align}
B=-\frac{i(1-\beta)}{2\omega(1+\beta)(\beta^2C_{23}\partial_{t_3}C_{32}^*-C_{23}^*\partial_{t_3}C_{32})}\bigg(&-S_{32}^*\partial_{t_3}C_{32}-\beta^2S_{32}\partial_{t_3}C_{32}^*+2\beta+C_{23}^*C_{32}\nonumber\\
\,\,&+\beta^2C_{23}C_{32}^*-\frac{(C_{23}^*+\beta C_{23})^2}{|C_{23}|^2}\bigg)\,,
\end{align}
\begin{align}
D&=-\frac{i(1-\beta)}{2\omega(1+\beta)\left(\beta^2C_{23}\partial_{t_3}C_{32}^*-C_{23}^*\partial_{t_3}C_{32}\right)}\bigg(-S_{32}^*\partial_{t_3}C_{32}-\beta^2S_{32}\partial_{t_3}C_{32}^*+2\beta+C_{23}^*C_{32}\nonumber\\
&\qquad\qquad+\beta^2C_{23}C_{32}^*-\frac{(1-\beta^2)(C_{23}^*+\beta C_{23})^2}{(1-\beta^2)|C_{23}|^2-\frac{S_{20}}{C_{02}}\left(\beta^2C_{23}\partial_{t_3}C_{32}^*-C_{23}^*\partial_{t_3}C_{32} \right)}\bigg)\,,
\end{align}
and
\begin{align}
C_{23}=C_{\Omega_2}(t_2,t_3)\:,\quad C_{32}=C_{\Omega_2}(t_3,t_2)\:,\quad S_{32}=S_{\Omega_2}(t_3,t_2)\:,\quad S_{20}=S_{\Omega_1}(t_2,t_0)\:,\quad C_{02}=C_{\Omega_1}(t_0,t_2)\,.
\end{align}
While eq.~\eqref{eq:ininhononfacexpl} may appear cumbersome, its numerical evaluation is not computationally demanding, as it involves only two time integrals.

In summary, in the large-$N_c$ limit and within the harmonic oscillator approximation for the dipole cross section, in-medium splittings are given by the sum of eqs.~\eqref{eq:inoutho}, \eqref{eq:ininhofac}, and \eqref{eq:ininhononfacexpl}, all of which can be evaluated numerically in a simple and efficient manner. We will refer to this result as the \emph{large-$N_c$-HO approach}. To the best of our knowledge,  this is the first time explicit expressions have been given for all the terms in the splitting spectrum within this approximation. Numerical results for this spectrum are presented in section~\ref{sec:evaluation}.

\section{Improved semi-hard approximation}
\label{sec:ISHA}

If we move beyond the harmonic approximation introduced in the previous section, analytical solutions for the path integrals are no longer available. This leaves us with two options: either recast the evaluation of the path integrals as a set of coupled differential equations, which requires a significantly more intensive numerical treatment \cite{Caron-Huot:2010qjx, Andres:2020vxs, Isaksen:2023nlr}, or introduce additional approximations allowing us to bypass the path integrals via a systematic series expansion. In this work, we follow the second strategy, exploiting the fact that for very energetic partons the path integrals are dominated by classical straight-line trajectories. The main advantage of this approach is that it yields simple analytical expressions within a kinematic regime that has been argued to be relevant for certain observables, such as energy correlators \cite{Andres:2022ovj,Andres:2023xwr,Andres:2023ymw}.

The \emph{semi-classical} or \emph{semi-hard} approximation (SHA) was first introduced in \cite{Dominguez:2019ges} and later developed in \cite{Isaksen:2020npj}, where all parton propagators in coordinate space are replaced by a vacuum propagator multiplied by a Wilson line evaluated along the straight-line path connecting the end-point transverse coordinates. Although its accuracy has not yet been systematically tested across a wide range of parameters, preliminary studies for $\gamma \to q \bar q$ splittings  suggest that it tends to overestimate the medium-induced enhancement of emissions in certain regions of phase space \cite{Isaksen:2020npj, Barata:2023bhh}.  In this section, we show how this approach can be extended to include the first corrections in an expansion in powers of $L\partial_{\perp}^2/(z(1-z)E)$, where $\partial_{\perp}^2$ denotes a partial derivative in transverse coordinates. While the original SHA of~\cite{Dominguez:2019ges,Isaksen:2020npj} was not formulated as an expansion in this parameter, the derivation presented here allows us to identify it as the zeroth-order result of such an expansion. Appendix~A of~\cite{Dominguez:2019ges} outlined a possible strategy for incorporating first-order corrections, but some contributions at this order were missing. To address this limitation, we develop a consistent expansion based on the formalism of~\cite{Altinoluk:2014oxa}, with modifications that allow us to retain the splitting-angle dependence. We refer to this extended approximation as the \emph{improved semi-hard approximation} (ISHA), and derive and evaluate the complete first-order correction within this approach.

Even though the SHA and ISHA  are valid for any dipole cross section, at the end of this section we will return to the harmonic oscillator approximation. This allows us to directly compare the results of both the SHA and ISHA with the full large-$N_c$-HO developed in the previous section. Through this quantitative comparison, we aim to identify the regions of phase space, if any, where either approximation can serve as a reliable substitute for the full path-integral evaluation.

\subsection{Expansion around the free path}
\label{subsec:expansionclassical}

Let us start by performing an expansion of the in-medium parton propagator in \eqref{eq:prop}. We parametrize the generic path $\vec{r}(t)$ as a perturbation around the free trajectory $\vec{x}_{\rm cl}(t)$ in the same way as done in section 2.2.1 of \cite{Altinoluk:2014oxa}, albeit in slightly different language. With the change of variables
\beq
\vec{r}(t) = \vec{x}_{\rm cl}(t) + \vec{u}(t)\,,\quad \vec{x}_{\rm cl}(t) = \frac{t_2-t}{t_2-t_1}\vec{x}_1 + \frac{t-t_1}{t_2-t_1}\vec{x}_2\,,
\eeq
eq.~\eqref{eq:prop} becomes
\beq
{\cal G}_R(\vec{x}_2,t_2;\vec{x}_1,t_1;E) ={\cal G}_0(\vec{x}_2,t_2;\vec{x}_1,t_1;E)\,{\cal R}_R(\vec{x}_2,t_2;\vec{x}_1,t_1;E)\,,\label{eq:G0R}
\eeq
where
\begin{equation}
{\cal R}_R(\vec{x}_2,t_2;\vec{x}_1,t_1;E) = \frac{2\pi i(t_2-t_1)}{E}\int_{\vec{u}(t_1)=\vec{0}}^{\vec{u}(t_2)=\vec{0}} {\cal D}\vec{u}\,\exp\left\{\frac{iE}{2}\int_{t_1}^{t_2}\td s\,\dot{\vec{u}}^2(s)\right\}\,U_R(t_2,t_1;[\vec{x}_{\rm cl}+\vec{u}])\,.
\end{equation}
Expanding the Wilson line in powers of $\vec{u}$ gives
\begin{align}
{\cal R}_R(\vec{x}_2,t_2;\vec{x}_1,t_1;E) &\simeq \frac{2\pi i(t_2-t_1)}{E}\int_{\vec{u}(t_1)=\vec{0}}^{\vec{u}(t_2)=\vec{0}} {\cal D}\vec{u}\,\exp\left\{\frac{iE}{2}\int_{t_1}^{t_2}\td s\,\dot{\vec{u}}^2(s)\right\}\,\bigg[U_R\left(t_2,t_1;[\vec{x}_{\rm cl}]\right)\nonumber\\
&\quad+\int_{t_1}^{t_2}\td s\,\vec{u}(s)\cdot\frac{\delta}{\delta\vec{x}_{\rm cl}(s)}U_R\left(t_2,t_1;[\vec{x}_{\rm cl}]\right)\nonumber\\
&\quad+\frac{1}{2}\int_{t_1}^{t_2}\td s\int_{t_1}^{t_2}\td s'\,\vec{u}^i(s)\vec{u}^j(s')\frac{\delta^2}{\delta\vec{x}^i_{\rm cl}(s)\delta\vec{x}^j_{\rm cl}(s')}U_R\left(t_2,t_1;[\vec{x}_{\rm cl}]\right)\bigg]\,.
\label{eq:Rexpclas}
\end{align}
For the first term in \eqref{eq:Rexpclas}, the path integral can be performed exactly, yielding
\begin{equation}
\int_{\vec{u}(t_1)=\vec{0}}^{\vec{u}(t_2)=\vec{0}} {\cal D}\vec{u}\,\exp\left\{\frac{iE}{2}\int_{t_1}^{t_2}\td s\,\dot{\vec{u}}^2(s)\right\} = {\cal G}_0(\vec{0},t_2;\vec{0},t_1;E) = \frac{E}{2\pi i(t_2-t_1)}\,.
\end{equation}

For the remaining two terms in \eqref{eq:Rexpclas}, it is useful to compute first the functional derivatives of the Wilson lines. Using~\eqref{eq:Wilson}, we find
\begin{align}
\frac{\delta}{\delta\vec{r}^i(s)}U_R\left(t_2,t_1;[\vec{r}]\right) &= U_R\left(t_2,s;[\vec{r}]\right)\,ig_s\,\partial^i_\perp A^-_R\left(s,\vec{r}(s)\right)\,U_R\left(s,t_1;[\vec{r}] \right)\,,
\label{eq:delta2U_1}
\end{align}
and
\begin{align}
\frac{\delta^2}{\delta\vec{r}^i(s)\delta\vec{r}^j(s')}U_R\left(t_2,t_1;[\vec{r}]\right) &= U_R(t_2,s';[\vec{r}])\,ig_s\,\partial^j_\perp A^-_R\left(s',\vec{r}(s')\right)\,U_R(s',s;[\vec{r}])\nonumber\\
&\qquad\times ig_s\,\partial^i_\perp A^-_R\left(s,\vec{r}(s)\right)\,U_R\left(s,t_1;[\vec{r}]\right)\theta(s'-s)\nonumber\\
&\quad + U_R(t_2,s;[\vec{r}])\,ig_s\,\partial^i_\perp A^-_R\left(s,\vec{r}(s)\right)\,U_R\left(s,s';[\vec{r}]\right)\nonumber\\
&\qquad\times ig_s\,\partial^j_\perp A^-_R\left(s',\vec{r}(s')\right)\,U_R\left(s',t_1;[\vec{r}]\right)\theta(s-s')\nonumber\\
&\quad + U_R\left(t_2,s;[\vec{r}]\right)\,ig_s\, \partial^i_\perp\partial^j_\perp A^-_R\left(s,\vec{r}(s)\right)\,U_R \left(s,t_1;[\vec{r}]\right)\,\delta(s'-s)\,.
\label{eq:delta2U_2}
\end{align}

For the second term in eq.~\eqref{eq:Rexpclas}, the explicit form of the functional derivative is not important, what matter is that it is non-singular. A simple manipulation shows that the path integral for this term can be written as
\begin{align}
&\int_{\vec{u}(t_1)=\vec{0}}^{\vec{u}(t_2)=\vec{0}} {\cal D}\vec{u}\,\exp\left\{\frac{iE}{2}\int_{t_1}^{t_2}\td s\,\dot{\vec{u}}^2(s)\right\}\int_{t_1}^{t_2}\td s \,\vec{u}^i(s)\,\frac{\delta}{\delta\vec{x}_{\rm cl}^i(s)}U_R\left(t_2,t_1;[\vec{x}_{\rm cl}]\right)\nonumber\\
&\qquad=\int_{t_1}^{t_2}\td s\int\td^2\vec{u}_s\,{\cal G}_0\left(\vec{0},t_2;\vec{u}_s,s;E\right)\,\vec{u}^i_s\,{\cal G}_0\left(\vec{u}_s,t_2;\vec{0},t_1;E\right)\,\frac{\delta}{\delta\vec{x}_{\rm cl}^i(s)}U_R\left(t_2,t_1;[\vec{x}_{\rm cl}]\right)\,,
\end{align}
where the integral over $\vec{u}_s$ can be performed explicitly, yielding
\beq
\int_{t_1}^{t_2}\td s\int\td^2\vec{u}_s\,{\cal G}_0\left(\vec{0},t_2;\vec{u}_s,s;E\right)\,\vec{u}^i_s\,{\cal G}_0\left(\vec{u}_s,t_2;\vec{0},t_1;E\right)=0\,,
\eeq
and therefore this term does not contribute. In fact, when expanding ${\cal R}_R$ in $\vec{u}$, any terms involving an odd number of powers of $\vec{u}$ vanish. Furthermore,  each power of $\vec{u}$ is accompanied by a transverse derivative acting on the Wilson line and each pair of powers of $\vec{u}$ gives a factor of order $(t_2-t_1)/E \leq L/E$. The expansion of ${\cal R}_R$ in powers of $\vec{u}$ thus corresponds to an expansion in powers of $(L/E)\partial_\perp^2$, with the last line in~\eqref{eq:Rexpclas} yielding the first correction in this expansion. To evaluate this correction term, we insert~\eqref{eq:delta2U_2} into the last line of~\eqref{eq:Rexpclas}   and, after performing a similar manipulation as for the second term  in~\eqref{eq:Rexpclas}, the relevant integrals in terms of the vacuum propagators are
\beq
\int\td^2\vec{u}_s\td^2\vec{u}_{s'}\,{\cal G}_0(\vec{0},t_2;\vec{u}_{s'},s';E)\,\vec{u}^j_{s'}\,{\cal G}_0(\vec{u}_{s'},s';\vec{u}_s,s;E)\,\vec{u}^i_{s}\,{\cal G}_0(\vec{u}_{s},s;\vec{0},t_1;E) = \frac{\delta^{ij}}{2\pi}\frac{(t_2-s')(s-t_1)}{(t_2-t_1)^2}\,,
\eeq
\beq
\int\td^2\vec{u}_s\,{\cal G}_0(\vec{0},t_2;\vec{u}_{s},s;E)\,\vec{u}^j_{s}\,\vec{u}^i_{s}\,{\cal G}_0(\vec{u}_{s},s;\vec{0},t_1;E) = \frac{\delta^{ij}}{2\pi}\frac{(t_2-s)(s-t_1)}{(t_2-t_1)^2}\,.
\eeq
Combining all these pieces, one arrives at 
\begin{align}
&{\cal R}_R(\vec{x}_2,t_2;\vec{x}_1,t_1;E) \simeq U_R\left(t_2,t_1;[\vec{x}_{\rm cl}]\right)\nonumber\\
&\qquad+i\frac{t_2-t_1}{2E}\int_{t_1}^{t_2}\td s\,\frac{(t_2-s)(s-t_1)}{(t_2-t_1)^2}U_R\left(t_2,s;[\vec{x}_{\rm cl}]\right)\,ig_s\,\partial^2_\perp A_R^-(s,\vec{x}_{\rm cl}(s))U_R\left(s,t_1;[\vec{x}_{\rm cl}]\right)\nonumber\\
&\qquad+i\frac{t_2-t_1}{E}\int_{t_1}^{t_2}\td s\int_{s}^{t_2}\td s'\,\frac{(t_2-s')(s-t_1)}{(t_2-t_1)^2}\,U_R\left(t_2,s';[\vec{x}_{\rm cl}]\right)\,ig_s\,\partial^i_\perp A_R^-\left(s',\vec{x}_{\rm cl}(s')\right)\,U_R\left(s',s;[\vec{x}_{\rm cl}]\right)\nonumber\\
&\qquad\qquad\qquad\qquad\times ig_s\,\partial^i_\perp A_R^-\left(s,\vec{x}_{\rm cl}(s)\right)\,U_R\left(s,t_1;[\vec{x}_{\rm cl}]\right)\,.
\label{eq:Rexpcor}
\end{align}
As expected, the correction terms in~\eqref{eq:Rexpcor} are suppressed by a factor $(t_2-t_1)/E \leq L/E$ and involve two transverse derivatives of the background field. The semi-hard approximation of \cite{Dominguez:2019ges,Isaksen:2020npj} keeps only the first term in this expansion (fist line in~\eqref{eq:Rexpcor}), whereas~\eqref{eq:Rexpcor} provides the complete expression for the first-order corrections  in powers of $L\partial_{\perp}^2/E$. 

To obtain the splitting spectrum, this expansion must be applied to each propagator, and its validity therefore requires all partons to be sufficiently energetic. As in next-to-eikonal expansions in small-$x$ physics, the result should be understood as an asymptotic expansion whose accuracy is governed by power counting.

\subsection{Fixing the free trajectory along the outgoing momentum}
\label{subsec:Fourierprop}

Let us now switch to transverse momentum space and expand around the free trajectory fixed by the outgoing momentum. We follow the approach of section 2.1.2 of \cite{Altinoluk:2014oxa}, but unlike there, we do not expand contributions proportional to the transverse momentum, which in our case are not small and allow us to keep track of the angle between the daughter partons. Taking the Fourier transform of eq.~\eqref{eq:G0R}, we obtain
\begin{align}
{\cal G}_R(\vec{p}_2,t_2;\vec{p}_1,t_1;E) &= e^{-\frac{i\vec{p}_2^2}{2E}(t_2-t_1)}\int\td^2\vec{x}_1\td^2\vec{x}_2\,e^{-i\vec{x}_1\cdot(\vec{p}_2-\vec{p}_1)}\nonumber\\
&\qquad \times\,\frac{E}{2\pi i(t_2-t_1)}\,e^{\frac{iE}{2(t_2-t_1)}\left(\vec{x}_2-\vec{x}_1-\frac{\vec{p}_2}{E}(t_2-t_1)\right)^2}\,{\cal R}_R \left(\vec{x}_2,t_2;\vec{x}_1,t_1;E\right)\,.
\end{align}
In the semi-hard approximation in \cite{Dominguez:2019ges, Isaksen:2020npj}, the factor multiplying $\cal R$ in the second line above is taken as a $\delta$-function, fixing the separation $\vec{x}_2-\vec{x}_1$. Here, we relax that approximation by performing the change of variables
\begin{equation}
\vec{x}_2-\vec{x}_1 = \frac{\vec{p}_2}{E}(t_2-t_1) + \sqrt{\frac{t_2-t_1}{E}}\vec{h}\,,
\end{equation}
and then expanding $\cal R$ in powers of $\vec{h}$. The propagator in momentum space then becomes
\begin{align}
{\cal G}_R(\vec{p}_2,t_2;\vec{p}_1,t_1;E) &\simeq e^{-\frac{i\vec{p}_2^2}{2E}(t_2-t_1)}\int\td^2\vec{x}_1\td^2\vec{h}\,e^{-i\vec{x}_1\cdot(\vec{p}_2-\vec{p}_1)}\,\frac{1}{2\pi i}\,e^{\frac{i}{2}\vec{h}^2}\nonumber\\
&\qquad\times\bigg[{\cal R}_R\left(\vec{x}_2,t_2;\vec{x}_1,t_1;E\right)\nonumber\\
&\qquad\qquad+\frac{t_2-t_1}{2E}\vec{h}^i\vec{h}^j\,\partial_{\vec{x}_2}^i\partial_{\vec{x}_2}^j{\cal R}_R(\vec{x}_2,t_2;\vec{x}_1,t_1;E)\bigg]_{\vec{x_2}=\vec{x}_1+\frac{\vec{p}_2}{E}((t_2-t_1)}\nonumber\\
&= e^{-\frac{i\vec{p}_2^2}{2E}(t_2-t_1)}\int\td^2\vec{x}_1\,e^{-i\vec{x}_1\cdot(\vec{p}_2-\vec{p}_1)}\,\bigg[{\cal R}_R\left(\vec{x}_1,t_2;\vec{x}_1,t_1;E\right)\nonumber\\
&\qquad\qquad+\frac{i(t_2-t_1)}{2E}\,\partial_{\vec{x}_2}^2{\cal R}_R(\vec{x}_2,t_2;\vec{x}_1,t_1;E)\bigg]_{\vec{x_2}=\vec{x}_1+\frac{\vec{p}_2}{E}((t_2-t_1)}\,.\label{eq:expandmom}
\end{align}
We now plug eq.~\eqref{eq:Rexpcor} into eq.~\eqref{eq:expandmom}. To the desired accuracy, only the zeroth-order term of eq.~\eqref{eq:Rexpcor} contributes to the last term in eq.~\eqref{eq:expandmom}. Then, the relevant second derivative is
\begin{align}
\partial_{\vec{x}_2}^2U_R(t_2,t_1;[\vec{x}_{\rm cl}]) &= \int_{t_1}^{t_2}\td s\Bigg[2\int_s^{t_2}\td s'\,\frac{(s'-t_1)(s-t_1)}{(t_2-t_1)^2}\,U_R(t_2,s';[\vec{x}_{\rm cl}])\,ig_s\partial^i_\perp A_R^-(s',\vec{x}_{\rm cl}(s'))\nonumber\\
&\qquad\qquad\qquad\times U_R(s',s;[\vec{x}_{\rm cl}])\,ig_s\partial^i_\perp A_R^-(s,\vec{x}_{\rm cl}(s))U_R(s,t_1;[\vec{x}_{\rm cl}])\nonumber\\
&\qquad+\left(\frac{s-t_1}{t_2-t_1}\right)^2\,U_R(t_2,s;[\vec{x}_{\rm cl}])\,ig_s\partial^2_\perp A_R^-(s,\vec{x}_{\rm cl}(s))U_R(s,t_1;[\vec{x}_{\rm cl}])\Bigg]\,.
\end{align}
Combining all terms, the in-medium propagator in transverse momentum space reads
\begin{align}
{\cal G}_R(\vec{p}_2,t_2;\vec{p}_1,t_1;E) &\simeq e^{-\frac{i\vec{p}_2^2}{2E}(t_2-t_1)}\int\td^2\vec{x}_1\,e^{-i\vec{x}_1\cdot(\vec{p}_2-\vec{p}_1)}\,\Bigg[U_R(t_2,t_1;[\vec{x}_{\rm cl}])\nonumber\\
&\qquad+\frac{i}{2E} (t_2-t_1)\int_{t_1}^{t_2}\td s\,\frac{s-t_1}{t_2-t_1}\bigg(2\int_s^{t_2}\td s'\,U_R(t_2,s';[\vec{x}_{\rm cl}])\,ig_s\partial^i_\perp A_R^-(s',\vec{x}_{\rm cl}(s'))\nonumber\\
&\qquad\qquad\qquad\times U_R(s',s;[\vec{x}_{\rm cl}])\,ig_s\partial^i_\perp A_R^-(s,\vec{x}_{\rm cl}(s))U_R(s,t_1;[\vec{x}_{\rm cl}])\nonumber\\
&\qquad+\,U_R(t_2,s;[\vec{x}_{\rm cl}])\,ig_s\partial^2_\perp A_R^-(s,\vec{x}_{\rm cl}(s))U_R(s,t_1;[\vec{x}_{\rm cl}])\bigg)\Bigg]\,,
\label{eq:propmomexp}
\end{align}
where the free trajectory is now given by
\begin{equation}
\vec{x}_{\rm cl}(t) = \vec{x_1} + \frac{\vec{p}_2}{E}(t-t_1)\,.
\end{equation}
At this stage, it is clear that the transverse derivatives in eq.~\eqref{eq:propmomexp} can be replaced by derivatives with respect to the initial position $\vec{x}_1$. Recognizing that the correction term can be written in terms of derivatives of Wilson lines, eq.~\eqref{eq:propmomexp} can be recast in the more compact and convenient form
\begin{align}
{\cal G}_R^{\rm ISHA}(\vec{p}_2,t_2;\vec{p}_1,t_1;E) &= e^{-\frac{i\vec{p}_2^2}{2E}(t_2-t_1)}\int\td^2\vec{x}_1\,e^{-i\vec{x}_1\cdot(\vec{p}_2-\vec{p}_1)}\,\Bigg[U_R(t_2,t_1;[\vec{x}_{\rm cl}])\nonumber\\
&\qquad+\frac{i}{2E}\int_{t_1}^{t_2}\td s\,\left[\partial_{\vec{x}_1}^2U_R(t_2,s;[\vec{x}_{\rm cl}])\right]U_R(s,t_1;[\vec{x}_{\rm cl}])\Bigg]\,.
\label{eq:propISH}
\end{align}
The first term in eq.~\eqref{eq:propISH} coincides with eq.~(10) in the original semi-hard derivation \cite{Dominguez:2019ges}. In that work, an additional approximation was made to this term by neglecting the dependence on the initial position of the free trajectory, leading to the propagator expression in their eq.~(11), which is the one used for all spectrum results in that paper. In contrast, we will keep the dependence on the initial position, as it yields additional corrections to the spectrum at the same order  of the ones arising from the second term in eq.~\eqref{eq:propISH}. 

\subsection{Splitting spectrum}
\label{subsec:ISHspectrum}

To obtain the splitting spectrum in this approximation, we need to calculate the 3- and 4-point functions. These can be computed either by inserting~\eqref{eq:propISH} into the definitions~\eqref{eq:K3}~and~\eqref{eq:S4} or by repeating the derivation of the previous subsections starting from eqs.~\eqref{eq:K3tilde} and \eqref{eq:S4v0}.

For the 3-point function, we find
\begin{align}
\tilde{\cal K}^{(3)}(\vec{l}_2,t_2;\vec{l}_1,t_1;z) &\simeq e^{-\frac{i\vec{l}_2^2}{2\omega}(t_2-t_1)}\int\td^2\vec{u}_1\,e^{-i\vec{u}_1\cdot(\vec{l}_2-\vec{l}_1)}\,\Bigg[{\cal C}^{(3)}\left(t_2,t_1;[\vec{u}_{\rm cl},\vec{0}]\right)\nonumber\\
&\qquad+\frac{i}{2\omega}\int_{t_1}^{t_2}\td s\,\left[\partial_{\vec{u}_1}^2{\cal C}^{(3)}(t_2,s;[\vec{u}_{\rm cl},\vec{0}])\right]{\cal C}^{(3)}\left(s,t_1;[\vec{u}_{\rm cl},\vec{0}]\right)\Bigg]\,,
\label{eq:K3ISH}
\end{align}
where $\vec{u}_{\rm cl}(t)=\vec{u}_1+\frac{\vec{l}_2}{\omega}(t-t_1)$. Plugging this expression into eq.~\eqref{eq:inout} gives the in-out contribution to the splitting spectrum
\begin{align}
\frac{\td I^{\text{in-out}}_{\rm ISHA}}{\td z \td^2\vec{l}} &= -\frac{\alpha_s}{2\pi^2 \omega\,\vec{l}^2}P_{a\to bc}(z)\,\text{Re}\,i\int_{\vec{l}_1}\int_{t_0}^L \td t_1\,(\vec{l}_1\cdot\vec{l})\,e^{-\frac{i\vec{l}^2}{2\omega}(L-t_1)}\int\td^2\vec{u}_1\,e^{-i\vec{u}_1\cdot(\vec{l}-\vec{l}_1)}\nonumber\\
&\qquad\times\Bigg[{\cal C}^{(3)}(L,t_1;[\vec{u}_{\rm cl},\vec{0}])+\frac{i}{2\omega}\int_{t_1}^{L}\td s\,\left[\partial_{\vec{u}_1}^2{\cal C}^{(3)}(L,s;[\vec{u}_{\rm cl},\vec{0}])\right]{\cal C}^{(3)}(s,t_1;[\vec{u}_{\rm cl},\vec{0}])\Bigg]\nonumber\\
&= -\frac{\alpha_s}{2\pi^2 \omega\,\vec{l}^2}P_{a\to bc}(z)\,\text{Re}\,i\int_{t_0}^L \td t_1\,e^{-\frac{i\vec{l}^2}{2\omega}(L-t_1)}\,\vec{l}\cdot(\vec{l}+i\partial_{\vec{u}_1})\nonumber\\
&\qquad\times\Bigg[{\cal C}^{(3)}(L,t_1;[\vec{u}_{\rm cl},\vec{0}])+\frac{i}{2\omega}\int_{t_1}^{L}\td s\,\left[\partial_{\vec{u}_1}^2{\cal C}^{(3)}(L,s;[\vec{u}_{\rm cl},\vec{0}])\right]{\cal C}^{(3)}(s,t_1;[\vec{u}_{\rm cl},\vec{0}])\Bigg]_{\vec{u}_1=0}\,.
\end{align}
Here, only the derivative in the factor $(\vec{l}+i\partial_{\vec{u}_1})$ acting on the first term needs to be kept, since acting on the correction term would yield higher-order contributions. This gives
\begin{align}
\frac{\td I^{\text{in-out}}_{\rm ISHA}}{\td z \td^2\vec{l}} &= -\frac{\alpha_s}{2\pi^2 \omega}P_{a\to bc}(z)\,\text{Re}\,i\int_{t_0}^L \td t_1\,e^{-\frac{i\vec{l}^2}{2\omega}(L-t_1)}\,\Bigg[{\cal C}^{(3)}(L,t_1;[\vec{u}_{\rm cl},\vec{0}])\nonumber\\
&\quad+\,i\frac{\vec{l}}{\vec{l}^2}\cdot\partial_{\vec{u}_1}{\cal C}^{(3)}(L,t_1;[\vec{u}_{\rm cl},\vec{0}])+\frac{i}{2\omega}\int_{t_1}^{L}\td s\,\left[\partial_{\vec{u}_1}^2{\cal C}^{(3)}(L,s;[\vec{u}_{\rm cl},\vec{0}])\right]{\cal C}^{(3)}(s,t_1;[\vec{u}_{\rm cl},\vec{0}])\Bigg]_{\vec{u}_1=0}\,.\label{eq:inoutISH}
\end{align}

Similarly, the 4-point function can be written as
\begin{align}
&\tilde{\cal S}^{(4)}(\vec{l},L;\vec{l}_2,\bar{\vec{l}}_2,t_2;z) \simeq \int \td^2\vec{u}_2\, \td^2\bar{\vec{u}}_2\,e^{-i\vec{u}_2\cdot(\vec{l}-\vec{l}_2)+i\bar{\vec{u}}_2\cdot(\vec{l}-\bar{\vec{l}}_2)}\,\bigg[{\cal C}^{(4)}(L,t_2;[\vec{u}_{\rm cl},\bar{\vec{u}}_{\rm cl},\vec{0}])\nonumber\\
&\qquad+\frac{i}{2\omega}\int_{t_2}^L\td s\,\left[(\partial^2_{\vec{u}_2}-\partial^2_{\bar{\vec{u}}_2})\,{\cal C}^{(4)}(L,s;[\vec{u}_{\rm cl},\bar{\vec{u}}_{\rm cl},\vec{0}])\right]\mathcal{C}_F^{(2)}(s,t_2;[\vec{u}_{\rm cl}])\nonumber\\
&\qquad\qquad\times \,{\cal C}_F^{(2)}(s,t_2;[\bar{\vec{u}}_{\rm cl}])\,\left[\mathcal{C}_F^{(2)}(s,t_2;[-z(\vec{u}_{\rm cl}-\bar{\vec{u}}_{\rm cl})])\right]^{n-1}\left[\mathcal{C}_F^{(2)}(s,t_2;[(1-z)(\vec{u}_{\rm cl}-\bar{\vec{u}}_{\rm cl})])\right]^{m-1}\bigg]\,,
\label{eq:S4ISH}
\end{align}
where now $\vec{u}_{\rm cl}(t)=\vec{u_2}+\frac{\vec{l}}{\omega}(t-t_2)$ and $\bar{\vec{u}}_{\rm cl}(t)=\bar{\vec{u}}_{2}+\frac{\vec{l}}{\omega}(t-t_2)$. Before inserting eqs.~\eqref{eq:K3ISH} and \eqref{eq:S4ISH} into the in-in contribution~\eqref{eq:inin}, it is instructive to consider the convolution over $\vec{l}_2$ for the zeroth order term
\begin{align}
&\int_{\vec{l}_2}\tilde{\cal S}^{(4)}(\vec{l},L;\vec{l}_2,\bar{\vec{l}}_2,t_2;z)\,\tilde{\cal K}^{(3)}(\vec{l}_2,t_2;\vec{l}_1,t_1;z)\big|_{0^{\rm th}} = \int_{\vec{l}_2}\int\td^2 \vec{u}_1\,\td^2 \vec{u}_2\,\td^2 \bar{\vec{u}}_2\,e^{-\frac{i\vec{l}_2^2}{2\omega}(t_2-t_1)}\,e^{-i\vec{u}_1\cdot(\vec{l}_2-\vec{l}_1)}\nonumber\\
&\qquad\times e^{-i\vec{u}_2\cdot(\vec{l}-\vec{l}_2)+i\bar{\vec{u}}_2\cdot(\vec{l}-\bar{\vec{l}}_2)}\,{\cal C}^{(4)}\left(L,t_2;\left[\vec{u}_2+\frac{\vec{l}}{\omega}(t-t_2),\bar{\vec{u}}_{\rm cl},\vec{0}\right]\right)\,{\cal C}^{(3)}\left(t_2,t_1;\left[\vec{u}_1+\frac{\vec{l}_ 2}{\omega}(t-t_1),\vec{0}\right]\right)\nonumber\\
&=\int_{\vec{l}_2}\int\td^2 \vec{u}_1\,\td^2 \vec{u}_2\,\td^2 \bar{\vec{u}}_2\,e^{-\frac{i\vec{l}^2}{2\omega}(t_2-t_1)}\,e^{-i\vec{u}_1\cdot(\vec{l}-\vec{l}_1)+i\bar{\vec{u}}_2\cdot(\vec{l}-\bar{\vec{l}}_2)}\,e^{-\frac{i\omega}{2(t_2-t_1)}\left(\vec{u}_2-\vec{u}_1-\frac{\vec{l}_2}{\omega}(t_2-t_1)\right)^2}\,e^{\frac{i\omega}{2(t_2-t_1)}\left(\vec{u}_2-\vec{u}_1-\frac{\vec{l}}{\omega}(t_2-t_1)\right)^2}\nonumber\\
&\qquad\times{\cal C}^{(4)}\left(L,t_2;\left[\vec{u}_2+\frac{\vec{l}}{\omega}(t-t_2),\bar{\vec{u}}_{\rm cl},\vec{0}\right]\right)\,{\cal C}^{(3)}\left(t_2,t_1;\left[\vec{u}_1+\frac{\vec{l}_2}{\omega}(t-t_1),\vec{0}\right]\right)\,.
\end{align}
As in section~\ref{subsec:Fourierprop}, we perform the $\vec{l}_2$ integral by expanding around $\vec{u}_2 - \vec{u}_1 = \frac{\vec{l}_2}{\omega} (t_2-t_1)$, and then the $\vec{u}_2$ integral around $\vec{u}_2 - \vec{u}_1 = \frac{\vec{l}}{\omega} (t_2-t_1)$. This leads to
\begin{align}
&\int_{\vec{l}_2}\tilde{\cal S}^{(4)}(\vec{l},L;\vec{l}_2,\bar{\vec{l}}_2,t_2;z)\,\tilde{\cal K}^{(3)}(\vec{l}_2,t_2;\vec{l}_1,t_1;z)\big|_{0^{\rm th}} = \int\td^2\vec{u}_1\td^2\bar{\vec{u}}_2\,e^{-\frac{i\vec{l}^2}{2\omega}(t_2-t_1)}\,e^{-i\vec{u}_1\cdot(\vec{l}-\vec{l}_1)+i\bar{\vec{u}}_2\cdot(\vec{l}-\bar{\vec{l}}_2)}\nonumber\\
&\qquad\times\,\bigg[{\cal C}^{(4)}(L,t_2;[\vec{u}_{\rm cl},\bar{\vec{u}}_{\rm cl},\vec{0}])\,{\cal C}^{(3)}(t_2,t_1;[\vec{u}_{\rm cl},\vec{0}])\nonumber\\
&\qquad\qquad+\frac{i}{\omega}\partial_{\vec{u}_1}^i{\cal C}^{(4)}(L,t_2;[\vec{u}_{\rm cl},\bar{\vec{u}}_{\rm cl},\vec{0}])\int_{t_1}^{t_2}\td s\,\left[\partial_{\vec{u}_1}^i{\cal C}^{(3)}(t_2,s;[\vec{u}_{\rm cl},\vec{0}])\right]\,{\cal C}^{(3)}(s,t_1;[\vec{u}_{\rm cl},\vec{0}])\nonumber\\
&\qquad\qquad+\frac{i(t_2-t_1)}{2\omega}\,\partial_{\vec{u}_1}^2{\cal C}^{(4)}(L,t_2;[\vec{u}_{\rm cl},\bar{\vec{u}}_{\rm cl},\vec{0}])\,{\cal C}^{(3)}(t_2,t_1;[\vec{u}_{\rm cl},\vec{0}])\bigg]\,,
\end{align}
where now
\begin{align}
\vec{u}_{\rm cl}(t) &= \vec{u}_1+\frac{\vec{l}}{\omega}(t-t_1) \quad\text{for}\quad t_1<t<L\,,\label{eq:classpath}\\ 
\bar{\vec{u}}_{\rm cl}(t) &= \bar{\vec{u}}_2+\frac{\vec{l}}{\omega}(t-t_2) \quad\text{for}\quad t_2<t<L\,.\label{eq:classpathbar}
\end{align}
This convolution has the same structure as in previous steps: a zeroth order term with no derivatives, plus corrections which are suppressed by inverse powers of the energy. The convolution over $\vec{l}_2$ also includes the remaining terms in eqs.~\eqref{eq:K3ISH} and \eqref{eq:S4ISH}. However, these contributions enter only at zeroth order, with no additional derivative terms, since they are already suppressed by powers of the energy. Their inclusion therefore amounts to simply replace the definition of the free path eq.~\eqref{eq:classpath} in all terms in eqs.~\eqref{eq:K3ISH} and \eqref{eq:S4ISH}.

The in-in contribution to the splitting spectrum is then given by
\begin{align}
\frac{\td I^{\text{in-in}}_{\rm ISHA}}{\td z\td^2\vec{l}} &= \frac{\alpha_s}{(2\pi)^2\omega^2}P_{a\to bc}(z)\,\text{Re}\int_{t_0}^L \td t_1\int_{t_1}^L \td t_2\, e^{-i\frac{\vec{l}^2}{2\omega}(t_2-t_1)}\,(\vec{l}+i\partial_{\vec{u}_1})\cdot(\vec{l}-i\partial_{\bar{\vec{u}}_2})\nonumber\\
&\quad\times \bigg[\mathcal{C}^{(4)}(L,t_2;[\vec{u}_{\rm cl},\bar{\vec{u}}_{\rm cl},\vec{0}])\,\mathcal{C}^{(3)}(t_2,t_1;[\vec{u}_{\rm cl},\vec{0}])\nonumber\\
&\quad+\frac{i}{2\omega}\mathcal{C}^{(4)}(L,t_2;[\vec{u}_{\rm cl},\bar{\vec{u}}_{\rm cl},\vec{0}])\,\int_{t_1}^{L}\td s\,\left[\partial_{\vec{u}_1}^2{\cal C}^{(3)}(L,s;[\vec{u}_{\rm cl},\vec{0}])\right]{\cal C}^{(3)}(s,t_1;[\vec{u}_{\rm cl},\vec{0}])\nonumber\\
&\quad+\frac{i}{2\omega}\int_{t_2}^L\td s\,\left[(\partial^2_{\vec{u}_1}-\partial^2_{\bar{\vec{u}}_2})\,{\cal C}^{(4)}(L,s;[\vec{u}_{\rm cl},\bar{\vec{u}}_{\rm cl},\vec{0}])\right]\mathcal{C}_F^{(2)}(s,t_2;[\vec{u}_{\rm cl}])\nonumber\\
&\qquad\qquad\times \,{\cal C}_F^{(2)}(s,t_2;[\bar{\vec{u}}_{\rm cl}])\,\left[\mathcal{C}_F^{(2)}(s,t_2;[-z(\vec{u}_{\rm cl}-\bar{\vec{u}}_{\rm cl})])\right]^{n-1}\left[\mathcal{C}_F^{(2)}(s,t_2;[(1-z)(\vec{u}_{\rm cl}-\bar{\vec{u}}_{\rm cl})])\right]^{m-1}\nonumber\\
&\qquad\qquad\times\mathcal{C}^{(3)}(t_2,t_1;[\vec{u}_{\rm cl},\vec{0}])\nonumber\\
&\quad+\frac{i}{\omega}\,\partial_{\vec{u}_1}^i{\cal C}^{(4)}(L,t_2;[\vec{u}_{\rm cl},\bar{\vec{u}}_{\rm cl},\vec{0}])\int_{t_1}^{t_2}\td s\,\left[\partial_{\vec{u}_1}^i{\cal C}^{(3)}(t_2,s;[\vec{u}_{\rm cl},\vec{0}])\right]\,{\cal C}^{(3)}(s,t_1;[\vec{u}_{\rm cl},\vec{0}])\nonumber\\
&\quad+\frac{i(t_2-t_1)}{2\omega}\,\partial_{\vec{u}_1}^2{\cal C}^{(4)}(L,t_2;[\vec{u}_{\rm cl},\bar{\vec{u}}_{\rm cl},\vec{0}])\,{\cal C}^{(3)}(t_2,t_1;[\vec{u}_{\rm cl},\vec{0}])\bigg]_{\substack{\vec{u}_1=0 \\ \bar{\vec{u}}_2=0}}\,,
\end{align}
where, as for the in-out contribution, the derivatives in the factors $(\vec{l}+i\partial_{\vec{u}_1})\cdot(\vec{l}-i\partial_{\bar{\vec{u}}_2})$ are kept only when acting over the first term inside the square brackets. The final expression for the in-in contribution then reads
\begin{align}
\frac{\td I^{\text{in-in}}_{\rm ISHA}}{\td z\td^2\vec{l}} &= \frac{\alpha_s\,\vec{l}^2}{(2\pi)^2\omega^2}P_{a\to bc}(z)\,\text{Re}\int_{t_0}^L \td t_1\int_{t_1}^L \td t_2\, e^{-i\frac{\vec{l}^2}{2\omega}(t_2-t_1)}\,\nonumber\\
&\quad\times \bigg[\left(1+i\frac{\vec{l}}{\vec{l}^2}\cdot\partial_{\vec{u}_1}\right)\left(1-i\frac{\vec{l}}{\vec{l}^2}\cdot\partial_{\bar{\vec{u}}_2}\right)\mathcal{C}^{(4)}(L,t_2;[\vec{u}_{\rm cl},\bar{\vec{u}}_{\rm cl},\vec{0}])\,\mathcal{C}^{(3)}(t_2,t_1;[\vec{u}_{\rm cl},\vec{0}])\nonumber\\
&\quad+\frac{i}{2\omega}\mathcal{C}^{(4)}(L,t_2;[\vec{u}_{\rm cl},\bar{\vec{u}}_{\rm cl},\vec{0}])\,\int_{t_1}^{L}\td s\,\left[\partial_{\vec{u}_1}^2{\cal C}^{(3)}(L,s;[\vec{u}_{\rm cl},\vec{0}])\right]{\cal C}^{(3)}(s,t_1;[\vec{u}_{\rm cl},\vec{0}])\nonumber\\
&\quad+\frac{i}{2\omega}\int_{t_2}^L\td s\,\left[(\partial^2_{\vec{u}_1}-\partial^2_{\bar{\vec{u}}_2})\,{\cal C}^{(4)}(L,s;[\vec{u}_{\rm cl},\bar{\vec{u}}_{\rm cl},\vec{0}])\right]\mathcal{C}_F^{(2)}(s,t_2;[\vec{u}_{\rm cl}])\nonumber\\
&\qquad\qquad\times \,{\cal C}_F^{(2)}(s,t_2;[\bar{\vec{u}}_{\rm cl}])\,\left[\mathcal{C}_F^{(2)}(s,t_2;[-z(\vec{u}_{\rm cl}-\bar{\vec{u}}_{\rm cl})])\right]^{n-1}\left[\mathcal{C}_F^{(2)}(s,t_2;[(1-z)(\vec{u}_{\rm cl}-\bar{\vec{u}}_{\rm cl})])\right]^{m-1}\nonumber\\
&\qquad\qquad\times\mathcal{C}^{(3)}(t_2,t_1;[\vec{u}_{\rm cl},\vec{0}])\nonumber\\
&\quad+\frac{i}{\omega}\,\partial_{\vec{u}_1}^i{\cal C}^{(4)}(L,t_2;[\vec{u}_{\rm cl},\bar{\vec{u}}_{\rm cl},\vec{0}])\int_{t_1}^{t_2}\td s\,\left[\partial_{\vec{u}_1}^i{\cal C}^{(3)}(t_2,s;[\vec{u}_{\rm cl},\vec{0}])\right]\,{\cal C}^{(3)}(s,t_1;[\vec{u}_{\rm cl},\vec{0}])\nonumber\\
&\quad+\frac{i(t_2-t_1)}{2\omega}\,\partial_{\vec{u}_1}^2{\cal C}^{(4)}(L,t_2;[\vec{u}_{\rm cl},\bar{\vec{u}}_{\rm cl},\vec{0}])\,{\cal C}^{(3)}(t_2,t_1;[\vec{u}_{\rm cl},\vec{0}])\bigg]_{\substack{\vec{u}_1=0 \\ \bar{\vec{u}}_2=0}}\,,
\label{eq:ininISH}
\end{align}
with $\vec{u}_{\rm cl}(t) = \vec{u}_1+\frac{\vec{l}}{\omega}(t-t_1)$ and $\bar{\vec{u}}_{\rm cl}(t) = \bar{\vec{u}}_2+\frac{\vec{l}}{\omega}(t-t_2)$.

The full in-medium splitting spectrum is then obtained by summing eqs.~\eqref{eq:inoutISH} and \eqref{eq:ininISH}, using ${\cal C}^{(2)}_F$ from~\eqref{eq:C2largeNc}, ${\cal C}^{(3)}$ from~\eqref{eq:C3largeNc}, and ${\cal C}^{(4)}$ from the sum of eqs.~\eqref{eq:C4fac} and \eqref{eq:C4nonfac}. For a given functional form of the dipole cross section $\sigma$, the transverse derivatives in~\eqref{eq:inoutISH} and \eqref{eq:ininISH} can be computed analytically, and the longitudinal integrals can be evaluated either analytically or numerically, depending on the specific choice of $\sigma$, without incurring significant computational cost.

It is worth noting that only the first terms in eqs.~\eqref{eq:inoutISH} and \eqref{eq:ininISH}, which do not involve  transverse derivatives, appear in the semi-hard approximation (SHA) of \cite{Dominguez:2019ges,Isaksen:2020npj}, i.e.,
\begin{equation}
\frac{\td I^{\text{in-out}}_{\rm SHA}}{\td z \td^2\vec{l}} 
= -\frac{\alpha_s}{2\pi^2 \omega}P_{a\to bc}(z)\,\int_{t_0}^L \td t_1\,\sin \left(\frac{\vec{l}^2}{2\omega}(L-t_1)\right)\,\left.\mathcal{C}^{(3)}(L,t_1;[\vec{u}_{\rm cl},\vec{0}])\right|_{\vec{u}_1=0}\,,
\label{eq:inoutsh}
\end{equation}
and
\begin{align}
\frac{\td I^{\text{in-in}}_{\rm SHA}}{\td z\td^2\vec{l}} 
&= \frac{\alpha_s\,\vec{l}^2}{(2\pi)^2\omega^2}P_{a\to bc}(z)\,\int_{t_0}^L \td t_1\int_{t_1}^L \td t_2\, \cos \left(\frac{\vec{l}^2}{2\omega}(t_2-t_1)\right)\nonumber\\
&\qquad\qquad\qquad\qquad\qquad\qquad\quad\times\left[\mathcal{C}^{(3)}(t_2,t_1;[\vec{u}_{\rm cl},\vec{0}])\,\mathcal{C}^{(4)}(L,t_2;[\vec{u}_{\rm cl},\bar{\vec{u}}_{\rm cl},\vec{0}])\right]_{\substack{\vec{u}_1=0 \\ \bar{\vec{u}}_2=0}}\,.
\end{align}
An important observation is that, within the semi-hard approximation, the initial transverse positions of the straight-line trajectories are usually neglected. Our ISHA framework shows that the leading term in a systematic high-energy expansion naturally sets the differences between these initial positions to zero, thereby reproducing the original SHA result in \cite{Dominguez:2019ges,Isaksen:2020npj}. While appendix~A of~\cite{Dominguez:2019ges} sketches how to incorporate corrections associated with these initial positions, we find that such contributions enter at the same order as those arising from the high-energy expansion of the propagators, and therefore should not be  considered separately.

Although the ISHA and SHA expressions derived here are general and valid for any dipole cross section, in the following we will adopt the harmonic oscillator approximation of eq.~\eqref{eq:sigmaho}. This choice enables a direct comparison with the results of section~\ref{sec:HO}, and thus facilitates a quantitative assessment robustness of the ISHA and SHA. Within the HO, the relevant Wilson line averages are
\begin{align}
{\cal C}^{(2)}_{F,{\rm HO}}(t_2,t_1;[\vec{r}]) &= \exp\left\{-\frac{1}{4}\int_{t_1}^{t_2}\td s\,\hat{q}(s)\,\vec{r}^2(s)\right\}\,,\\
{\cal C}^{(3)}_{\rm HO}(t_2,t_1;[\vec{u},\vec{0}]) &= \exp\left\{-\frac{1}{4}\int_{t_1}^{t_2}\td s\,\hat{q}(s)\,\left(p+(n-1)z^2+(m-1)(1-z)^2\right)\,\vec{u}^2(s)\right\}\,,\\
{\cal C}^{(4)}_{\rm fac,HO}(L,t_2;[\vec{u},\bar{\vec{u}},\vec{0}]) &= \exp\left\{-\frac{1}{4}\int_{t_1}^{t_2}\td s\,\hat{q}(s)\,\left(nz^2+m(1-z)^2\right)\,\left(\vec{u}(s)-\bar{\vec{u}}(s)\right)^2\right\}\,,\\
{\cal C}^{(4)}_{\rm non\text{-}fac,HO}(L,t_2;[\vec{u},\bar{\vec{u}},\vec{0}]) &= -\int_{t_2}^L\td t_3\,\frac{z(1-z)}{2}\hat{q}(t_3)\left(\vec{u}(t_3)-\bar{\vec{u}}(t_3)\right)^2\nonumber\\
&\qquad\times\exp\left\{-\frac{1}{4}\int_{t_3}^{L}\td s\,\hat{q}(s)\,\left(nz^2+m(1-z)^2\right)\,\left(\vec{u}(s)-\bar{\vec{u}}(s)\right)^2\right\}\nonumber\\
&\qquad\times\exp\left\{-\frac{1}{4}\int_{t_3}^{L}\td s\,\hat{q}(s)\,\left(\vec{u}^2(s)+\bar{\vec{u}}^2(s)\right.\right.\nonumber\\
&\qquad\qquad\quad\left.\left.+((n-1)z^2+(m-1)(1-z)^2)\left(\vec{u}(s)-\bar{\vec{u}}(s)\right)^2\right)\,\right\}\,.
\end{align}

%%%%%%%%%%%%%%%%%%%%%%%%%%%%%%%%%%%%%%%%%%%%%
%%%%%%%%%%%%%%%%%%%%%%%%%%%%%%%%%%%%%%%%%%%%%

\section{Numerical Results}
\label{sec:evaluation}

\begin{figure}
\centering
\includegraphics[scale=0.40]{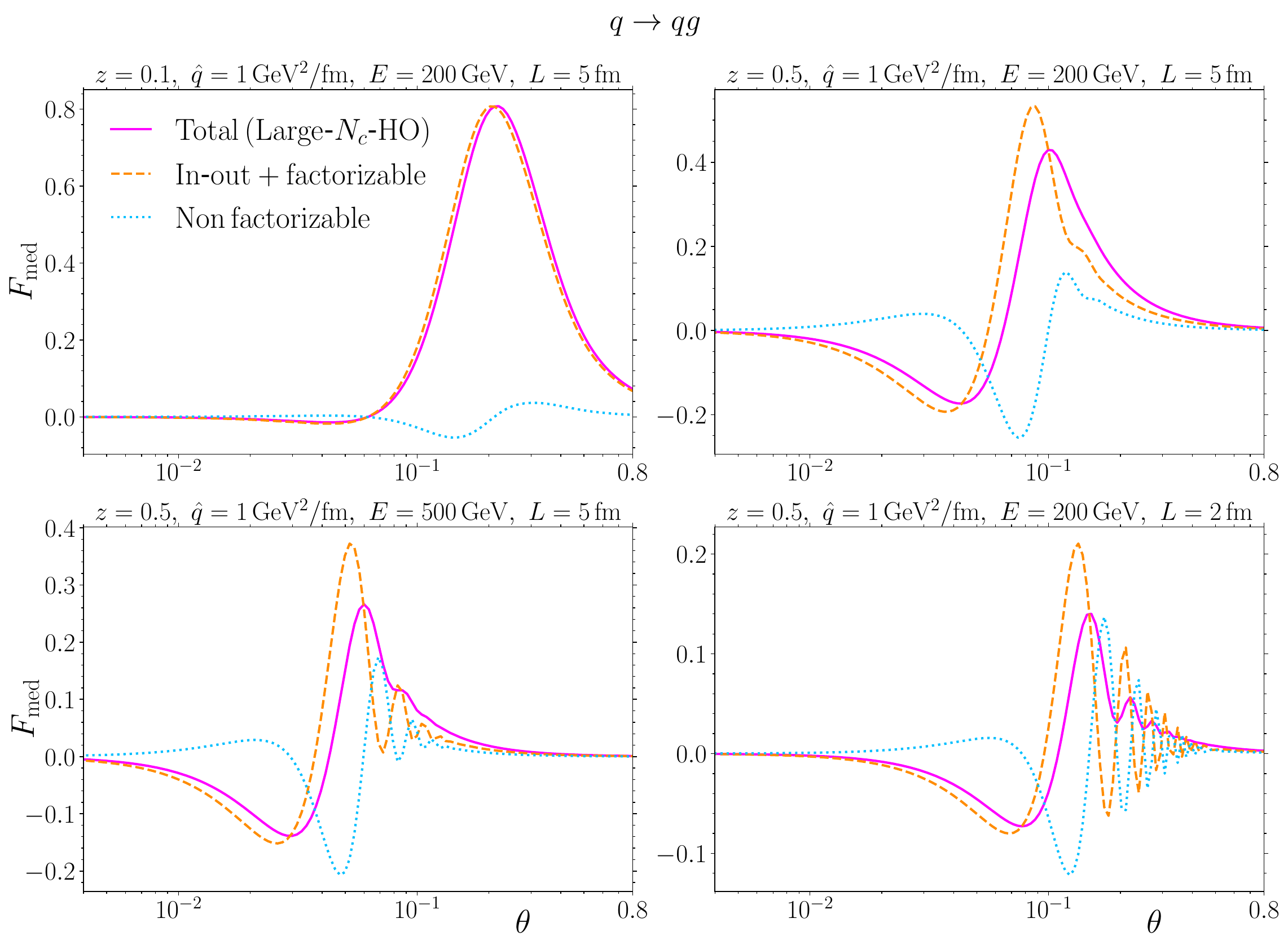}
\caption{$F_{\rm med}$ for $q \to qg$ as a function of $\theta$ in the  large-$N_c$-HO approach (see section~\ref{sec:HO}). Solid magenta curves show the full large-$N_c$–HO result, orange dashed curves correspond to the result without the non-factorizable term, and  blue dotted curves show the non-factorizable term alone. The top-left and top-right panels correspond to $z=0.1$ and $z=0.5$, respectively, for the same emitter energy and medium parameters (indicated in the figure). The bottom panels show $z=0.5$ for a higher emitter energy (left) and a shorter medium length (right).}
\label{fig:Fmed_qg_LNC}
\end{figure}

In this section, we present the numerical results for the in-medium $q \to qg$ splitting. The picture  emerging from these results is the same as for the  $g \to gg$ channel, for this reason, the results for the latter are deferred to appendix~\ref{app:gg}. To isolate the medium-induced contribution to the splitting, we introduce the modification factor $F_{\rm med}$, defined as
\beq
F_{\rm med}(z, \theta) = \left(\frac{\td I^{\text{in-in}}}{\td z \td^2 \vec{l}}+\frac{\td I^{\text{in-out}}}{\td z \td^2 \vec{l}}\right)\Bigg / \frac{\td I^{\text{out-out}}}{\td z \td^2 \vec{l}}\,,
\label{eq:Fmed}
\eeq
where $\theta=\frac{|\vec{l}|}{\omega}$ as defined in eq.~\eqref{eq:theta} and the out-out term corresponds to the vacuum contribution given in eq.~\eqref{eq:outout}. The quantity $F_{\rm med}$ thus quantifies how much the in-medium emission spectrum is modified relative to its vacuum counterpart. In the following, we will present results for $F_{\rm med}$ in the  large-$N_c$-HO approach, as well as within the ISHA and the SHA, all evaluated with the same HO dipole cross section. These correspond to different evaluations of the numerator in~\eqref{eq:Fmed}. In most of this section, we choose to show results for a static medium of length $L$ with constant $\hat{q}$ (a “brick”), which makes the dependence on the medium parameters most transparent. Finally, to illustrate the applicability of all three formalisms to longitudinally expanding media, we include as the last figure of this section a comparison made for a trajectory sampled from a hydrodynamic simulation of the QGP.

We begin with the large-$N_c$-HO approach, introduced in section~\ref{sec:HO}, which is currently the most general framework available for these splittings when resumming all in-medium interactions.  It also allows for (semi-)analytical control and numerically efficient evaluation. Finite-$N_c$ corrections are known to scale as $1/N_c^2$, but their size remains to be fully quantified. It was shown in \cite{Isaksen:2023nlr} that for the $\gamma \to q \bar q$ channel in a thin medium these corrections are very small across the entire ($\theta$, $z$)-plane, typically below  1\%. Their impact in other channels is, however, less well understood, as finite-$N_c$ results for $q \to qg$ and $g \to gg$ splittings are only available under further approximations. Existing finite-$N_c$ calculations in these channels have been performed within the SHA, where finite-$N_c$  were found to be below 5\% over most of the phase space. Only at very large angles do these corrections become more sizable, reaching up to $\sim10\%$ \cite{Isaksen:2020npj}. A complete $N_c$ calculation for this channels would therefore be  valuable to accurately assess the size of the  $1/N_c^2$-terms. 

The results for $F_{\rm med}$ in the $q \to qg$ channel, computed in the large-$N_c$-HO approach, are shown in figure~\ref{fig:Fmed_qg_LNC}. The blue dotted curves show the contribution to $F_{\rm med}$ obtained by retaining only the non-factorizable part of the in-in term in the numerator of eq.~\eqref{eq:Fmed}, namely eq.~\eqref{eq:ininhononfacexpl}. The orange curves correspond to $F_{\rm med}$ including the contribution from the in-out term, eq.~\eqref{eq:inoutho}, together with the factorizable part of the in-in term, eq.~\eqref{eq:ininhofac}. The magenta solid curves represent the full result for $F_{\rm med}$, obtained by including all contributions in the numerator, i.e. eqs.~\eqref{eq:inoutho},~\eqref{eq:ininhofac}, and~\eqref{eq:ininhononfacexpl}. The top-left panel, corresponding to $z=0.1$, confirms that the non-factorizable piece is negligible in the soft ($z \ll 1 $) limit. This is expected, as the non-factorizable term~\eqref{eq:ininhononfacexpl} is proportional to $z(1-z)$ and therefore vanishes as  $z \to 0$ and $z \to 1$. Away from these limits, however, this contribution can have a sizable impact. To illustrate this, the remaining panels show results for $z=0.5$. We observe that the non-factorizable contribution plays a significant role for symmetric splittings, and neglecting it leads to unphysical oscillations in the splitting spectrum. Its relative importance further increases with increasing emitter energy or/and decreasing medium length.  This observation is particularly relevant given that many analytical studies neglect the non-factorizable term to simplify the treatment of the quadrupole.  Our findings, together with those in \cite{Isaksen:2020npj} for the $\gamma \to q \bar q$ channel, demonstrate that the non-factorizable contribution can have a sizable impact beyond the soft limit. Finally, we note that, as shown in figure~\ref{fig:Fmed_qg_LNC_qhat} of appendix~\ref{app:qhat}, the relative importance of the non-factorizable contribution slightly decreases with increasing $\hat{q}$ when all other parameters are held fixed. Overall, our results are consistent with the statement in \cite{Blaizot:2012fh} that the non-factorizable contribution is small for small $t_{\rm br}=\sqrt{\omega/\hat{q}}$. However, away from this limit is becomes sizable and can no longer be neglected.

\begin{figure}[t]
\centering
\includegraphics[scale=0.43]{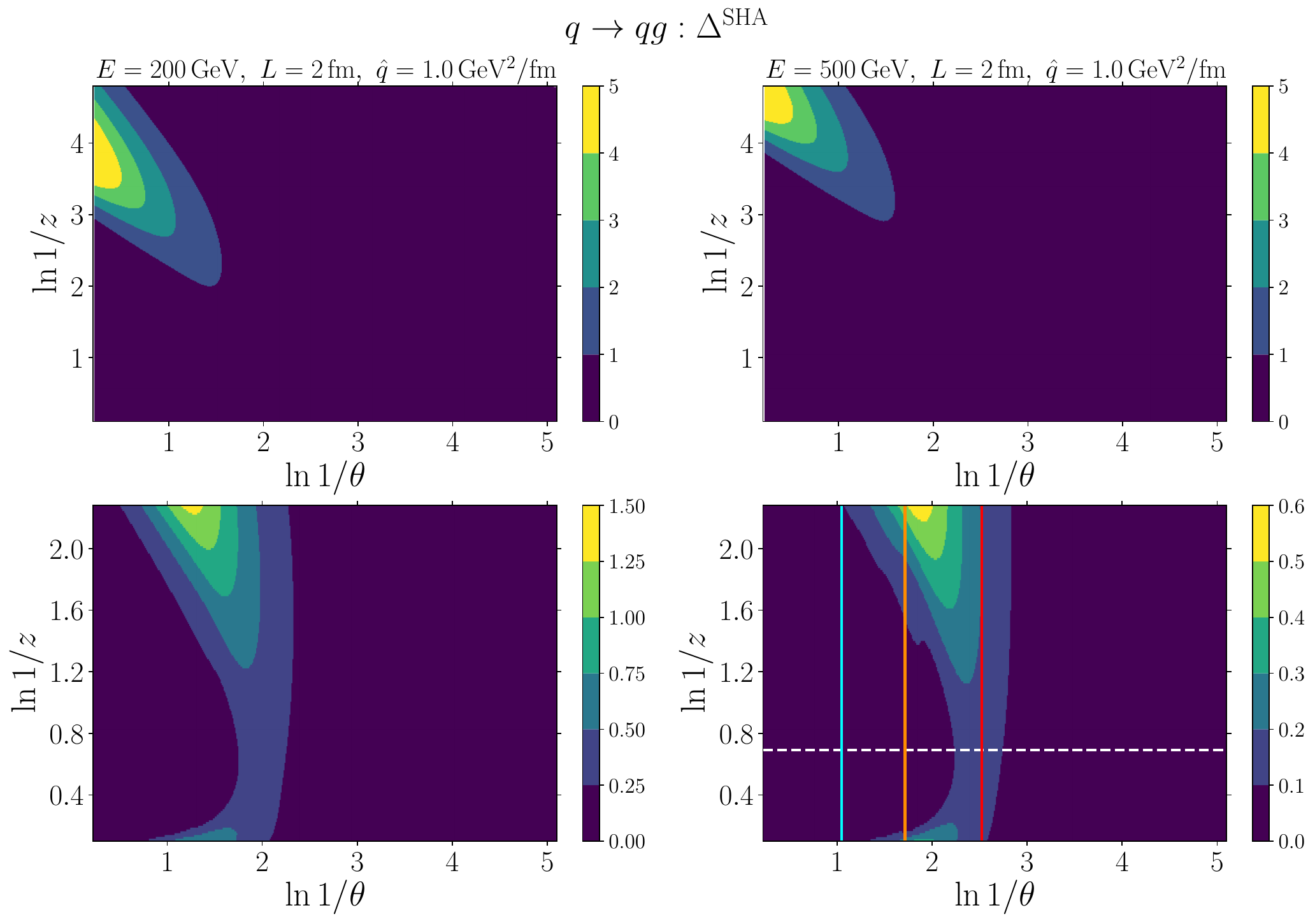}
\caption{Difference between $1+ F_{\rm med}$ for $q \to qg$ calculated using the SHA and large-$N_c$-HO approaches, as defined in Eq.~\eqref{eq:SHA_diff}. Both columns use $L=2$\,fm and $\hat q=1\, {\rm GeV^2/fm}$, with $E = 200$\,GeV (left) and $E = 500$\,GeV (right). The bottom row shows a zoom of the top row in the region $0.1 \leq z \leq 0.9$ (or equivalently $0.10 \leq  \ln 1/z \leq 2.3$). On the bottom-right panel, vertical lines indicate the $\theta$ values used for the projections in figure~\ref{fig:Fmed_qg_vs_z}: $\theta =0.08$ (red), $\theta = 0.18$ (orange), and $\theta = 0.35$ (cyan), while the dashed horizontal line marks $z = 0.5$.}
\label{fig:LP_SHA}
\end{figure}

We now turn to a comparison between the semi-hard (SHA) and improved semi-hard (ISHA) approximations, introduced in section~\ref{sec:ISHA}, and the complete large-$N_c$-HO results. In all figures in this manuscript, the SHA and ISHA are evaluated using the harmonic oscillator form of the dipole cross section, enabling a direct, “apples-to-apples” comparison with the large-$N_c$-HO result, which is only analytic under this approximation. We emphasize, however, that within the SHA and ISHA frameworks, other choices for the dipole cross section, such as Yukawa \cite{Wang:1992qdg} or HTL forms \cite{Aurenche:2002pd}, can be used without major difficulties.

\begin{figure}
\centering
\includegraphics[scale=0.42]{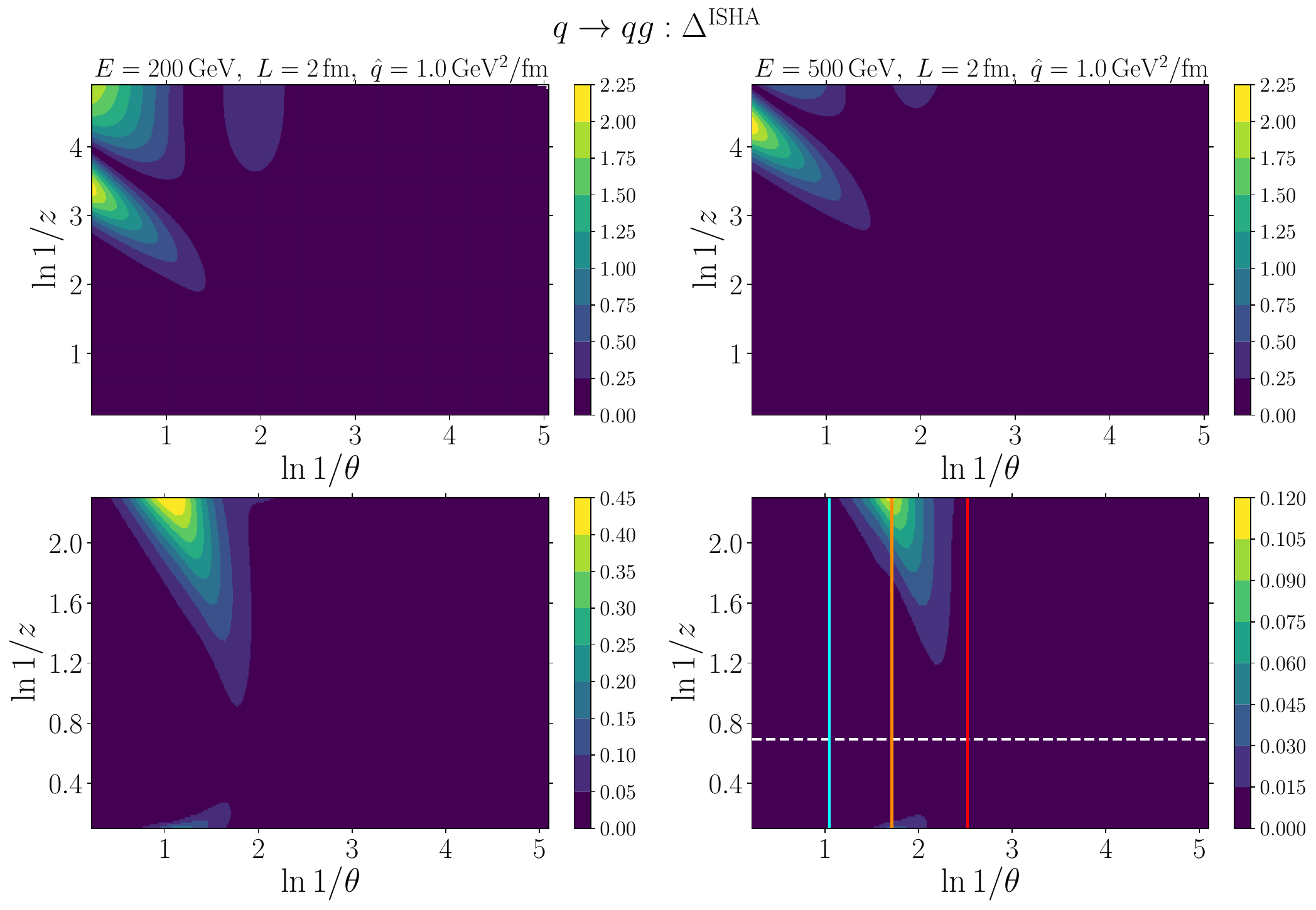}
\caption{Difference between $1+ F_{\rm med}$ for $q\to qg$ calculated using the ISHA and large-$N_c$-HO approaches, as defined in eq.~\eqref{eq:ISHA_diff}. The medium parameters and energies are the same as in Fig.~\ref{fig:LP_SHA}: $L = 2$\,fm, $\hat{q} = 1\,{\rm GeV^2/fm}$, and $E = 200$ GeV (left column) and $E = 500$ GeV (right column). The bottom row shows a zoom of the top row in the region $0.1 \leq z \leq 0.9$ (or equivalently $0.10 \leq  \ln 1/z \leq 2.3$). On the bottom-right panel, vertical lines indicate the $\theta$ values used for the projections in figure~\ref{fig:Fmed_qg_vs_z}: $\theta =0.08$ (red), $\theta = 0.18$ (orange), and $\theta = 0.35$ (cyan), while the dashed horizontal line marks $z = 0.5$.}
\label{fig:LP_ISHA}
\end{figure}

We begin by comparing the ($\ln 1/z$, $\ln 1/\theta$)-planes or Lund planes of the SHA with those obtained from the large-$N_c$-HO distribution. To make the comparisons clearer, we define the normalized difference 
\beq
\Delta^{{\rm SHA}}= \frac{\left|F_{\rm med}^{{\rm Large}\text{-}N_c\text{-}{\rm HO}}-F_{\rm med}^{\rm SHA}\right|}{\left|1+ F_{\rm med}^{{\rm Large\text{-}}N_c\text{-}{\rm HO}}\right|}\,,
\label{eq:SHA_diff}
\eeq
which serves as a proxy for the deviation of $1+F_{\rm med}$ between the SHA and the large-$N_c$-HO approach.  This quantity is shown  in the two-dimensional Lund plane in figure~\ref{fig:LP_SHA} for two different energies. As observed in the figure, the differences between the SHA and large-$N_c$-HO are not negligible, becoming particularly large for small $z$.  Increasing the emitter's energy (right column) reduces the size of the deviations\footnote{Note that the range of the color bar varies from panel to panel in  both figures \ref{fig:LP_SHA} and \ref{fig:LP_ISHA}.}. This behavior is expected, since the SHA is not reliable when either $zE$ or $(1-z)E$ becomes small. Therefore, at fixed $z$, increasing $E$, should reduce the discrepancies between the SHA and large-$N_c$-HO result.  Finally, we note that the emitter energies considered in this figure ($E=200$\,GeV and $E=500$\,GeV) are relatively large, given that the SHA approximation is not well suited for low energies. Consequently, at smaller energies, the discrepancies with respect to the large-$N_c$-HO result would be even more pronounced. 

In analogy to eq.~\eqref{eq:SHA_diff}, we define
\beq
\Delta^{{\rm ISHA}}= \frac{\left|F_{\rm med}^{{\rm Large}\text{-}N_c\text{-}{\rm HO}}-F_{\rm med}^{\rm ISHA}\right|}{\left |1+ F_{\rm med}^{{\rm Large}\text{-}N_c\text{-}{\rm HO}} \right|}\,,
\label{eq:ISHA_diff}
\eeq
as a proxy for the differences between the $1 + F_{\rm med}$ in the ISHA  and the large-$N_c$-HO approach. This quantity is presented in figure~\ref{fig:LP_ISHA}, for the same medium parameters and emitter energies as in figure~\ref{fig:LP_SHA}, allowing for a direct comparison between the SHA and ISHA deviations relative to the large-$N_c$-HO benchmark. As observed in figure~\ref{fig:LP_ISHA}, the differences between the ISHA and large-$N_c$-HO are significantly smaller than those between the SHA and large-$N_c$-HO  shown in  figure~\ref{fig:LP_SHA}, and remain very small across most of the phase space.  However, at small $z$, noticeable deviations persist for certain values of $\theta$. This is expected, since the SHIA framework is not anticipated to be reliable for  splittings {where the energy of at least one of the partons is not sufficiently high. Restricting to $0.1 \leq z \leq 0.9$ (bottom panels) provides a clearer view of the small discrepancies at intermediate $z$ values and shows that increasing the emitter energy (moving from the left to the right panel) further reduces both the phase-space region with sizable deviations and their overall magnitude.

\begin{figure}
\centering
\includegraphics[scale=0.40]{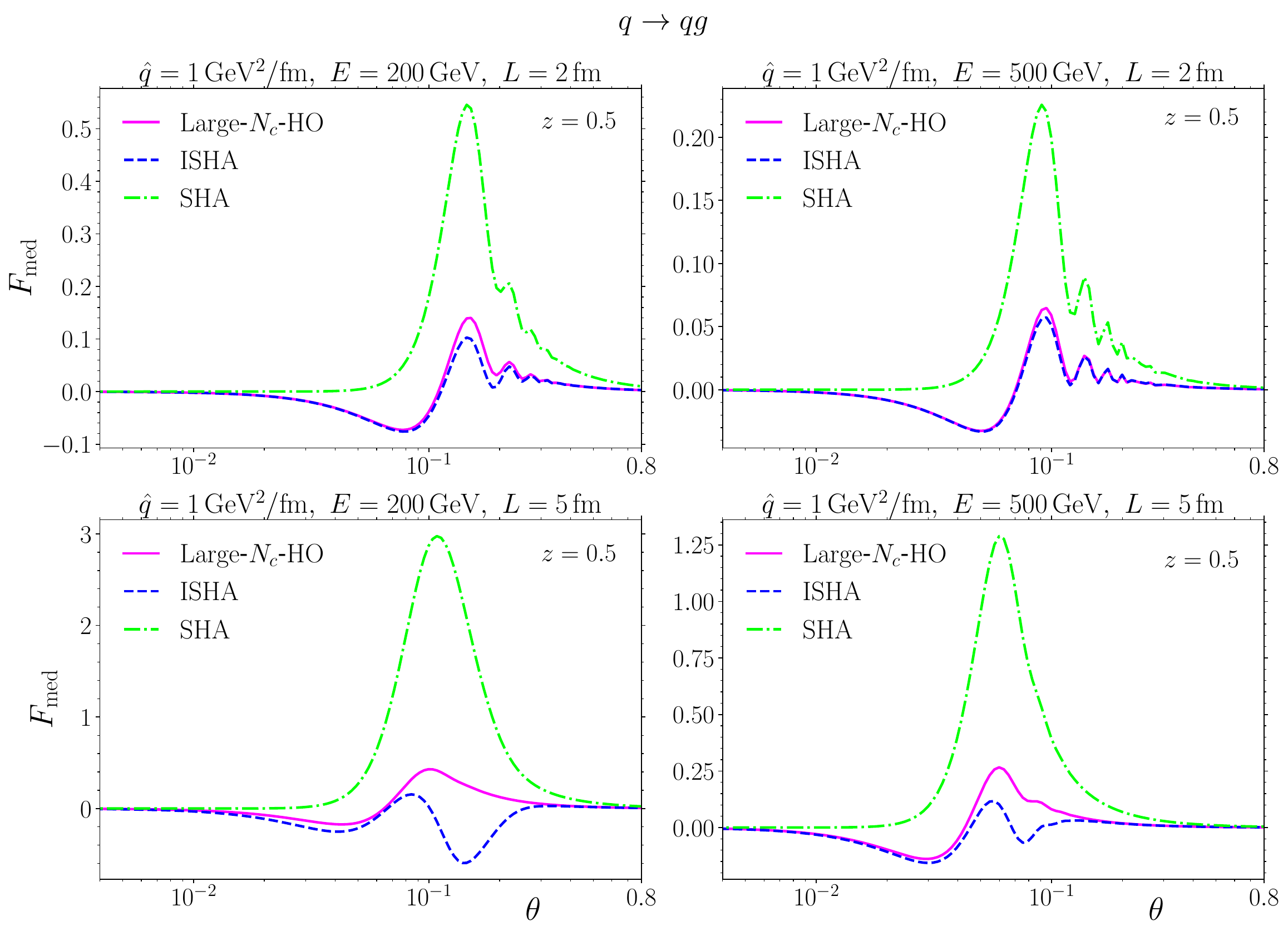}
\caption{$F_{\rm med}$ for the $q\to qg$ splitting as a function of $\theta$, computed using the large-$N_c$-HO (magenta solid), ISHA (blue dashed) and SHA (green dash-dotted)  for $z=0.5$. Each panel shows a different combination of emitter energy $E$ and medium length $L$, as indicated.}
\label{fig:Fmed_qg_vs_theta}
\end{figure}

Since understanding the differences between the various approaches directly from the two-dimensional distributions can be challenging, instead of presenting additional Lund planes for different parameter variations, we now show one-dimensional projections of $F_{\rm med}$. In particular, we compare the different approaches as a function of $z$ at fixed $\theta$, and as a function of $\theta$ at fixed $z$. 

We begin by figure~\ref{fig:Fmed_qg_vs_theta}, which shows $F_{\rm med}$ as a function of $\theta$ at fixed $z$, computed in the SHA (green dash-dotted), ISHA (blue dashed) and large-$N_c$-HO approach (magenta solid). As noted above, neither the SHA nor the ISHA is reliable for highly asymmetric splittings, we therefore focus here on $z=0.5$, deferring the discussion for other values of $z$ to the following figure. The different panels correspond to varying emitter energies and medium lengths. We observe that the deviations between the SHA and the large-$N_c$-HO results are significant across all panels, indicating that the SHA does not provide quantitatively reliable results even when $zE$ and $(1-z)E$ are relatively large. In contrast, for $z=0.5$, the ISHA closely approximates the large-$N_c$-HO results for high emitter energies and short medium lengths. All results here were obtained for $\hat q = 1\,{\rm GeV^2/fm}$, but the same general picture still holds when varying $\hat q$, as illustrated in figure~\ref{fig:Fmed_qg_qhat} of appendix~\ref{app:qhat}. There, one observes small discrepancies emerging at larger values $\hat q$, as expected, since deviations from the straight-line trajectories become more important when the  kicks kicks from the medium are stronger.

We now compare in figure~\ref{fig:Fmed_qg_vs_z}, $F_{\rm med}$ for the three approaches as a function of $z$ for three fixed values of $\theta$. The emitter energy and medium parameters in this figure where chosen to coincide with those on the right column of the Lund planes in figures~\ref{fig:LP_SHA}~and~\ref{fig:LP_ISHA}. The $\theta$ values used in figure~\ref{fig:Fmed_qg_vs_z} are indicated by vertical lines in those Lund planes. They are selected such that the two smallest angles (left and center column in figure~\ref{fig:Fmed_qg_vs_z}) correspond to a region with significant deviations between the SHA and the large-$N_c$-HO results, while the largest angle corresponds to a region where these deviations are as minimal as possible. As a first observation, the SHA shows sizable deviations from the large-$N_c$-HO result even at intermediate values of $z$, as seen in the bottom panels. In contrast, the ISHA provides a good approximation to the large-$N_c$-HO result in this intermediate-$z$ region. As expected, neither the SHA nor the ISHA accurately reproduces the large-$N_c$-HO result at very small or very large $z$ (see top panels). However, in these regions, the discrepancies of the ISHA relative to the large-$N_c$-HO remain significantly smaller than those of the SHA.

\begin{figure}
\centering
\includegraphics[scale=0.33]{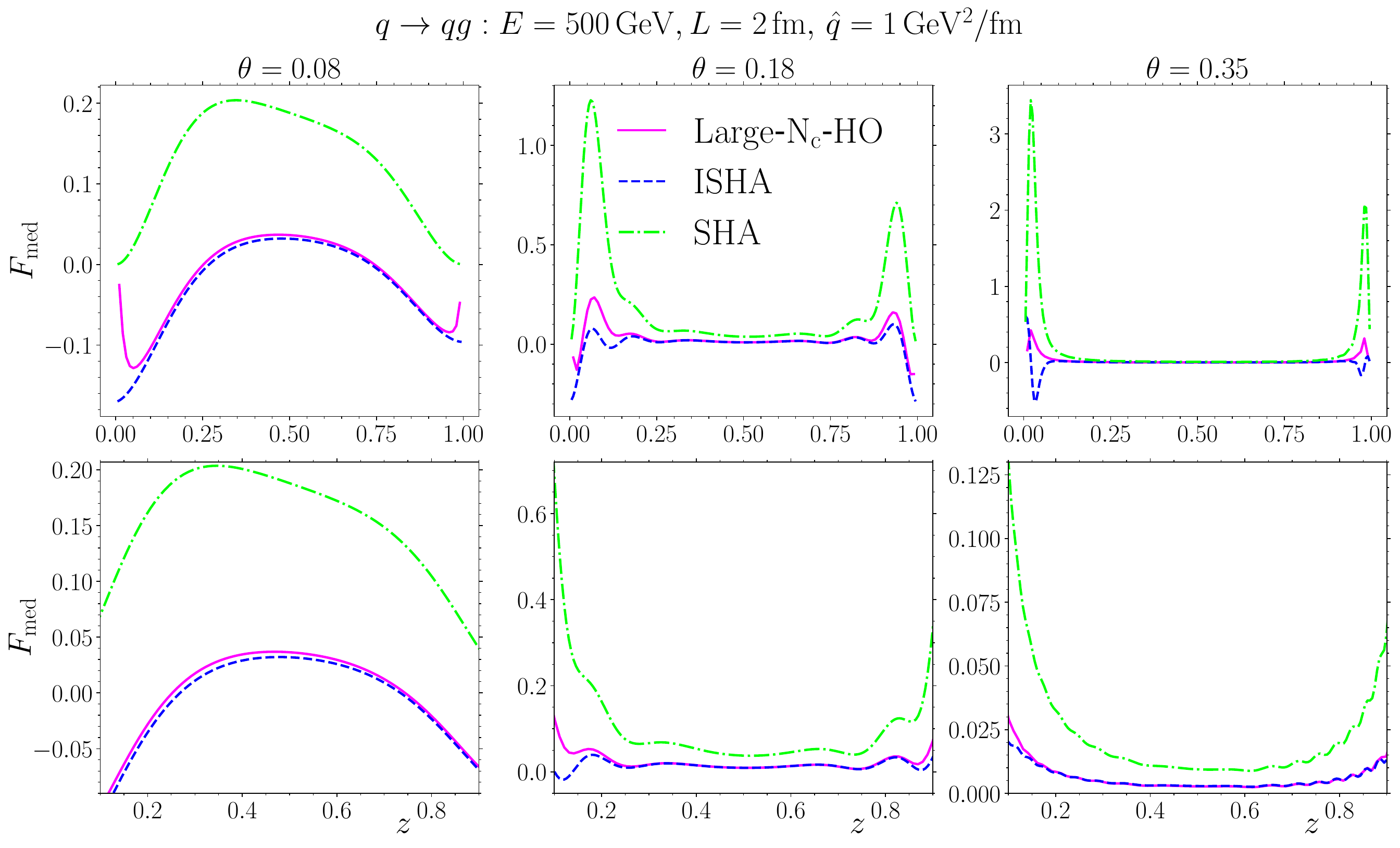}
\caption{$F_{\rm med}$ for the $q\to qg$ splitting as a function of $z$, computed using the large-$N_c$-HO (magenta solid), ISHA (blue dashed) and SHA (green dash-dotted),  for $E=500 $\, GeV, $L=2$\,fm, and $\hat {q}=1\,{\rm GeV^2/fm}$. The left, center, and right columns correspond to $\theta = 0.08$, $0.18$, and $0.35$, respectively. The bottom row shows a zoom of the top row in the region $0.1 \leq z \leq 0.9$.}
\label{fig:Fmed_qg_vs_z}
\end{figure}

To conclude this section, we illustrate the applicability of  ISHA and large-$N_c$-HO approaches developed in this manuscript to longitudinally evolving media. Within the HO approximation \eqref{eq:sigmaho}, the information about the medium entering $F_{\rm med}$ (in all approaches considered in this manuscript: SHA, ISHA and large-$N_c$-HO) is encoded in the length $L$ of the trajectory $\xi(t)$, sampled from a hydrodynamic simulation, and in the local jet quenching parameter $\hat q(t)$. Computing  $F_{\rm med}$ therefore requires specifying  the relation between $\hat q(t)$ and the local value of the temperature along the trajectory $T(\xi)$. We adopt the widely-used parametrization \cite{Baier:2002tc}
\beq
\hat{q}(t)= k \,T^3(\xi(t))\,,
\eeq
where $k$ is a free parameter and  the local  $T(\xi)$  will be extracted from the smooth-averaged 2+1 viscous hydrodynamic simulation developed in \cite{Luzum:2008cw,Luzum:2009sb}.

In figure~\ref{fig:Fmed_qg_evolv}, we present results for $F_{\rm med}$ within the three formalisms considered in the manuscript (SHA, ISHA and large-$N_c$-HO) evaluated along a typical straight-line trajectory sampled from the 0-10$\%$ centrality class in $\sqrt{s_{ \rm NN}}=5.02$ TeV Pb-Pb collisions, shown in the inset. We set $k=2$ and the emitter energy to $E=200$ GeV. The left panel shows $F_{\rm med}$ as a function of $\theta$ for fixed $z=0.5$. As in the static media case, the SHA deviates significantly from the  large-$N_c$-HO, while the ISHA provides a much more reliable approximation. We note that the agreement between ISHA and large-$N_c$-HO in this figure is worse than that shown on the top row of figure~\ref{fig:Fmed_qg_vs_theta}. This difference arises because the present setup corresponds to the lowest of the energies in figure~\ref{fig:Fmed_qg_vs_theta} ($E=200$ GeV) and a substantially longer trajectory than the $L=2$ fm case used in the top panel of \ref{fig:Fmed_qg_vs_theta}. Indeed, at fixed emitter energy and $k$, shorter trajectories (for instance, most trajectories in more peripheral collisions)  lead to better agreement between ISHA and large-$N_c$-HO. On the right hand panel, we show $F_{\rm med}$ as a function of $z$ for fixed $\theta=0.2$. We note that the $x$-axis ranges from 0.1 to 0.9. As expected, the SHA exhibits sizable deviations from the large-$N_c$-HO result, while the ISHA provides a good approximation in the intermediate-$z$ region. We further note that the deviations between ISHA and large-$N_c$-HO approaches are  larger than those observed in figure~\ref{fig:Fmed_qg_vs_theta} due to the longer trajectory length and smaller emitter energy used in the present figure.

\begin{figure}
\centering
\includegraphics[scale=0.39]{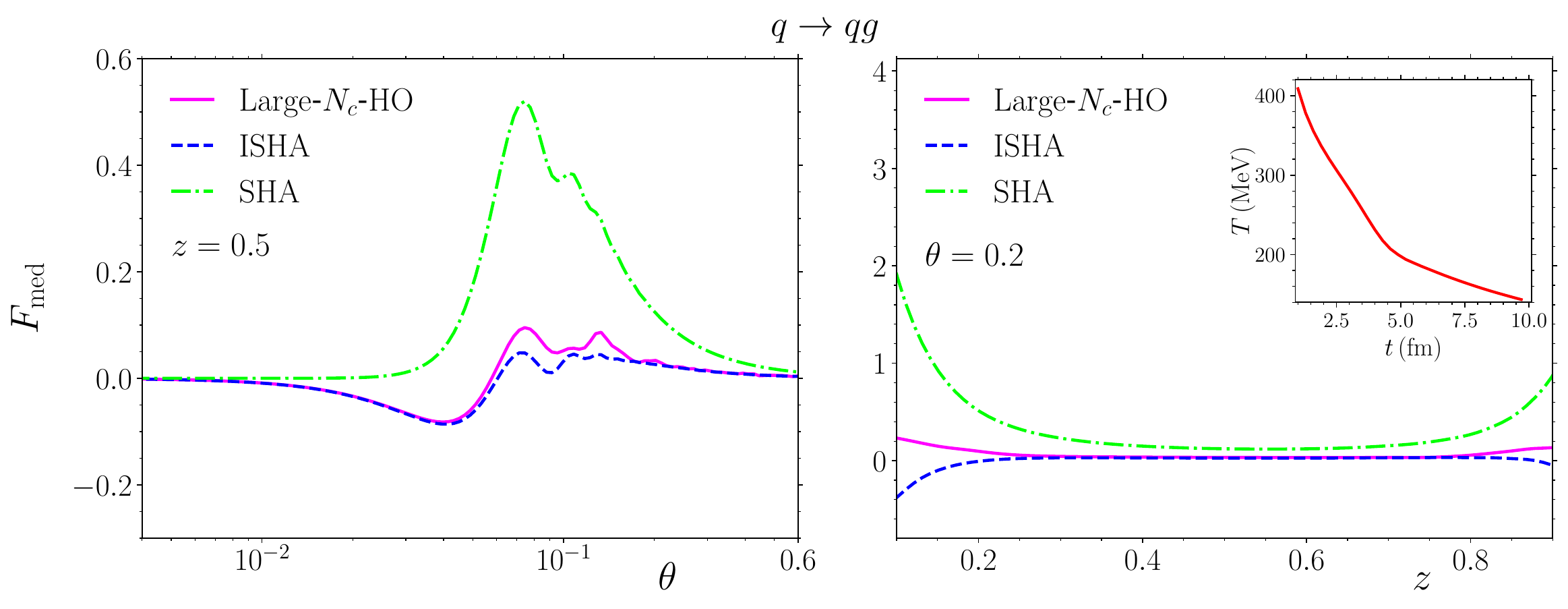}
\caption{Left panel: $F_{\rm med}$ for the $q\to qg$ splitting as a function of $\theta$ at fixed $z=0.5$, for $E=200$ GeV and $k=2$. Right panel: same as left panel as a function of $z$ at fixed $\theta=0.2$. The magenta solid curve corresponds to the large-$N_c$-HO, the blue dashed curve to the ISHA, and the green dash-dotted curve to the SHA, evaluated along the trajectory shown in the inset figure, which was sampled with a central production
point over the 0-10$\%$ centrality class in  $\sqrt{s_{ \rm NN}}=5.02$ TeV Pb-Pb collisions at the LHC.}
\label{fig:Fmed_qg_evolv}
\end{figure}

\section{Conclusions}
\label{sec:conclusions}

In this work, we have studied medium-induced $1 \to 2$  parton splittings with full dependence on both the energy fraction $z$ and the splitting angle $\theta$, within the BDMPS-Z framework for all-order resummation of multiple scatterings. Working in the large-$N_c$ limit, we presented general expressions for the double-differential in-medium splitting spectrum and showed that, under the harmonic oscillator approximation for the dipole cross section, all path integrals can be evaluated analytically for any splitting channel. This yields a computationally efficient semi-analytical result for the medium-induced modification of  $1 \to 2$ splittings, which we refer to as large-$N_c$-HO result and corresponds to the sum of eqs.~\eqref{eq:inoutho}, \eqref{eq:ininhofac}~and~\eqref{eq:ininhononfacexpl}. 
This approach thus provides a general and analytically tractable description of double-differential in-medium $1 \to 2$ splittings with multiple scatterings, going beyond the well-known soft ($z \to 0$) limit.

The medium-induced component of the splitting spectrum can be divided into the in-out and in-in contributions, with the latter commonly separated into factorizable and non-factorizable pieces \cite{Blaizot:2012fh}.  In agreement with previous results for the $\gamma \to q\bar{q}$ channel  \cite{Isaksen:2023nlr}, our analysis shows  that for $q \to qg$ and $g \to gg$ splittings, at large-$N_c$ and under the HO approximation, the non-factorizable contribution to the in-in term~\eqref{eq:ininhofac}, neglected in some analytical and phenomenological studies, plays a significant role away from the soft limit. In particular, for balanced splittings ($z \sim 0.5$), neglecting this term leads to unphysical oscillations in the spectrum, for both the $q \to qg$ and $g \to gg$ channels analyzed in this manuscript. These results, consistent with earlier findings for the $\gamma \to q\bar{q}$ channel obtained at finite $N_c$, but in a very restricted phase-space and parameters regime \cite{Isaksen:2023nlr}, show that neglecting the non-factorizable term may not be justified, especially when dealing with observables that are not soft-sensitive.

 We have also developed the improved semi-hard approximation (ISHA), which extends the SHA by consistently incorporating the leading corrections in inverse powers of the parton energies, following a strategy similar to that of~\cite{Altinoluk:2014oxa}. Appendix~A of~\cite{Dominguez:2019ges} outlined a possible strategy to incorporate some of these additional corrections, which were subsequently presented in~\cite{Andres:2024ksi} and applied to the calculation of the modification of the two-point energy correlator arsing from medium-induced radiation. However, we show in this manuscript that additional terms, entering at the same order as those considered there, were overlooked.  The ISHA consistently accounts for all contributions at this order, while its zeroth-order result recovers the original SHA presented in the main text of~\cite{Dominguez:2019ges,Isaksen:2020npj}

Through a comparison with the large-$N_c$-HO results  both for the $q \to qg$ and $g \to gg$ channels, we showed that, while the SHA significantly overestimates the medium-induced spectrum of emissions across most of phase space, even for relatively large emitter energies and symmetric splittings, the ISHA provides a much more robust approximation.  Its validity requires all partons involved in the splitting to be sufficiently energetic. For the HO form of the dipole cross section, the condition $\hat q L^2 \ll \min(z,1-z)E$ must be satisfied. In section 5, we showed that the ISHA  closely reproduces the large-$N_c$-HO results for $z_{\rm min} \lesssim z \lesssim 1 - z_{\rm min}$, with $z_{\rm min} \approx \hat q L^2/E$. Deviations become noticeable  for highly asymmetric splittings, where the validity condition is no longer satisfied and neither the SHA nor the ISHA is expected to provide a reliable approximation. Nevertheless, even in this regime, the deviations of the ISHA from the large-$N_c$ HO results are considerably smaller than those of the SHA.

The semi-analytical frameworks developed in this paper offer several advantages. The large-$N_c$-HO approach, being fully analytical under the harmonic oscillator approximation, can be directly applied to longitudinally evolving media by allowing $\hat q$ to vary with time, for arbitrary time dependence. The ISHA (and SHA) can similarly accommodate longitudinally expanding media. The principal advantage of the ISHA, however, is that it provides semi-analytical results for arbitrary choices of the dipole cross section, unlike the large-$N_c$-HO approach whose analyticity is tied to the harmonic oscillator form. This opens the door to perform phenomenological studies using more realistic interaction models, such as Yukawa \cite{Wang:1992qdg} or HTL \cite{Aurenche:2002pd} forms, as well as potentials extracted from lattice calculations at high temperatures \cite{Moore:2019lgw,Moore:2021jwe,Schlichting:2021idr}. Adopting a Yukawa form~\cite{Wang:1992qdg} would also enable a direct comparison with existing single-scattering results for in-medium splittings in~\cite{Ovanesyan:2011kn}. Furthermore, the ISHA can in principle be extended to include higher-order corrections in the energy expansion, providing a systematic pathway to further improve its accuracy. Although incorporating higher-order terms will inevitably make the expressions for the spectrum more cumbersome, we do not expect this to lead to a significant increase in computational cost.

It would nonetheless be valuable to develop a fully numerical approach at finite $N_c$, along the lines of ref.~\cite{Isaksen:2023nlr}, which is currently restricted to the $\gamma \to q \bar q$ splitting, in order to quantify the size of finite-$N_c$ corrections. In that work, such corrections were found to be very small for the  $\gamma \to q \bar q$ splitting, below $1\%$ across the entire $(\theta, z)$-plane, but their size in other channels remains less clear, and, for instance, within the restrictive SHA at finite $N_c$ they can reach up to $\sim 10\%$ for $g \to gg$ in certain regions of  phase space \cite{Isaksen:2020npj}. During the revision of this manuscript, a full numerical approach at finite $N_c$, extending~\cite{Isaksen:2023nlr} to all splitting channels, was developed \cite{Leitao:2026fgh}. This study finds that finite-$N_c$  corrections to our semi-analytical large-$N_c$-HO calculation remain below or at most reach $6\%$ across all splitting channels.

We finally note that the results developed here can have direct implications for energy correlator observables in high-energy nuclear collisions \cite{Moult:2025nhu,Andres:2022ovj, Barata:2023bhh,Yang:2023dwc,Xing:2024yrb,Bossi:2024qho,Singh:2024vwb,Apolinario:2025vtx,CMS:2025ydi,Andres:2024xvk,Fu:2024pic,Barata:2025fzd,Ke:2025ibt,Barata:2025uxp,Kudinoor:2026wcs}. Previous semi-analytical calculations of medium-induced contributions to EECs were either performed in the single-scattering limit  \cite{Andres:2022ovj, Andres:2023xwr,Andres:2023ymw,Fu:2024pic,Ke:2025ibt} or, when resumming multiple scatterings, relied on the SHA \cite{Barata:2024wsu,Andres:2022ovj, Andres:2023xwr,Andres:2023ymw} or its incomplete corrections \cite{Andres:2024ksi}, which, as we have shown, can significantly overestimate the medium-induced enhancement across large regions of phase space. Our large-$N_c$-HO results provide a more reliable baseline for multiple-scattering approaches, while the ISHA offers a practical framework to extend them to more realistic parton-medium interaction models, provided enough terms in this expansion are included to make sure is robust in the kinematic regime analyzed.

\acknowledgments
We thank Cyrille Marquet, Guilherme Milhano and Alba Soto-Ontoso for useful discussions. This project received funding from the Ecole Polytechnique Foundation. This work is funded by the European Union (ERC, QGPthroughEECs, grant agreement No.~101164102). Views and opinions expressed are however those of the authors only and do not necessarily reflect those of the European Union or the European Research Council. Neither the European Union nor the granting authority can be held responsible for them. This research was also in part supported by European Research Council project ERC-2018-ADG-835105 YoctoLHC, by Xunta de Galicia (CIGUS Network of Research Centres), by European Union ERDF, and by the Spanish Research State Agency under projects PID20231527\-62NB---I00 and CEX2023-001318-M financed by MICIU/AEI/10.13039/501100011033.   CA is grateful to the Kavli Institute for Theoretical Physics  for its  hospitality and support while this work was being completed. This research was supported in part by grant NSF PHY-2309135 to the Kavli Institute for Theoretical Physics (KITP).

\appendix
\section{$q \to qg$: variation with $\hat{q}$}
\label{app:qhat}

\begin{figure}[t]
\centering
\includegraphics[scale=0.39]{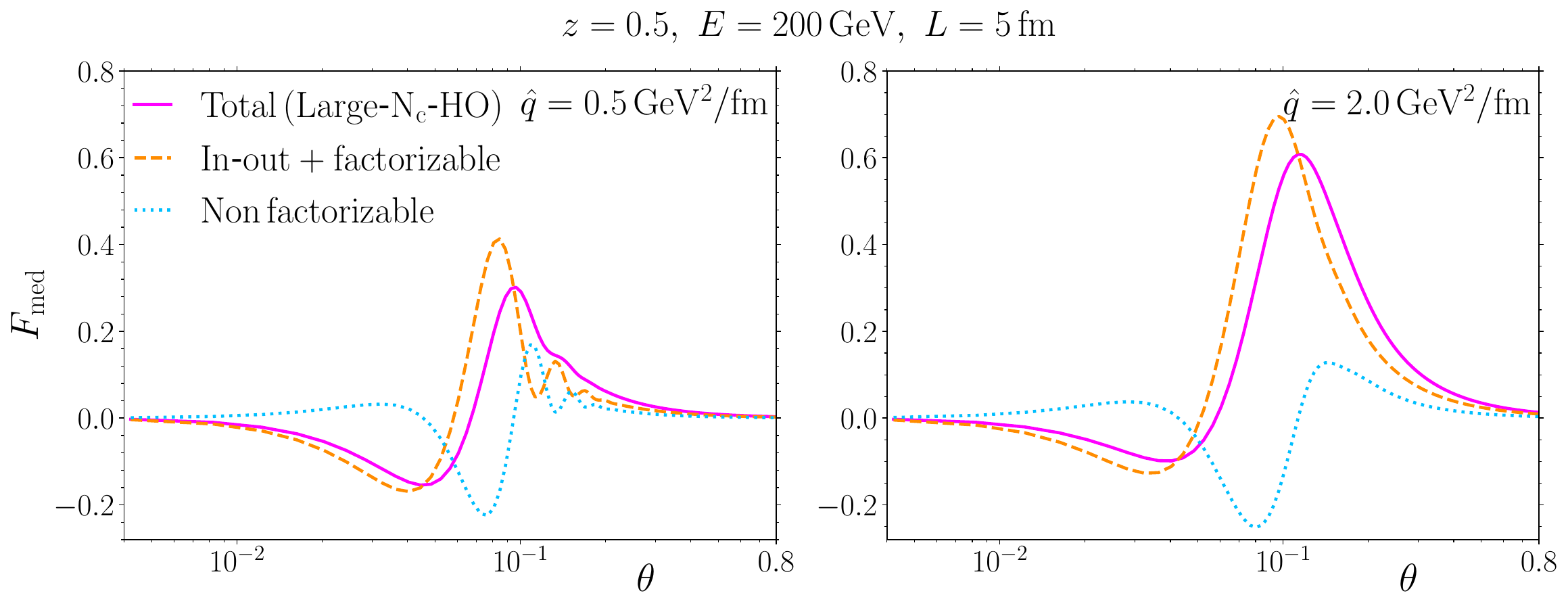}
\caption{$F_{\rm med}$ for the $q\to qg$ splitting as a function of $\theta$, computed in the large-$N_c$-HO approach (see section~\ref{sec:HO}). Solid magenta curves show the full large-$N_c$–HO result, orange dashed curves correspond to the result without the non-factorizable term, and blue dotted curves show the non-factorizable term alone. Both panels correspond to $E=200$\,GeV, $z=0.5$, and $L=5$\,fm. The left (right) panel corresponds to $\hat{q}=0.5\,\mathrm{GeV}^2/\mathrm{fm}$ ($\hat{q}=2.0\,\mathrm{GeV}^2/\mathrm{fm}$).}
\label{fig:Fmed_qg_LNC_qhat}
\end{figure}

\begin{figure}[h]
\centering
\includegraphics[scale=0.38]{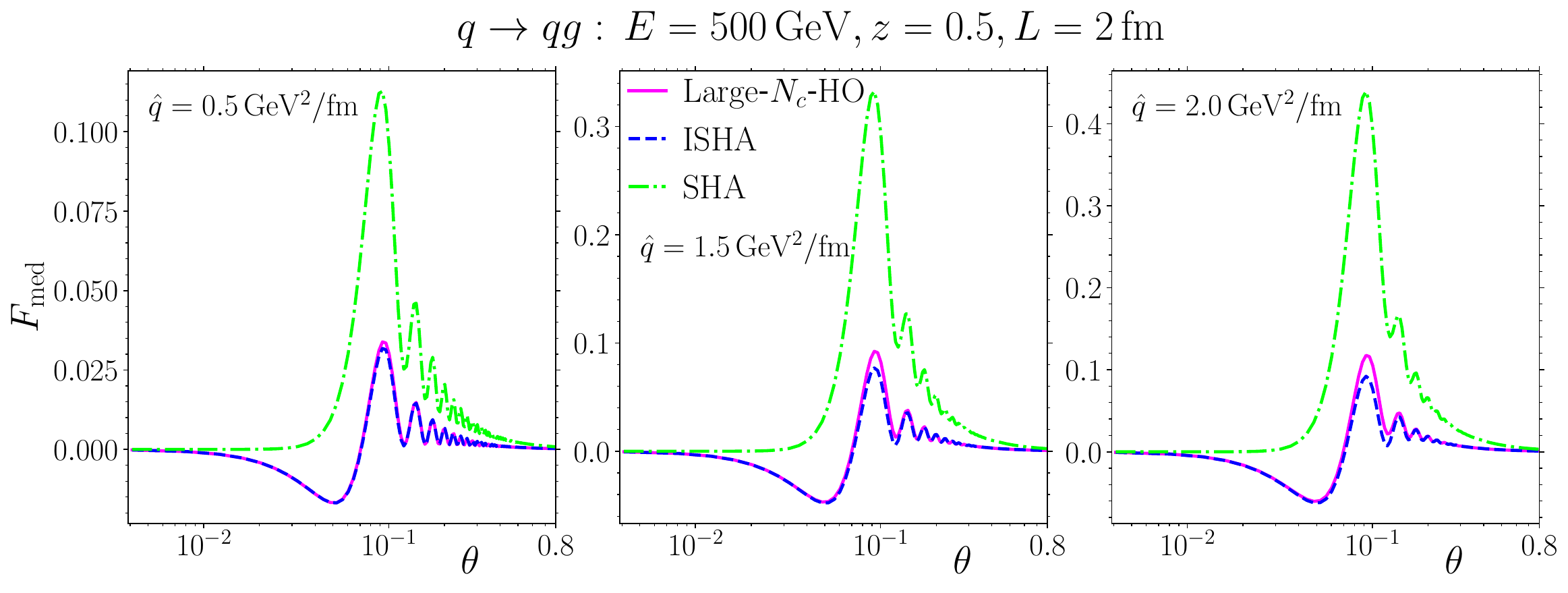}
\caption{$F_{\rm med}$ for the $q\to qg$ as a function of $\theta$, computed in the large-$N_c$-HO approach (magenta solid), ISHA (blue dashed), and SHA (green dash-dotted). All panels correspond to $E=500$\, GeV, $z=0.5$, and $L=2$\,fm. The left, center, and right panels correspond to $\hat{q}=0.5$, $1.5$, and $2.0,\mathrm{GeV}^2/\mathrm{fm}$, respectively.}
\label{fig:Fmed_qg_qhat}
\end{figure}

In the main text, we present results for $F_{\rm med}$ in the $q \to qg$ channel, computed in the SHA, ISHA, and large-$N_c$-HO approaches for various emitter energies and medium lengths. All results shown there are obtained for $\hat{q} = 1\,\mathrm{GeV}^2/\mathrm{fm}$. In this appendix, we provide complementary results for different values of $\hat{q}$.

In figure~\ref{fig:Fmed_qg_LNC_qhat}, we show $F_{\rm med}$ for the $q\to qg$ in the large-$N_c$-HO approach (magenta solid) and its decomposition into  the in-out plus factorizable in-in contribution (orange dashed) and the non-factorizable in-in term (blue dotted), as in figure~\ref{fig:Fmed_qg_LNC}. Both panels correspond to $z=0.5$ and share the same emitter energy and medium length, differing only in the value of $\hat{q}$. As $\hat{q}$ increases (moving from the left panel to the right one ), the relative weight of the non-factorizable contribution decreases, although it remains non-negligible even for the largest $\hat{q}$-value considered.

In figure~\ref{fig:Fmed_qg_qhat}, we present $F_{\rm med}$ as a function of $\theta$ for $z = 0.5$, comparing the large-$N_c$-HO approach (magenta solid), ISHA (blue dashed), and SHA (green dash-dotted). Each panel corresponds to a different value of $\hat{q}$, with all other parameters held fixed. In agreement with the main text (see figure~\ref{fig:Fmed_qg_vs_theta}), the SHA is not a reliable approximation of the large-$N_c$-HO result, even for high emitter energies and symmetric splittings. In contrast, the ISHA closely reproduces the large-$N_c$-HO result for this set of parameters, with only minor deviations that grow slightly as $\hat{q}$ increases.

\section{$g \to gg$ splitting}
\label{app:gg}

\begin{figure}
\centering
\includegraphics[scale=0.40]{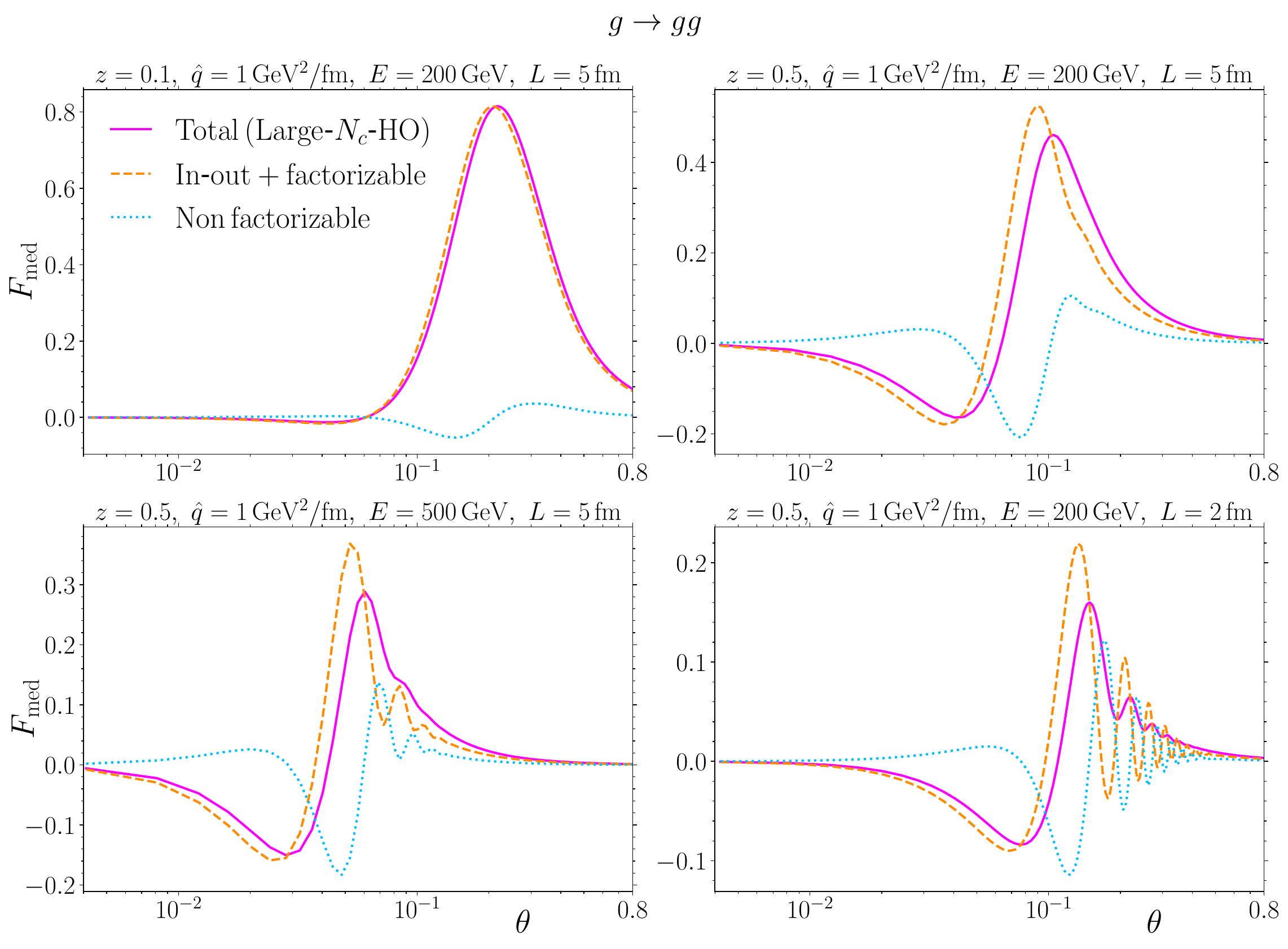}
\caption{$F_{\rm med}$ for $g \to gg$ in the  large-$N_c$-HO approach (see section~\ref{sec:HO}). Solid magenta curves show the full large-$N_c$–HO result, orange dashed curves correspond to the result without the non-factorizable term, and blue dotted curves show the non-factorizable term alone. The top-left and top-right panels correspond to $z=0.1$ and $z=0.5$, respectively, for the same emitter energy and medium parameters (indicated in the figure). The bottom panels show $z=0.5$ for a higher emitter energy (left) and a shorter medium length (right).}
\label{fig:Fmed_gg_LNC}
\end{figure}

While the main text focuses on the $q \to qg$ splitting, this appendix presents representative results for the $g \to gg$ channel, showing that the different approximation schemes behave in the same way across both channels.

Figure~\ref{fig:Fmed_gg_LNC} is the analogue of  figure~\ref{fig:Fmed_qg_LNC} for the $g \to gg$ case. As in the $q \to qg$ results,the top-left panel of figure~\ref{fig:Fmed_gg_LNC} shows that the non-factorizable contribution is negligible in the soft limit ($z \ll 1$). However, for $z = 0.5$, the non-factorizable term becomes significant, and its relative importance grows with increasing emitter energy or decreasing medium length.

\begin{figure}
\centering
\includegraphics[scale=0.40]{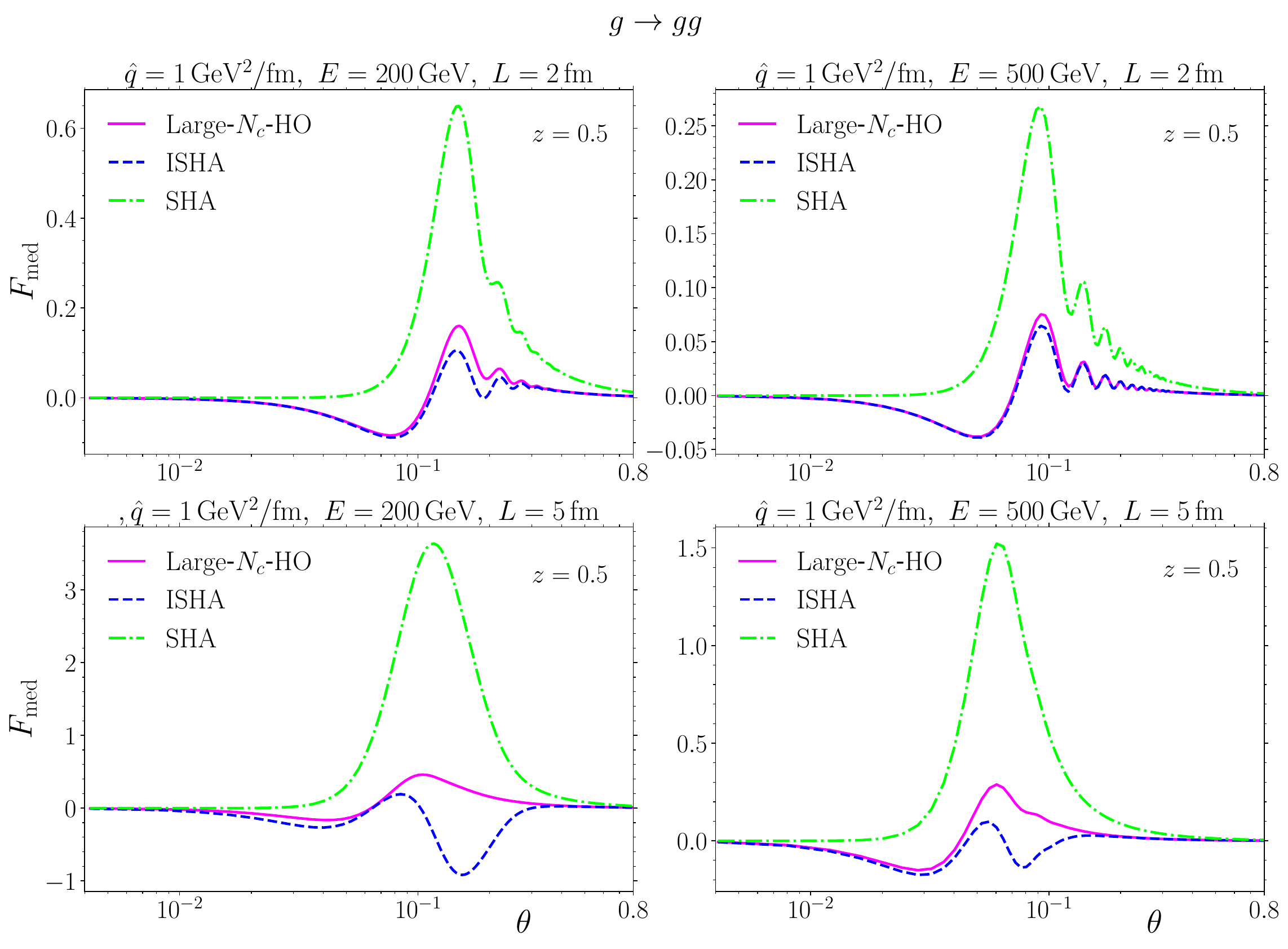}
\caption{$F_{\rm med}$ for the $g \to gg$ splitting as a function of $\theta$ for $z=0.5$, computed in the large-$N_c$-HO (magenta solid), ISHA (blue dashed), and SHA (green dash-dotted) approaches. Different panels correspond to different emitter energies $E$ and medium lengths $L$, as indicated in each of them.}
\label{fig:Fmed_gg_vs_theta}
\end{figure}

Figure~\ref{fig:Fmed_gg_vs_theta} shows $F_{\rm med}$ for the $g \to gg$ splitting as a function of $\theta$ at fixed $z=0.5$. All the parameters in this figure are chosen to match those in \ref{fig:Fmed_qg_vs_theta}. The conclusions mirror those for the $q \to qg$ channel: the SHA fails to reproduce the large-$N_c$-HO result across all panels, while the ISHA closely approximates the large-$N_c$-HO result for sufficiently high emitter energies or short medium lengths.

\begin{figure}
\centering
\includegraphics[scale=0.34]{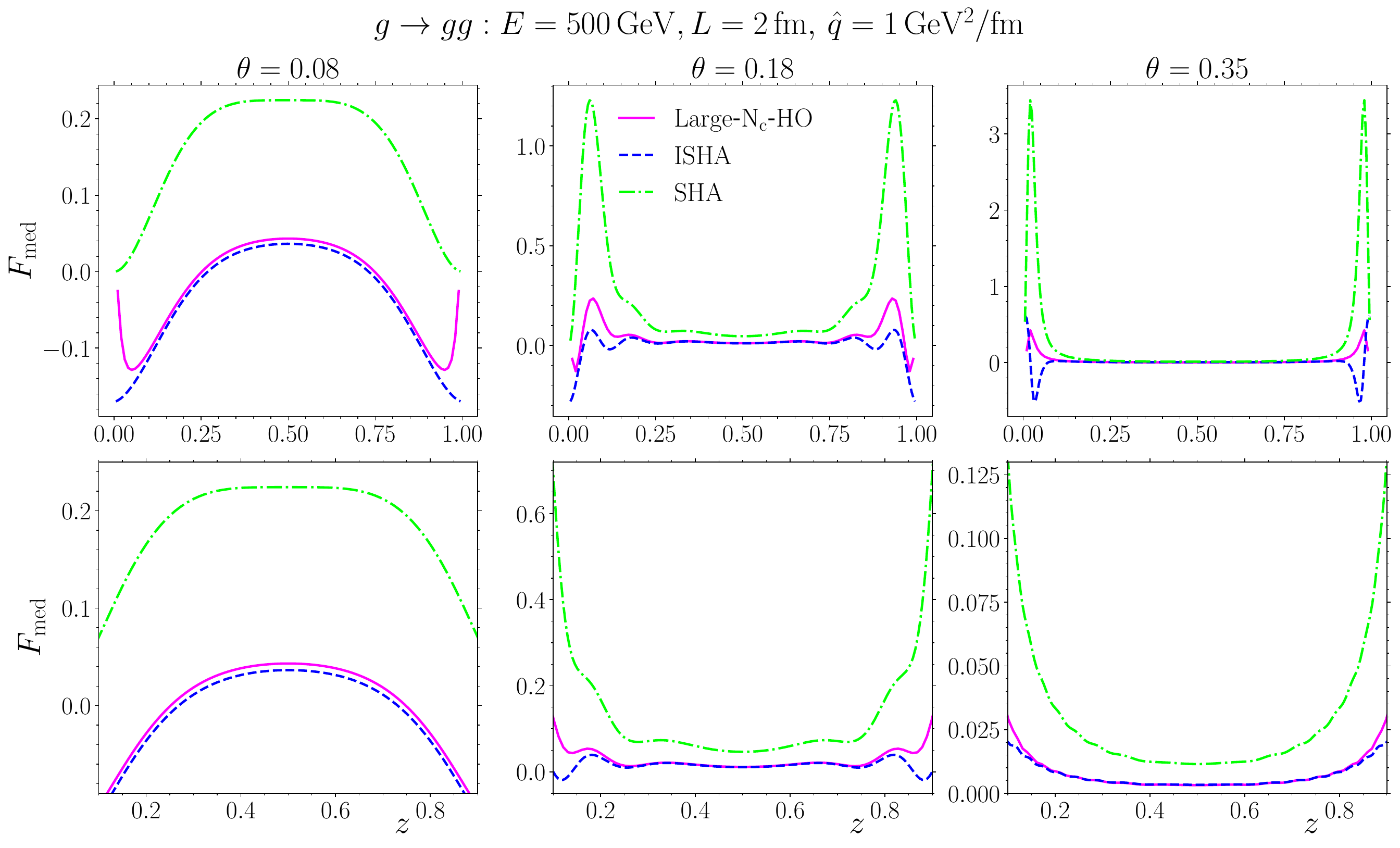}
\caption{$F_{\rm med}$ for the $g\to gg$ splitting as a function of $z$,  computed using the large-$N_c$-HO (magenta solid), ISHA (blue dashed), and SHA (green dash-dotted),  for $E=500 $\, GeV, $L=2$\,fm, and $\hat {q}=1\,{\rm GeV^2/fm}$. The left, center, and right columns correspond to $\theta=0.08$, $0.18$, and $0.35$, respectively. The bottom row shows a zoom of the top row in the region $0.1 \leq z \leq 1$.}
\label{fig:Fmed_gg_vs_z}
\end{figure}

Figure~\ref{fig:Fmed_gg_vs_z} shows $F_{\rm med}$ for the $g \to gg$ splitting as a function of $z$ for three values of $\theta$. All parameters and $\theta$ values are chosen to match those of the corresponding figure for the $q \to qg$ channel (see figure~\ref{fig:Fmed_qg_vs_z}). The same behavior emerges in both splittings: the SHA fails to provide a good description of the large-$N_c$-HO result, even for intermediate values of $z$, whereas the ISHA offers a good approximation except for very asymmetric splittings. We note that, unlike in the $q \to qg$ case, $F_{\rm med}$ for $g \to gg$ is symmetric in $z$, as expected.

For completeness, figure~\ref{fig:Fmed_gg_qhat} shows the $\hat{q}$ dependence of $F_{\rm med}$ for $g \to gg$ splittings. The figure displays $F_{\rm med}$ as a function of $\theta$ at $z = 0.5$, comparing the large-$N_c$-HO approach (magenta solid), ISHA (blue dashed), and SHA (green dash-dotted). The emitter energy and medium length are kept fixed across all panels, with each panel corresponding to a different value of $\hat{q}$. All parameters are chosen to match those used in figure~\ref{fig:Fmed_qg_qhat} for the $q \to qg$ channel. As in the $q \to qg$ case discussed in appendix~\ref{app:qhat}, the agreement between the large-$N_c$-HO and the ISHA slightly deteriorates as $\hat{q}$ increases.

\begin{figure}
\centering
\includegraphics[scale=0.38]{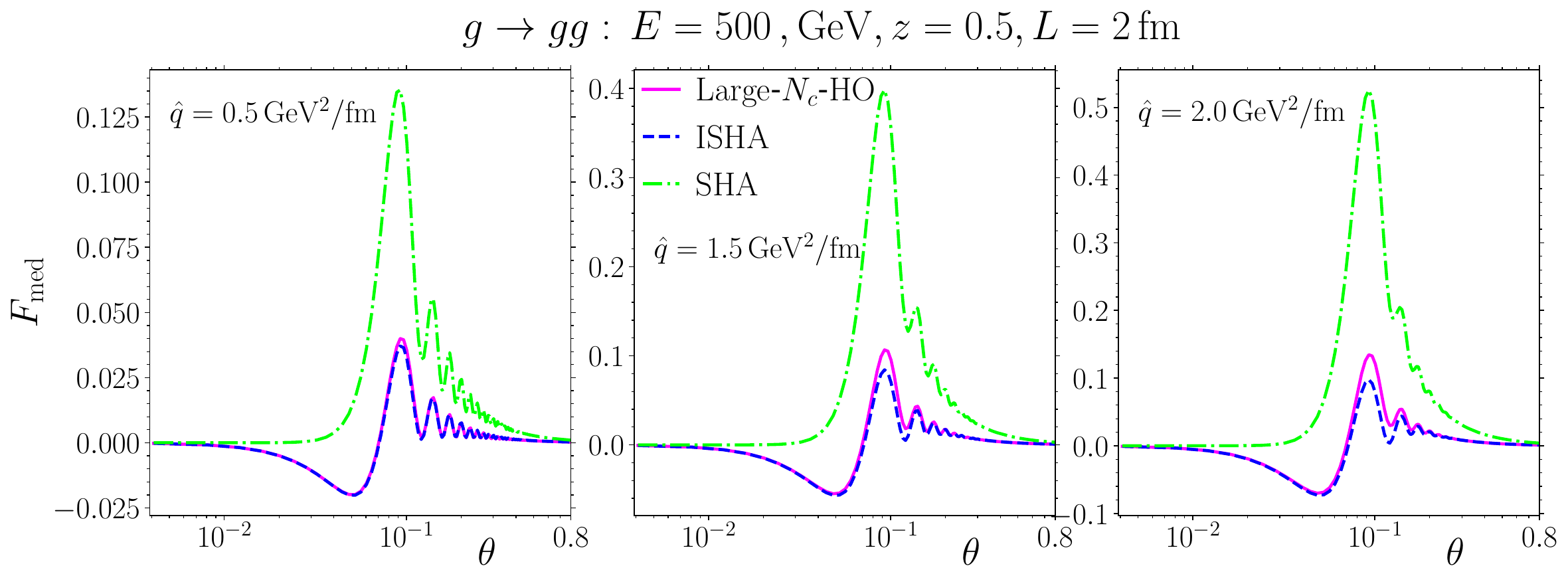}
\caption{$F_{\rm med}$ for the $g \to gg$ splitting as a function of $\theta$, computed using the large-$N_c$-HO (magenta solid), ISHA (blue dashed), and SHA (green dash-dotted) approaches, for $E=500$\,GeV, $z=0.5$, and $L=2$\,fm. The left, center, and right panels correspond to $\hat{q}=0.5$, $1.5$, and $2.0\,{\rm GeV^2/fm}$, respectively.}
\label{fig:Fmed_gg_qhat}
\end{figure}

Finally, in figure~\ref{fig:Fmed_gg_evolv}, we present
$F_{\rm med}$ for the $g \to gg$ splitting within the SHA, ISHA and large-$N_c$-HO,  evaluated along the same trajectory used in figure~\ref{fig:Fmed_qg_evolv}. This trajectory is sampled from the 0-10$\%$ centrality class in $\sqrt{s_{ \rm NN}}=5.02$ TeV Pb-Pb collisions and is shown in the inset. As in figure~\ref{fig:Fmed_qg_evolv}, we set $k=2$ and the emitter energy to $E=200$ GeV. The left panel presents $F_{\rm med}$ as a function of $\theta$ at fixed $z=0.5$, while the right panel $F_{\rm med}$ as a function of $z$ at fixed $\theta=0.2$ As in the $q\to qg$ case, the SHA shows sizable deviations from the large-$N_c$-HO result in both panels, whereas the ISHA provides a much more robust approximation at intermediate values of $z$.

\begin{figure}
\centering
\includegraphics[scale=0.39]{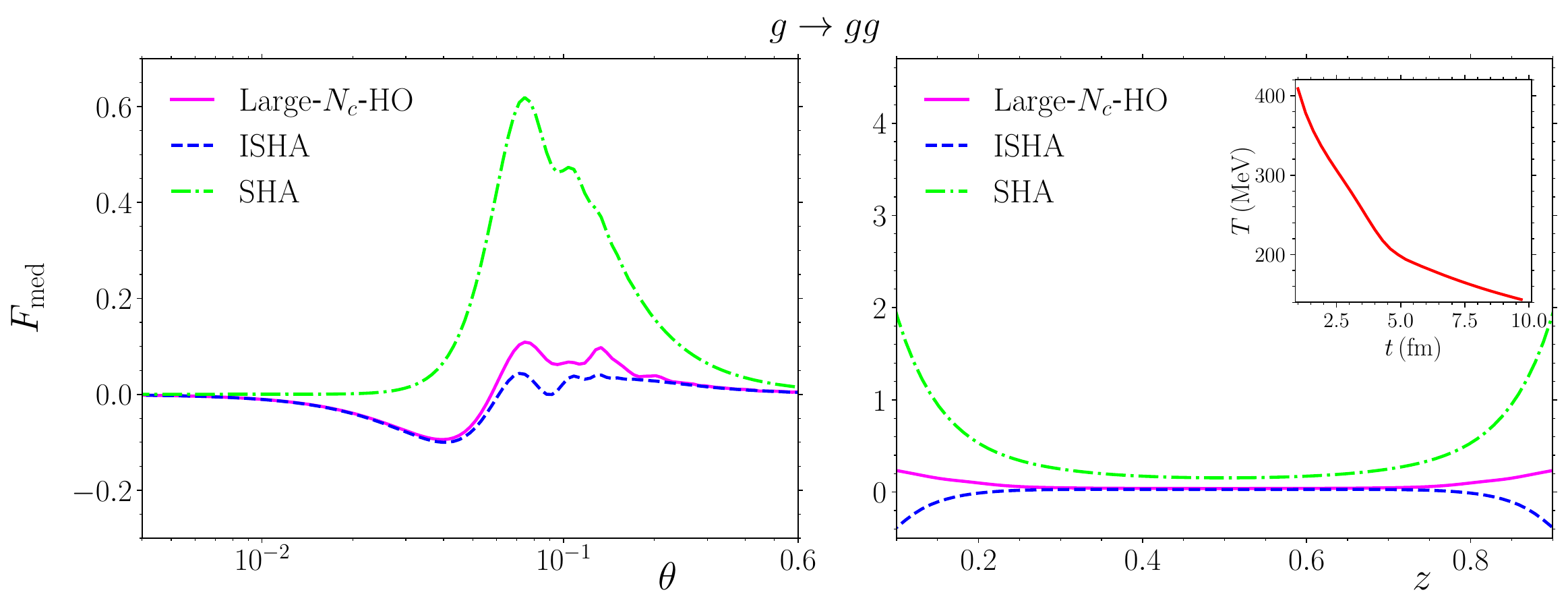}
\caption{Left panel: $F_{\rm med}$ for the $g\to gg$ splitting as a function of $\theta$ at fixed $z=0.5$, for $E=200$ GeV and $k=2$. Right panel: same as left panel as a function of $z$ at fixed $\theta=0.2$. The magenta solid curve corresponds to the large-$N_c$-HO, the blue dashed curve to the ISHA, and the green dash-dotted curve to the SHA, evaluated along the trajectory shown in the inset figure, which was sampled with a central production
point over the 0-10$\%$ centrality class in  $\sqrt{s_{ \rm NN}}=5.02$ TeV Pb-Pb collisions at the LHC.}
\label{fig:Fmed_gg_evolv}
\end{figure}

\bibliographystyle{JHEP}
\bibliography{references}

\end{document}